\documentclass[12pt]{article}
\usepackage{epsf}
\usepackage{bm}
\usepackage{cite}
\usepackage{color}
\usepackage{amsmath,amssymb}
\usepackage{graphicx}
\usepackage{physics}

\usepackage{comment}
\usepackage{float}
\usepackage{dsfont}

\usepackage{subcaption}

\newcommand{\setR}{\mathbb{R}}

\definecolor{CopenhagueRed}{RGB}{236, 125, 128}
\definecolor{CopenhagueBlue}{RGB}{168, 207, 239}
\definecolor{CopenhagueYellow}{RGB}{249, 215, 148}

\definecolor{CopenhagueGreen}{RGB}{89, 149, 152}
\definecolor{CopenhagueDarkBlue}{RGB}{43, 103, 162}

\definecolor{CopenhagueSand}{RGB}{219, 221, 208}
\definecolor{CopenhagueOrange}{RGB}{243, 170, 138}

\definecolor{CopenhagueSaturatedYellow}{RGB}{249, 199, 100}
\definecolor{CopenhagueSaturatedBlue}{RGB}{118, 185, 239}

\definecolor{CopenhagueSaturatedSand}{RGB}{207, 212, 174}
\definecolor{CopenhagueSaturatedOrange}{RGB}{242, 156, 119}

\usepackage[hidelinks,bookmarks=true,bookmarksdepth=4]{hyperref} 
\hypersetup{
  colorlinks=true,
    citecolor=CopenhagueSaturatedBlue, 
    linkcolor=blue,
    urlcolor=CopenhagueRed}

\renewcommand{\thefootnote}{\fnsymbol{footnote}}
\def\thefootnote{\fnsymbol{footnote}}

\makeatletter

\@addtoreset{equation}{section}
\makeatother

\begin{document}

\begin{titlepage}

\begin{center}

\vskip .45in

{\Large \bf Quantum state of interacting primordial \vspace{5pt} \\
inhomogeneities: de-squeezing and decoherence}

\vskip .65in

{\large 
Amaury~Micheli$^{1}$,
Yuto~Oshima$^{2}$, and 
Tomo~Takahashi$^{3}$ \vspace{2mm} \\
}
\vskip 0.2in

{\em 
$^{1}$RIKEN Center for Interdisciplinary Theoretical and Mathematical Sciences (iTHEMS), Wako, Saitama 351-0198, Japan
\vspace{2mm}\\
$^{2}$Graduate School of Science and Engineering, Saga University, Saga 840-8502, Japan
\vspace{2mm}\\
$^{3}$Department of Physics, Saga University, Saga 840-8502, Japan 

}

\end{center}
\vskip .5in

\begin{abstract}
We investigate how interactions affect the quantum state of scalar perturbations during inflation and the quantum correlations they may exhibit. Focusing on the case of scalar perturbations in single-field inflation, we model interactions using a Lindblad equation with a non-unitary contribution quadratic in the scalar perturbations, and of parametrisable amplitude and time dependence. We compute the quantum state of these interacting perturbations, which is fully described by its purity and squeezing parameters. First, we show that, in most of the parameter space, not only the purity but also the squeezing parameter is significantly reduced by interactions. Second, we show that this de-squeezing induced by the interactions, on top of the purity loss, causes a further suppression of quantum correlations. We thus emphasise that the quantum or classical character of the correlations exhibited by the perturbations cannot be correctly determined by computing the effect of interactions on the purity alone. Since the phenomenological framework adopted in this paper encompasses a wide class of possible interactions, our results provide general insights into the nature of decoherence processes in primordial fluctuations.
\end{abstract}

\end{titlepage}

\renewcommand{\thepage}{\arabic{page}}
\setcounter{page}{1}
\renewcommand{\thefootnote}{\#\arabic{footnote}}
\setcounter{footnote}{0}

\section{Introduction} \label{sec:intro}

The gravitational structures of the Universe, including galaxy distributions and cosmic microwave background (CMB) anisotropies, are believed to originate from vacuum fluctuations amplified during inflation~\cite{Guth:1980zm,Mukhanov:1981xt}. Although these initial fluctuations are quantum, the inhomogeneities they generate are treated as mere classical perturbations in every subsequent analyses, e.g., in CMB anisotropy predictions. However, their classicalisation process remains elusive, and a lot of effort has been made to understand how their quantum origin can manifest~\cite{Albrecht:1992kf}. Recent approaches have attempted to quantify the degree of quantum correlations in the distribution of perturbations during inflation. They used tools borrowed from quantum information, such as non-separability~\cite{Campo:2005sy,Martin:2022kph}, Bell inequalities~\cite{Campo:2005sv,Martin:2017zxs,Sou:2024tjv}, quantum discord~\cite{Martin:2015qta,Martin:2021znx}, and others~\cite{Campo:2005sy,Nambu:2008my}. Evaluating these quantities requires computing the quantum state of the perturbations. In the simplest inflationary dynamics of single-field slow-roll inflation~\cite{Bassett:2005xm}, both scalar and tensor perturbations are a collection of independent pairs of opposite momenta $\pm \bm{k}$ perturbations in two-mode squeezed vacuum states~\cite{Grishchuk:1990bj}, fully characterised by their squeezing parameters $r_k$ and $\varphi_k$. However, this picture neglects the effects of the perturbations' interactions with themselves and other degrees of freedom, which are expected to play a key role in their classicalisation process. Interactions will correlate the otherwise independent $\pm \bm{k}$ pairs with other degrees of freedom, referred to as the environment, thereby weakening their bi-partite quantum correlations, a process known as decoherence~\cite{Zurek:1981xq,Zurek:1982ii,Joos:1984uk}. To describe the decoherence of the perturbations without having to solve their joint dynamics with a specific environment, we can treat the perturbations as an \textit{open} quantum system with non-unitary dynamics. Numerous works have explored different environments for the cosmological perturbations and different types of interactions~\cite{Matacz:1992tp,Kiefer:1998qe,Campo:2005sy,Burgess:2006jn,Burgess:2022nwu,Sou:2022nsd,Lopez:2025arw,Burgess:2025dwm}. Beyond the specifics of the decoherence processes considered, as long as the state remains close to Gaussian, homogeneous, and isotropic~\cite{Campo:2005sy}, the state of a pair $\pm \bm{k}$ of perturbations can still be fully described using the squeezing parameters $r_k$, the squeezing angle $\varphi_k$, and the purity $p_k$\cite{Martin:2021qkg}. We refer collectively to these quantities as the effective squeezing parameters. In\cite{Martin:2021znx}, the authors quantified the quantum correlations of the scalar perturbations by expressing their quantum discord as a function of these parameters. They then analysed how the specific combination of parameters entering the discord is affected by decoherence in a phenomenological model of interactions based on the Lindblad equation~\cite{Lindblad:1975ef,Breuer:2015zlm,Manzano:2020yyw}. In this paper, building on the framework developed in~\cite{Martin:2021znx}, we explore how the effective squeezing parameters themselves are affected by interactions modeled with the equations. Interestingly, we first demonstrate that while interactions generically reduce the purity $p_k$ as expected, they also typically reduce the squeezing parameter $r_k$, i.e., they de-squeeze the perturbations. Using these results, we clarify how the aforementioned quantum measures are affected by decoherence through these two separate effects: a decreased purity that classicalises correlations and a reduced squeezing parameter that diminishes the overall level of correlation. As an illustration, we compute the values of several quantifiers of correlations using the effective squeezing parameters $r_k$, $\varphi_k$, and $p_k$ that we derived for the model of~\cite{Martin:2021znx}. We then compare these values to those obtained if we assume that interactions only affect purity $p_k$ while leaving the squeezing parameters $r_k$ and $\varphi_k$ unchanged, and show they differ significantly.

The organisation of this paper is as follows. In Sec.~\ref{sec:formalism}, we review the effective squeezing parameters and covariance matrix formalism used to describe the state of the curvature perturbations, as well as the Lindblad equation we use to model their dynamics. In Sec.~\ref{sec:deco_squeezing}, we derive, both numerically and analytically, the evolution of the effective squeezing parameters for this open-system dynamics, which constitutes the first main result of this paper. In Sec.~\ref{sec:quantum_corr}, based on these expressions, we compare the degree of quantum correlations as measured by three different quantifiers (quantum discord, Bell operator, and logarithmic negativity) extending~\cite{Martin:2021znx}, which forms the second main result of this paper. We end by discussing some possible extensions of our analyses.

Throughout this paper, we adopt natural units and set $c= \hbar = 1$, unless otherwise stated.

\section{Quantum state of curvature perturbations in inflation \label{sec:formalism}}

In this section, we recap the necessary background for the discussion of quantum correlations of a field in curved spacetime, applied to cosmological perturbations during inflation.

\subsection{Free perturbations}

We are interested in the evolution of the quantum state of cosmological perturbations during a period of slow-roll inflation. We focus on scalar perturbations in this work, although the same formalism can be equally well applied to tensor perturbations; see, e.g.~\cite{Micheli:2022tld} for a review. We assume a spatially flat Universe and describe the inflationary Universe at the background level with a flat FLRW metric,
\begin{equation}
\dd s^2 = a\left( \eta \right)^2 \left( - \dd \eta^2 + \dd \bm{x}^2 \right) \, ,
\end{equation}
where $\eta$ denotes conformal time and $a(\eta)$ the scale factor. Working in perturbation theory, in single-field slow-roll inflation, the scalar part of the metric and inflaton perturbations can be fully packaged into a single gauge-invariant scalar field $\hat{v}$, the Mukhanov--Sasaki field~\cite{Mukhanov:1990me}. To describe the dynamics using the Hamiltonian formalism, we choose its canonically conjugate field to be $\hat{p} = \partial_{\eta} \hat{v}$, such that $[\hat{v}(\bm{x}), \hat{p}(\bm{x}^{\prime})] = i \delta (\bm{x} - \bm{x}^{\prime})$\footnote{
This choice is a matter of convention: it does not change the quantum expectation values of observables. However, it does change how these values are expressed in terms of squeezing parameters, and thus the values of these parameters; see Sec.~2.2 of~\cite{Micheli:2023qnc}. Here we follow the conventions of~\cite{Martin:2021znx,Burgess:2022nwu}.
}.
Since the dynamics is homogeneous and isotropic, it is convenient to work in Fourier space. The Fourier transform of $\hat{v}$ is defined by
\begin{align}
\label{eq:MSfourier}
\hat{v}\left(\bm{x}\right) = \frac{1}{\left(2\pi\right)^{3/2}}
\int_{\setR^3}\dd^3\bm{k}\, e^{-i\bm{k}\cdot\bm{x}} \hat{v}_{\bm{k}} \, ,
\end{align}
and similarly for $\hat{p}\left(\bm{x}\right)$. At first order in perturbation theory, the dynamics of the perturbation is linear~\cite{Mukhanov:1990me} and is thus described by a quadratic Hamiltonian, which in our conventions reads~\cite{Martin:2021znx}
\begin{align}
\label{eq:Hsystem}
\hat{H} = \int_{\setR^{3+}} \dd^3 \bm{k} \, \hat{\mathcal{H}}_{\pm \bm{k}}
= \int_{\setR^{3+}} \dd^3 \bm{k} \left[
  \hat{p}_{\bm{k}} \hat{p}_{\bm{k}}^{\dagger}
  + \omega^2\left(k, \eta \right)
  \hat{v}_{\bm{k}} \hat{v}_{\bm{k}}^{\dagger} \right] \, ,
\end{align}
with
\begin{align}
\label{eq:defomega}
\omega^2\left(k,\eta\right) =
k^2 - \frac{\partial_{\eta}^2 \left(a \sqrt{\epsilon_1}\right)}
{a \sqrt{\epsilon_1}} \, ,
\end{align}
where $\setR^{3+} = \{ \bm{k} \in \setR^{3} \mid k_{z} \geq 0 \}$, $\epsilon_1 = 1 - \partial_{\eta} \mathcal{H} / \mathcal{H}^2$ is the first slow-roll parameter controlling how close we are to a de Sitter phase, and $\mathcal{H} = (\partial_{\eta} a)/a$. Let us comment on the form of the Hamiltonian. First, note that since $\hat{v}(\bm{x})$ and $\hat{p}(\bm{x})$ are Hermitian, their Fourier amplitudes for $\pm \bm{k}$ are not independent, and $\hat{v}^{\dagger}_{\bm{k}} = \hat{v}_{- \bm{k}}$, $\hat{p}^{\dagger}_{\bm{k}} = \hat{p}_{- \bm{k}}$. Thus, the sum is only performed over half of the entire space of modes, $\mathbb{R}^{3+}$, and $\hat{\mathcal{H}}_{\pm \bm{k}}$, which can be checked to be Hermitian, encodes the dynamics of both modes $\pm \bm{k}$. Second, $\hat{H}$ is a sum of Hamiltonians for the different opposite-momentum pairs $\pm \bm{k}$, the dynamics of each of these pairs is thus independent of the others. Finally, the form of Eq.~\eqref{eq:Hsystem} is, up to canonical transformations, completely determined by the requirements of linearity, homogeneity, and isotropy~\cite{Martin:2021znx}.
The density matrix of the system, $\hat{\rho}(\eta)$, then obeys the Liouville--von Neumann equation
\begin{equation}
    i \frac{\mathrm{d} \hat{\rho}}{\mathrm{d} \eta} = \left[ \hat{H}, \hat{\rho} \right] \, .
\end{equation}

\subsubsection{Dynamics and partitions}
\label{sec:dyn_part}

Once an initial state is chosen, the state of the cosmological perturbations is fully determined by solving the Schr\"{o}dinger equation for the Hamiltonian~\eqref{eq:Hsystem}. However, discussing the presence of quantum correlations in this state requires us to further specify which degrees of freedom we intend to study the correlations of. At the quantum level, this corresponds to partitioning the Hilbert space of the system into a direct sum of Hilbert subspaces. Each of these subspaces is characterised by a pair of canonically conjugate Hermitian operators. There are infinitely many ways to do so, and the appropriate ones ultimately depend on the observables measured~\cite{Martin:2021znx,Agullo:2022ttg}. Yet, some partitions play an important role either because of their physical meaning or because they allow for simple computations.

Here, let us describe three such partitions. First, we have the defining real space partition $\mathrm{H} = \oplus_{\setR^3} \mathrm{H}_{\bm{x}}$, where the field is viewed as a collection of $\setR^3$ local bosonic modes represented by the pairs $(\hat{v}(\bm{x}),\hat{p}(\bm{x}))$.
Since the dynamics is best described in Fourier space, where it separates into independent sectors of opposite-momentum pairs $\pm \bm{k}$, we would like to write an adapted partition of the Hilbert space. Yet, the relations $\hat{v}^{\dagger}_{\bm{k}} = \hat{v}_{- \bm{k}}$, $\hat{p}^{\dagger}_{\bm{k}} = \hat{p}_{- \bm{k}}$ imply that $\hat{v}_{\bm{k}}$ are not Hermitian operators and not independent degrees of freedom~\cite{Burgess:2022nwu}. A first way to define canonical pairs from these operators is to consider their Hermitian and anti-Hermitian parts (up to a factor). Following~\cite{Martin:2021znx}, we define
\begin{align}
\label{eq:vRI:def}
\hat{v}_{\bm{k}}^{\mathrm{R}} &=
\frac{\hat{v}_{\bm{k}}+\hat{v}_{\bm{k}}^\dagger}{\sqrt{2}} \, ,\qquad
\hat{v}_{\bm{k}}^{\mathrm{I}} =
\frac{\hat{v}_{\bm{k}}-\hat{v}_{\bm{k}}^\dagger}{\sqrt{2} i} \, , \\
\hat{p}_{\bm{k}}^{\mathrm{R}} &=
\frac{\hat{p}_{\bm{k}}+\hat{p}_{\bm{k}}^\dagger}{\sqrt{2}} \, ,\qquad
\hat{p}_{\bm{k}}^{\mathrm{I}} =
\frac{\hat{p}_{\bm{k}}-\hat{p}_{\bm{k}}^\dagger}{\sqrt{2} i} \, .
\end{align}
One can check that these operators are Hermitian and that the sectors $\mathrm{R},\mathrm{I}$ are indeed independent, as $[\hat{v}_{\bm{k}}^{\mathrm{R}}, \hat{p}_{\bm{k}^{\prime}}^{\mathrm{I}}] = [\hat{v}_{\bm{k}}^{\mathrm{R}}, \hat{v}_{\bm{k}^{\prime}}^{\mathrm{I}}] = [\hat{p}_{\bm{k}}^{\mathrm{R}}, \hat{p}_{\bm{k}^{\prime}}^{\mathrm{I}}] = 0$. Both satisfy canonical commutation relations,
$[\hat{v}_{\bm{k}}^{\mathrm{s}}, \hat{p}_{\bm{k}^{\prime}}^{\mathrm{s}^{\prime}}] = i \delta(\bm{k} - \bm{k}') \delta_{s,s'}$
with $s, s' = \mathrm{R,I}$. The Hilbert space can then be decomposed as $\mathrm{H} = \oplus_{\setR^{3+}} \mathrm{H}_{\bm{k},\mathrm{R}} \oplus \mathrm{H}_{\bm{k},\mathrm{I}}$, where $\mathrm{H}_{\bm{k},\mathrm{s}}$ is the Hilbert space on which the pair $\hat{v}_{\bm{k}}^{\mathrm{s}}, \hat{p}_{\bm{k}}^{\mathrm{s}}$ acts. We call this partition of the Hilbert space $\mathrm{R}/\mathrm{I}$. This decomposition allows us to write the dynamics in its simplest form, in which Eq.~\eqref{eq:Hsystem} reads
\begin{align}
\label{eq:Hamiltonian:RI}
  \hat{H} &= \int_{\setR^{3+}} \dd^3 \bm{k} \sum_{s=\mathrm{R,I}}
  \hat{\mathcal{H}}_{\bm{k}}^s
  = \int_{\setR^{3+}} \dd^3 \bm{k} \sum_{s=\mathrm{R,I}}
  \frac{(\hat{p}_{\bm{k}}^s)^2 + \omega^2(k, \eta) (\hat{v}_{\bm{k}}^s)^2}{2} \, ,
\end{align}
casting the field dynamics as that of an infinite set of $2 \times \setR^{3+} = \setR^3$ decoupled modes. Note that this is the correct number of degrees of freedom for a real scalar field. Finally, a third partition is given by considering the standard quantum field-theoretic procedure to define the particle content of a field. We introduce the standard creation and annihilation operators $\hat{c}_{\bm{k}}^{(\dagger)}$, defined by
\begin{align}
\label{eq:creation_operators_pmk}
\hat{c}_{\bm{k}} = \frac{1}{\sqrt{2}} \left( \sqrt{k} \hat{v}_{\bm{k}} + i \frac{\hat{p}_{\bm{k}}}{\sqrt{k}} \right) \, .
\end{align}
These operators satisfy the expected commutation relations $[\hat{c}_{\bm{k}}, \hat{c}_{\bm{k}'}] = 0$ and $[\hat{c}_{\bm{k}}, \hat{c}_{\bm{k}'}^\dagger] = \delta(\bm{k} - \bm{k}')$, so that they act on separate Hilbert spaces. This gives the proper definition of Fourier modes of the field and leads to the following Hilbert-space partition, $\mathrm{H} = \oplus_{\setR^{3+}} \mathrm{H}_{\bm{k}} \oplus \mathrm{H}_{-\bm{k}}$, where $\mathrm{H}_{\bm{k}}$ is the Hilbert space on which the pair $\hat{c}_{\bm{k}}, \hat{c}_{\bm{k}}^{\dagger}$ acts. We call this partition $\pm \bm{k}$. We can then straightforwardly define canonically conjugate bosonic pairs from $\hat{c}_{\pm \bm{k}}$ as
\begin{align}
\label{eq:position_operators_pmk}
\hat{q}_{\pm \bm{k}} &= \frac{1}{\sqrt{2k}}
\left(\hat{c}_{\pm \bm{k}} + \hat{c}_{\pm \bm{k}}^\dagger \right)
= \frac{1}{\sqrt{2}} \hat{v}_{\bm{k}}^{\mathrm{R}} \mp \frac{1}{\sqrt{2}k} \hat{p}_{\bm{k}}^{\mathrm{I}} \, , \\
\hat{\pi}_{\pm \bm{k}} &=
- i \sqrt{\frac{k}{2}}
\left(\hat{c}_{\pm \bm{k}} - \hat{c}_{\pm \bm{k}}^\dagger \right)
= \frac{1}{\sqrt{2}} \hat{p}_{\bm{k}}^{\mathrm{R}} \pm \frac{k}{\sqrt{2}} \hat{v}_{\bm{k}}^{\mathrm{I}} \, .
\end{align}
The bosonic operators $\hat{q}_{\pm \bm{k}}$ and $\hat{\pi}_{\pm \bm{k}}$ characterise $\mathrm{H}_{+ \bm{k}}$ and $\mathrm{H}_{-\bm{k}}$. It is instructive to rewrite the Hamiltonian in terms of these operators,
\begin{align}
\begin{split}
\label{eq:Hamiltonian:pmk}
\hat{\mathcal{H}}_{\pm \bm{k}} 
 =
 \frac{1}{2} & \left[ 1 + k^{-2} \omega^2(k, \eta) \right] 
 \left[ \hat{\pi}_{\bm{k}}^2 + k^2 \hat{q}_{\bm{k}}^2 + \hat{\pi}_{-\bm{k}}^2 + k^2 \hat{q}_{-\bm{k}}^2 \right] \\
& + \left[ 1 - k^{-2} \omega^2(k, \eta) \right]
\left[ \hat{\pi}_{\bm{k}} \hat{\pi}_{-\bm{k}} - \hat{q}_{\bm{k}} \hat{q}_{-\bm{k}} \right] \, .
\end{split}
\end{align}

The first two terms are, up to an overall factor, the Hamiltonians of two free oscillators for the modes $+ \bm{k}$ and $- \bm{k}$. The last term gives a time-dependent coupling between the $\pm \bm{k}$ modes. Thus, while the dynamics leaves the degrees of freedom in $\mathrm{H}_{\pm \bm{k},\mathrm{R}}$ and $\mathrm{H}_{\pm \bm{k},\mathrm{I}}$ decoupled, it does generate entanglement between degrees of freedom in $\mathrm{H}_{+ \bm{k}}$ and $\mathrm{H}_{- \bm{k}}$. In the $\pm \bm{k}$ partition, the dynamics corresponds to a two-mode squeezing, as first noted by~\cite{Grishchuk:1990bj}, whereas in the $\mathrm{R}/\mathrm{I}$ partition it is the direct product of two identical one-mode squeezing operations. In the $\pm \bm{k}$ partition, the dynamics takes the initial vacuum state to a two-mode squeezed vacuum state.

\subsubsection{Correlations in cosmological perturbations 
\label{sec:which_correlations}}

Short of a precise experimental protocol to select a partition in which to study correlations, in this work we take guidance from the flat-spacetime situation. In Minkowski space, the Poincar\'{e} symmetries select a preferred partition, or equivalently a preferred vacuum, from which the notion of a particle is defined~\cite{Weinberg:1995mt}. The essential feature of quantum field theory in curved spacetime is that this notion of vacuum, and thus that of particle, loses its global character~\cite{Birrell:1982ix}. The non-coincidence of the notion of particles in two different spacetime regions is what leads to cosmological particle production and, in our case, to the generation of cosmological inhomogeneities during inflation. Yet, in the asymptotic past $\eta \to -\infty$, all modes under consideration are sub-Hubble and we have $\omega^2(k, \eta) \to k^2$. The frequency is time-independent and we are effectively in a flat-spacetime situation, with a preferred partition given by the usual flat-spacetime methods. This is precisely how we built the $\pm \bm{k}$ partition in the previous section. Equation~\eqref{eq:Hamiltonian:pmk} then shows that, in the limit $\omega^2(k, \eta) \to k^2$, the $\pm \bm{k}$ modes behave as independent free oscillators of frequency $k$. The vacuum of each mode is uniquely defined as the state $\lvert 0 \rangle_{\bm{k}}$ annihilated by $\hat{c}_{\bm{k}}(\eta \to -\infty)$. Applying this constraint to all modes $\bm{k}$ defines the Bunch--Davies vacuum $\lvert 0 \rangle_{\mathrm{BD}} = \prod_{\bm{k} \in \setR^{3}} \lvert 0 \rangle_{\bm{k}}$. The Bunch--Davies vacuum thus gives a ``minimal'' degree of fluctuations in the asymptotic past, and we will assume in the rest of this work that this was the initial state of the perturbations. At an arbitrary later time $\eta$, we are left with no well-defined notion of vacuum and, as such, no preferred partition. This arbitrariness was emphasised in~\cite{Agullo:2022ttg} to argue that there is no meaningful way to discuss the correlations of cosmological perturbations. A first way out is to consider correlations between field values in real space. This route was followed in~\cite{Martin:2021qkg,Espinosa-Portales:2022yok,Agullo:2022ttg}, which showed that there is essentially no entanglement in real space when considering simple mode decompositions. A second way out is to consider the correlations in the $\pm \bm{k}$ partition as potential correlations that would be actualised when the mode $\bm{k}$ becomes sub-Hubble again after inflation. We then have $\omega(k, \eta) \simeq k$, and the Minkowski treatment applies again, so that the correlations are well defined. We follow this second approach in this paper and discuss correlations in the $\pm \bm{k}$ partition.

\subsubsection{Reduced dimension dynamics}

One can define creation and annihilation operators for the $\mathrm{R}/\mathrm{I}$ partition related to $\hat{v}_{\bm{k}}^{\mathrm{s}}, \hat{p}_{\bm{k}}^{\mathrm{s}}$ by the same relations as those linking $\hat{q}_{\pm \bm{k}}, \hat{\pi}_{\pm \bm{k}}$ to $\hat{c}^{(\dagger)}_{\pm \bm{k}}$ in~\eqref{eq:position_operators_pmk}. Their associated vacuum matches the Bunch--Davies one, so that $\lvert 0 \rangle_{\mathrm{BD}} = \prod_{\bm{k} \in \setR^{3+}, \, \mathrm{s} \in \{\mathrm{R},\mathrm{I}\}} \lvert 0 \rangle_{\bm{k},\mathrm{s}}$. The initial state of the cosmological perturbations is thus a direct product in the $\mathrm{R}/\mathrm{I}$ partition, and the separation of the Hamiltonian~\eqref{eq:Hamiltonian:RI} ensures that this factorisation persists at any later time. On the other hand, in the $\pm \bm{k}$ partition the dynamics entangles opposite-momentum modes, so that we can only write the state as a direct product of two-mode subsystems. Using the density-matrix formalism, we have at all times $\eta$
\begin{equation}
\label{eq:density_matrix}
\hat{\varrho} \left( \eta \right) =
\underset{\bm{k} \in \setR^{3+}, \, \mathrm{s} \in \{\mathrm{R},\mathrm{I}\}}{\bigotimes}
\hat{\varrho}_{\bm{k}, \mathrm{s}} \left( \eta \right)
=
\underset{\bm{k} \in \setR^{3+}}{\bigotimes}
\hat{\varrho}_{\pm \bm{k}} \left( \eta \right) \, ,
\end{equation}
where
$\hat{\varrho}_{\pm \bm{k}} = \hat{\varrho}_{\bm{k}, \mathrm{R}} \otimes \hat{\varrho}_{\bm{k}, \mathrm{I}} \neq \hat{\varrho}_{\bm{k}} \otimes \hat{\varrho}_{-\bm{k}}$ in general. Thanks to this factorisation, in the rest of this work we will focus on the evolution of modes in the two-dimensional Hilbert subspace $H_{\pm \bm{k}} = H_{+\bm{k}} \oplus H_{-\bm{k}} = H_{\bm{k}, \mathrm{R}} \oplus H_{\bm{k}, \mathrm{I}}$. The Liouville--von Neumann equation then separates on these different Hilbert subspaces,
\begin{equation}
\label{eq:Liouville_free_RI}
\frac{\mathcal{V}}{(2\pi)^3} \, \partial_{\eta} \left( \hat{\varrho}_{\bm{k}, \mathrm{s}} \right)
= - i \left[ \hat{\mathcal{H}}_{\bm{k}}^{\mathrm{s}}, \hat{\varrho}_{\bm{k}, \mathrm{s}} \right] \, ,
\end{equation}
where $\mathcal{V}$ is the volume of the finite box in real space (e.g. $\mathcal{V} = L^3$) that we must use to make computations well defined and take to infinity eventually. Its inclusion is carefully explained in Appendix~D of~\cite{Burgess:2022nwu} and comes from taking the time derivative of a tensor product
\begin{equation}
\partial_{\eta} \hat{\varrho}
=
\frac{\mathcal{V}}{(2\pi)^3}
\int_{\setR^{3+}} \dd^3 \bm{k}
\sum_{s=\mathrm{R,I}} \hat{\varrho}_{\bm{k}, \mathrm{s}} \, .
\end{equation}
We note that, although this factor was missed in the rougher treatment of~\cite{Martin:2021znx}, it does not affect the subsequent analysis based on transport equations for the elements of the covariance matrix. As will be shown below, Eq.~\eqref{eq:Liouville_free_RI} allows one to describe the state separately on each $H_{\bm{k}, \mathrm{s}}$.

\subsubsection{Covariance matrix}

Because the initial Bunch--Davies state is Gaussian, and the state is evolved under a quadratic Hamiltonian~\eqref{eq:Hamiltonian:RI}, it remains Gaussian at all times $\eta$. Gaussian states are fully characterised by their covariance matrix, the $2N \times 2N$ matrix of the two-point correlation functions of canonical pairs, where $N$ is the number of modes considered, and the $2N$-vector of expectation values of these pairs. Our state is centred and these expectation values vanish, so we only need to consider the covariance matrix. The covariance matrix of a single mode associated with $\bm{k}, \mathrm{s}$ and of state $\hat{\varrho}_{\bm{k}, \mathrm{s}}(\eta)$, reads
\begin{equation}
\label{eq:covariance_RI_1mode}
\gamma_{\bm{k}}^{\mathrm{s}} = \begin{pmatrix}
\gamma_{11} & \gamma_{12} \\
\gamma_{12} & \gamma_{22}
\end{pmatrix} \, ,
\end{equation}
where the quantities $\gamma_{ij}$ are the finite part of the adimensionalised two-point functions\footnote{
Note that different conventions exist here. For instance, Ref.~\cite{Burgess:2022nwu} does not consider symmetrised and adimensionalised anticommutators, so that their expectation values differ by numerical factors from ours. In our convention, their power spectra $P_{vv}$ and $P_{pp}$ are related to our covariance-matrix elements as $\gamma_{11} = 2 k P_{vv}$, $\gamma_{12} = 2 P_{vv}$, and $\gamma_{22} = 2 P_{pp}/k$.
}
\begin{subequations}
\begin{align}
2 k \left\langle \hat{v}_{\bm{k}}^{\mathrm{s}} \hat{v}_{\bm{k}^{\prime}}^{\mathrm{s^{\prime}}} \right\rangle
&= \gamma_{11} \delta_{\mathrm{s},\mathrm{s^{\prime}}} \delta \left( \bm{k} - \bm{k}^{\prime} \right) \, , \\
\left\langle \hat{v}_{\bm{k}}^{\mathrm{s}} \hat{p}_{\bm{k}^{\prime}}^{\mathrm{s^{\prime}}}
+ \hat{p}_{\bm{k}}^{\mathrm{s}} \hat{v}_{\bm{k}^{\prime}}^{\mathrm{s^{\prime}}} \right\rangle
&= \gamma_{12} \delta_{\mathrm{s},\mathrm{s^{\prime}}} \delta \left( \bm{k} - \bm{k}^{\prime} \right) \, , \\
\frac{2}{k} \left\langle \hat{p}_{\bm{k}}^{\mathrm{s}} \hat{p}_{\bm{k}^{\prime}}^{\mathrm{s^{\prime}}} \right\rangle
&= \gamma_{22} \delta_{\mathrm{s},\mathrm{s^{\prime}}} \delta \left( \bm{k} - \bm{k}^{\prime} \right) \, .
\end{align}
\end{subequations}
Since the density matrix factorises in this partition, the covariance matrix of $\hat{\varrho}_{\pm \bm{k}}$ is simply given by the $4 \times 4$ matrix $\gamma_{\bm{k}}^{\mathrm{R}} \oplus \gamma_{\bm{k}}^{\mathrm{I}}$. At this stage, we emphasise that, since different partitions use different canonical pairs to describe the same state, the covariance matrix assumes a different form in each partition, although it contains the same information. The transformation rules between two partitions are derived in detail in~\cite{Martin:2021znx}. It is illustrative to give the form obtained by transforming Eq.~\eqref{eq:covariance_RI_1mode} to the $\pm \bm{k}$ partition. We have~\cite{Martin:2021znx}
\begin{equation}
\label{eq:covariance_pmk_2mode}
\gamma_{\pm \bm{k}} =
\begin{pmatrix}
\tilde{\gamma}_{11} & \tilde{\gamma}_{12} \\
\tilde{\gamma}_{12} & \tilde{\gamma}_{11}
\end{pmatrix}
\end{equation}
with
\begin{equation}
\tilde{\gamma}_{11} = \frac{\gamma_{11} + \gamma_{22}}{2} I_2 \, , \qquad
\tilde{\gamma}_{12} = \frac{1}{2}
\begin{pmatrix}
\gamma_{11} - \gamma_{22} & 2 \gamma_{12} \\
2 \gamma_{12} & - \left( \gamma_{11} - \gamma_{22} \right)
\end{pmatrix} \, .
\end{equation}

\begin{figure}[ht]
  \centering
  \includegraphics[width=0.5\textwidth]{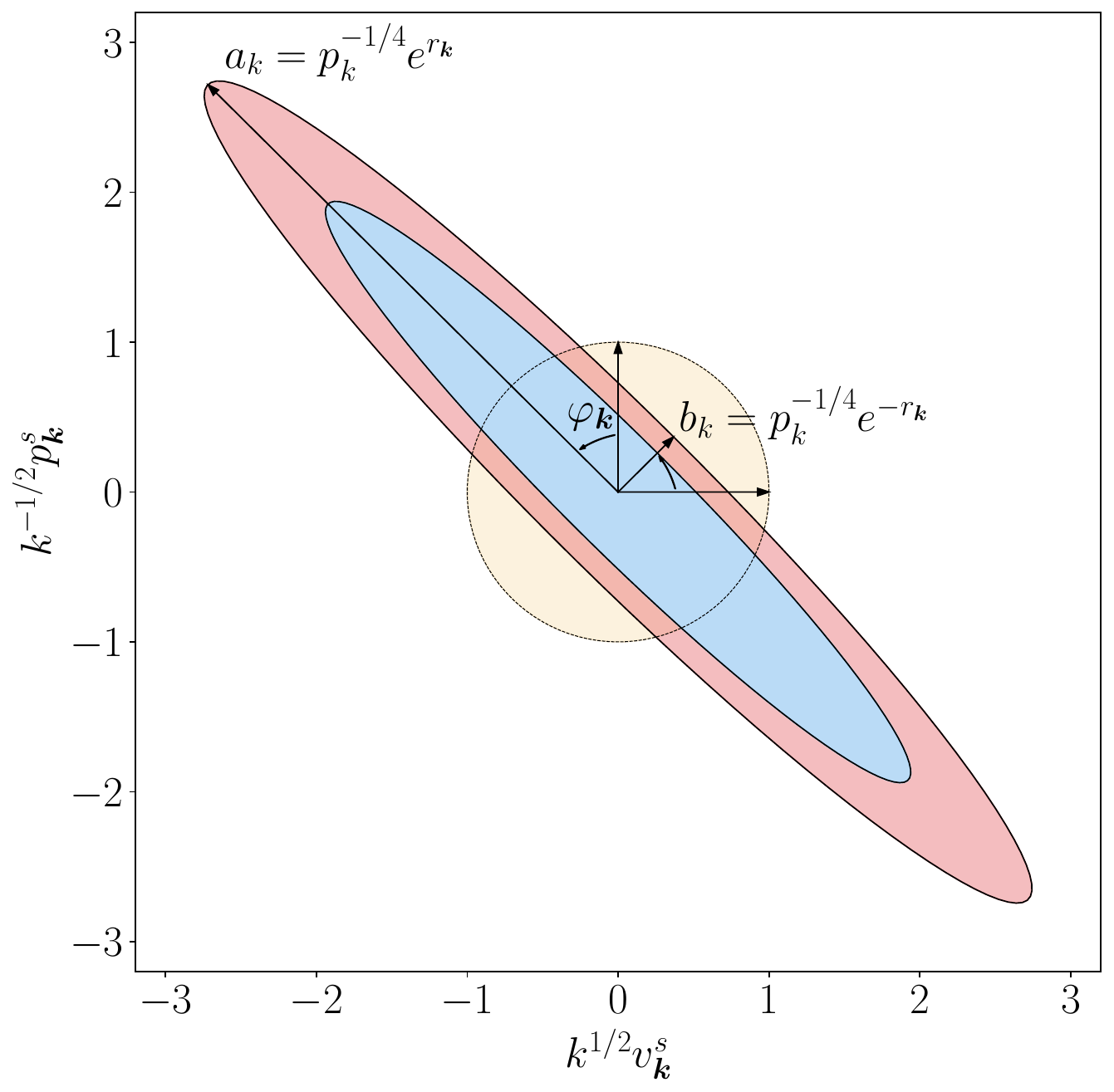}
  \caption{ Squeezing ellipses representing the points where the Wigner function of the mode $\bm{k}, \mathrm{s}$ given in Eq.~\eqref{def:Wigner_1mode} reaches $e^{-1/2}$ times its maximum value. The yellow circle corresponds to the vacuum state for whihc $r_k= \varphi_k=0$ and $p_k=1$. The blue ellipse is a 2-mode squeezed vacuum state with $r_k=1$, $ \varphi_k= \pi/4$ and $p_k=1$. The red ellipse is a 2-mode squeezed thermal state with $r_k=1$, $ \varphi_k= \pi/4$ and $p_k=1/4$.
  \label{fig:squeezing_ellipse}
  }
\end{figure}

Note that, in this case, the matrix is not block diagonal, and the off-diagonal blocks encode the $\pm \bm{k}$ correlations. All quantum correlation measures between these two modes can be computed from the data of Eq.~\eqref{eq:covariance_pmk_2mode}. We can always express the three covariance-matrix elements $\gamma_{ij}$ in terms of three real quantities: $r_k \, (\geq 0)$, the squeezing parameter; $\varphi_k$, the squeezing angle; and a third parameter $p_k$. Together, they are referred to as the effective squeezing parameters. We can write the elements of the covariance matrix as~\cite{Martin:2022kph}
\begin{subequations}
\label{def:squeezing_para}
\begin{align}
\gamma_{11} &= p_k^{-1/2} \left[ \cosh(2 r_k) - \cos(2 \varphi_k) \sinh(2 r_k) \right] \, , \\
\gamma_{22} &= p_k^{-1/2} \left[ \cosh(2 r_k) + \cos(2 \varphi_k) \sinh(2 r_k) \right] \, , \\
\gamma_{12} &= - p_k^{-1/2} \sin(2 \varphi_k) \sinh(2 r_k) \, .
\end{align}
\end{subequations}
We can check that $p_k^{-1}$ is given by the determinant of the covariance matrix,
\begin{equation}
\label{eq:purity_determinant}
p_k^{-1} = \det \left( \gamma_{\bm{k}}^{\mathrm{s}} \right)
= \gamma_{11} \gamma_{22} - \gamma_{12}^2 \, .
\end{equation}
Since, for Gaussian states~\cite{Adesso:2014npz}, we have
$\mathrm{Tr} \left[ \hat{\varrho}_{\pm \bm{k}}^2 \right] = \det \left( \gamma_{\bm{k}}^{\mathrm{s}} \right)^{-1}$,
$p_k$ is the purity of the state $\hat{\varrho}_{\pm \bm{k}}$ and thus satisfies $0 \leq p_k \leq 1$. Moreover, the effective squeezing parameters can be given a geometrical interpretation by considering the Wigner function $W$ associated with the state $\hat{\varrho}_{\bm{k}, \mathrm{s}}(\eta)$, which is a centred Gaussian with covariance matrix $\gamma_{\bm{k}}^{\mathrm{s}}$~\cite{Martin:2022kph}. We have
\begin{equation}
\label{def:Wigner_1mode}
W\left( v_{\bm{k}}^{\mathrm{s}}, p_{\bm{k}}^{\mathrm{s}} \right)
= \frac{1}{\sqrt{2 \pi \det \gamma}}
\exp \left( - \frac{V^{T} \gamma^{-1} V}{2} \right) \, ,
\end{equation}
with
\begin{equation}
V =
\begin{pmatrix}
v_{\bm{k}}^{\mathrm{s}} \\
p_{\bm{k}}^{\mathrm{s}}
\end{pmatrix} \, .
\end{equation}
Then, because $W$ is a centred Gaussian, the locus of points where $W$ reaches a fraction $0 \leq \beta \leq 1$ of its maximum, i.e.
\begin{equation}
W\left( v_{\bm{k}}^{\mathrm{s}}, p_{\bm{k}}^{\mathrm{s}} \right)
= \beta \, \underset{v_{\bm{k}}^{\mathrm{s}}, p_{\bm{k}}^{\mathrm{s}}}{\mathrm{max}}
\left[ W\left( v_{\bm{k}}^{\mathrm{s}}, p_{\bm{k}}^{\mathrm{s}} \right) \right]
\end{equation}
is given by an ellipse centred at $(0,0)$. For $\beta = e^{-1/2}$, the lengths of the semi-major axis $a_k$ and of the semi-minor axis $b_k$ are simply given by
\begin{equation}
\label{def:axes_ellipse}
a_k = p_k^{-1/4} e^{r_k} \, , \qquad
b_k = p_k^{-1/4} e^{-r_k} \, ,
\end{equation}
and the squeezing angle $\varphi_k$ is the angle between the semi-minor axis and the $v_{\bm{k}}^{\mathrm{s}}$-axis. Thus, the squeezing parameter $r_k$ controls the aspect ratio of the ellipse, $a_k / b_k = e^{2 r_k}$, while the purity $p_k$ gives its area, $S_k = \pi \, p_k^{-1/2}$. We plot several such squeezing ellipses in Fig.~\ref{fig:squeezing_ellipse} as an illustration.

Finally, the Hamiltonian dynamics~\eqref{eq:Liouville_free_RI} can be cast simply as a set of three coupled ordinary differential equations for the $\gamma_{ij}$s, or equivalently for $r_k, \varphi_k$ and $p_k$, as given in~\cite{Martin:2021znx}. We make a few general remarks on their evolution. First, since the perturbations are initially in their vacuum, which is a pure state, and since unitary dynamics preserve the purity of the state, we have $p_k = 1$ at all times for a two-mode squeezed vacuum state. Second, when two quantum modes are in a pure state and correlated, they are necessarily entangled, i.e. all correlations are quantum in this case; see, e.g.,~\cite{Martin:2022kph}. Thus, as we will discuss in Sec.~\ref{sec:quantum_corr}, the $\pm \bm{k}$ modes are entangled by the dynamics, and their bipartite correlations violate Bell inequalities and exhibit a large quantum discord. Yet, this equivalence between classical and quantum correlations, as well as that between the different quantumness criteria, breaks down in the presence of interactions which takes the state to a mixed one~\cite{Martin:2022kph}. We now study the evolution of perturbations in this context.

\subsection{Interacting perturbations}

Describing the evolution of the perturbations by the closed-system dynamics given by the Hamiltonian~\eqref{eq:Hamiltonian:RI} is an idealisation. For instance, the cosmological perturbations would be weakly coupled to other subdominant fields present during inflation~\cite{Martin:2018zbe}, and computing the dynamics at next order in perturbation theory would lead to cubic self-interactions of the perturbations~\cite{Maldacena:2002vr}, which couple any $\pm \bm{k}$ pair of modes to all other modes of the curvature perturbation~\cite{Burgess:2006jn}. Retaining the approach of the previous section, we focus on a single $\pm \bm{k}$ pair and ask how such interactions would affect their quantum state, as described by the effective squeezing parameters. We rely on an open quantum system approach, in which the degrees of freedom with which the pair interacts are traced out, and the reduced dynamics of the $\pm \bm{k}$ pair is described by a master equation. Rather than deriving this equation within a specific interaction model, we adopt the approach of~\cite{Martin:2018zbe} and assume from the outset that the interactions result in a dynamics described by a specific Lindblad equation, see Eq.~\eqref{eq:lindblad_Caldeira_Leggett} below. This allows us to keep simple and explicit expressions for the different parameters, making the effect of interactions transparent.

\subsubsection{Caldeira--Leggett model}

We consider the curvature perturbations to couple to their environment through a coupling proportional to $\hat{v}$. Such a linear coupling has two virtues~\cite{Martin:2021znx,Burgess:2022nwu}: it leads to a master equation that preserves both the separation of the dynamics into independent one-mode systems $\mathrm{H}_{\pm \bm{k},\mathrm{s}}$ and the Gaussianity of the state. The resulting states for pairs of modes $\pm \bm{k}$ are two-mode squeezed thermal states, described by the effective squeezing parameters defined in Eq.~\eqref{def:squeezing_para}. We further assume that the evolution of the density matrix is governed by a Lindblad equation~\cite{Lindblad:1975ef} of the form obtained in~\cite{Martin:2018zbe}:
\begin{align}
\begin{split}
\label{eq:lindblad_Caldeira_Leggett}
\frac{\mathcal{V}}{(2\pi)^3}
\frac{\partial \hat{\varrho}_{ \bm{k},\mathrm{s}}}{\partial \eta}
& =  - i \left[ \hat{\mathcal{H}}_{\bm{k}}^{\mathrm{s}} ,
\hat{\varrho}_{ \bm{k},\mathrm{s}} \right]
- \mathfrak{F}_{\bm{k}}(\eta)
\left[ k \, \hat{v}_{\bm k}^{\mathrm{s}} , \big[ k \, \hat{v}_{\bm k}^{\mathrm{s}}, \hat{\varrho}_{ \bm{k},\mathrm{s}} \big] \right] \, ,
\end{split}
\end{align}
where
\begin{equation}
\label{def:coupling}
\mathfrak{F}_{\bm{k}}(\eta) =
\frac{1}{2} \left( \frac{k_{\Gamma}}{k} \right)^2
\left( \frac{a}{a_{\star}} \right)^{p-3}
\Theta \left( 1 - \frac{k \ell_{E} }{a} \right) \, ,
\end{equation}
$\Theta$ is the Heaviside step function and $k_\Gamma~(>0)$, $p$, $a_{\star}$ and $\ell_{E}~(>0)$ are real parameters. Such a Lindblad equation, with a coupling linear in the system degree of freedom, corresponds to the Caldeira--Leggett model~\cite{Caldeira:1981rx}. The first term corresponds to the free Hamiltonian dynamics given by the Liouville--von Neumann equation~\eqref{eq:Liouville_free_RI}. The second term, which is not of Hamiltonian type and leads to non-unitary evolution, encodes the action of the environment on the system via the decoherence kernel $\mathfrak{F}_{\bm{k}}$%
\footnote{
Using that $\hat{v}(\bm{x})$ has the dimension of an energy per unit length, one can check that $k_{\Gamma}$ must have the dimension of a wavenumber for the equation to be dimensionally consistent.
},
where $k_\Gamma$ sets the overall amplitude of the kernel. Its time dependence is given by a power-law in the scale factor, parametrised by $p$, with reference scale $a_{\star}$. The parameter $\ell_{E}~(>0)$ is the physical auto-correlation length of the environment, below which the interaction is inefficient. The assumptions required to derive Eq.~\eqref{eq:lindblad_Caldeira_Leggett}, and the relation between the phenomenological parameters of the kernel and specific interaction Hamiltonians, are discussed in Appendix~\ref{app:scope}. Starting from a given interaction model, the range of parameters for which Eq.~\eqref{eq:lindblad_Caldeira_Leggett} is valid is in general limited, for instance by requiring the system--environment interaction to be perturbative. Nevertheless, since we take here an agnostic view of the microphysical origin of the dissipation of the perturbations, we analyse Eq.~\eqref{eq:lindblad_Caldeira_Leggett} as it stands.

Finally, for practical purposes it is useful to consider the evolution of the expectation value of an operator $\hat{O}$ acting on $\mathrm{H}_{\pm \bm{k},\mathrm{s}}$. Using $\langle \hat{O} \rangle (\eta) = \mathrm{Tr} [ \hat{\varrho}_{ \bm{k},\mathrm{s}} (\eta) \hat{O} ]$, we obtain
\begin{align}
\begin{split}
\label{eq:ev_operator}
\frac{\mathcal{V}}{(2\pi)^3}
\frac{\dd \langle \hat{O} \rangle}{\dd \eta}
& = - i \left \langle \left[ \hat{O} , \hat{\mathcal{H}}_{\bm{k}}^{\mathrm{s}} \right] \right \rangle
- \mathfrak{F}_{\bm{k}}(\eta)
\left \langle \left[ \left[ \hat{O} , k \, \hat{v}_{\bm k}^{\mathrm{s}} \right], k \, \hat{v}_{\bm k}^{\mathrm{s}} \right]  \right \rangle \, .
\end{split}
\end{align}

\section{Squeezing parameters in presence of decoherence}
\label{sec:deco_squeezing}

By choosing a coupling linear in the system variable, we obtained the Lindblad equation~\eqref{eq:lindblad_Caldeira_Leggett}, that can still be separated into single-mode density matrices $\hat{\varrho}_{ \bm{k},\mathrm{s}}$ in the $\mathrm{R}/\mathrm{I}$ partition, and that is quadratic in $\hat{v}$. The equation therefore admits Gaussian state solutions. Assuming a Bunch--Davies initial state, the quantum state can still be fully described by covariance matrices of the same form as Eq.~\eqref{eq:covariance_pmk_2mode}, and parametrised by its purity and squeezing parameters, as in Eqs.~\eqref{def:squeezing_para}. In this section, we discuss how these parameters are affected by the presence of interactions modelled by Eq.~\eqref{def:coupling}. In particular, we study numerically the case of a de Sitter background.

\subsection{Transport equations in the Caldeira--Leggett model}

First, we derive evolution equations for the covariance-matrix elements using Eq.~\eqref{eq:ev_operator}. Note that all volume factors will cancel out. As an illustration, to obtain the evolution equation for $\gamma_{22}$, we apply the equation to $\hat{O} = \hat{p}_{\bm k}^{\mathrm{s}} \hat{p}_{\bm k^{\prime}}^{\mathrm{s^{\prime}}}$. Then, on the right-hand side, commuting operators give rise to a product of a volume factor and Dirac delta function. For instance, we have
\begin{equation}
\left \langle \left[ \hat{p}_{\bm k}^{\mathrm{s}} \hat{p}_{\bm k^{\prime}}^{\mathrm{s^{\prime}}} , \hat{\mathcal{H}}_{\bm{k}}^{\mathrm{s}} \right] \right \rangle
= - i \gamma_{12} \, \delta_{\mathrm{s},\mathrm{s^{\prime}}} \,
\delta \left( \bm k - \bm k^{\prime} \right)
\frac{\mathcal{V}}{(2\pi)^3} \, .
\end{equation}
On the left-hand side, the expectation value appearing in $\langle \hat{p}_{\bm k}^{\mathrm{s}} \hat{p}_{\bm k^{\prime}}^{\mathrm{s^{\prime}}} \rangle$ leads to a Dirac delta function in addition to the already present volume factor. These volume factors and Dirac deltas then factor out. The final equations, already derived in~\cite{Martin:2021znx,Burgess:2022nwu}, read
\begin{subequations}
\label{eq:transport_cov}
\begin{eqnarray}
\label{eq:transport_gamma11} 
k^{-1} \frac{\partial \gamma_{11}\left(k,\eta \right)}{\partial \eta}
&=& 2 \gamma_{12} \left(k,\eta \right) \,, \\ [8pt] 
\label{eq:transport_gamma12}
k^{-1} \frac{\partial \gamma_{12}\left(k,\eta \right)}{\partial \eta}
&=& \gamma_{22} \left(k,\eta \right)
- k^{-2} \omega^2(k,\eta) \, \gamma_{11} \left(k,\eta \right) \, , \\ [8pt] 
\label{eq:transport_gamma22}
k^{-1} \frac{\partial \gamma_{22}\left(k,\eta \right)}{\partial \eta}
&=& - 2 k^{-2} \omega^2(k,\eta) \, \gamma_{12} \left(k,\eta \right)
+ 4 \mathfrak{F}_{\bm{k}}(\eta) \, .
\end{eqnarray}
\end{subequations}

The purity $p_k$ is a key quantity for describing the efficiency of the decoherence process. It is thus useful to derive a transport equation for it. Using Eq.~\eqref{eq:purity_determinant} together with the differential equations~\eqref{eq:transport_cov}, we get
\begin{equation}
\label{eq:transport_purity}
k^{-1} \frac{\partial \left( p_k^{-1} \right)}{\partial \eta}
= 4 \mathfrak{F}_{\bm{k}}(\eta) \, \gamma_{11} \, .
\end{equation}

\subsection{Covariance matrix and effective squeezing parameters in presence of decoherence}

The solutions of Eqs.~\eqref{eq:transport_cov} were obtained in~\cite{Martin:2021znx}. They read
\begin{subequations}
\label{eq:gij:exact}
\begin{eqnarray}
\label{eq:g11:exact}
\gamma_{11} \left(k,\eta \right) & = & \left| u_{\bm{k}}(\eta) \right|^2 + \mathcal{I}_k \left( \eta \right) \, , \\ [6pt]
\label{eq:g12:exact}
\gamma_{12} \left(k,\eta \right)  & = & k^{-1} \, \mathrm{Re} \left[ \partial_\eta u_{\bm{k}}(\eta)\, u_{\bm{k}}^*(\eta) \right] + \mathcal{J}_k \left( \eta \right) \, , \\ [6pt]
\label{eq:g22:exact}
\gamma_{22} \left(k,\eta \right)  & = & k^{-2} \left| \partial_\eta u_{\bm{k}}(\eta) \right|^2 + \mathcal{K}_k \left( \eta \right) \, ,
\end{eqnarray}
\end{subequations}
with
\begin{subequations}
\label{def:integrals_gammaij}
\begin{eqnarray}
\mathcal{I}_k \left( \eta \right) & = & 4 k \int_{\eta_{\mathrm{in}}}^\eta \mathfrak{F}_{\bm{k}}\left(\eta' \right)
\, \mathrm{Im}^2 \left[ u_{\bm{k}}(\eta')\, u_{\bm{k}}^*(\eta) \right] \, \dd \eta' \, , \\
\mathcal{J}_k \left( \eta \right)  & = & 4 \int_{\eta_{\mathrm{in}}}^\eta 
\mathfrak{F}_{\bm{k}}\left(\eta' \right)
\, \mathrm{Im} \left[ u_{\bm{k}}(\eta')\, u_{\bm{k}}^*(\eta) \right]
\, \mathrm{Im} \left[ u_{\bm{k}}(\eta')\, \partial_\eta u_{\bm{k}}^{*}(\eta) \right] \dd \eta' \, , \\
\mathcal{K}_k \left( \eta \right) & = & 4 k^{-1} \int_{\eta_{\mathrm{in}}}^\eta 
\mathfrak{F}_{\bm{k}}(\eta')
\, \mathrm{Im}^2 \left[ u_{\bm{k}}(\eta')\, \partial_\eta u_{\bm{k}}^{*}(\eta) \right] \dd \eta' \, .
\end{eqnarray}
\end{subequations}
Here $u_{\bm{k}}$ is a solution of the Mukhanov--Sasaki equation
\begin{equation}
\label{eq:MS}
\frac{\partial^2 u_k}{\partial \eta^2} + \omega^2 \left( k , \eta \right) u_k = 0 \, ,
\end{equation}
with Wronskian normalised to
$W = u_k \, ( \partial_\eta u_k^{\star}) - (\partial_\eta u_k ) \, u_k^{\star} = 2 i k$.
Note that $W$ is constant for any solution, and with this normalisation $u_k$ is dimensionless.

Using Eqs.~\eqref{def:squeezing_para}, we can then compute the effective squeezing parameters. However, a problem already encountered in~\cite{Martin:2021znx} is that, at late times, it is not numerically viable to compute the purity by evaluating the integrals appearing in the covariance-matrix elements in Eqs.~\eqref{eq:g11:exact}--\eqref{eq:g22:exact} and inserting the result into Eq.~\eqref{eq:purity_determinant}. The final step involves exact cancellations between exponentially large numbers, which fail numerically. We therefore derive an exact expression for the purity $p_k$ that allows a reliable estimate of its late-time behaviour. There are two ways to obtain such an expression. One can either insert the exact expression~\eqref{eq:g11:exact} for $\gamma_{11}$ into the transport equation~\eqref{eq:transport_purity} and integrate it, or perform algebraic manipulations starting from Eq.~\eqref{eq:purity_determinant} and inserting the exact solutions~\eqref{eq:g11:exact}--\eqref{eq:g22:exact}. Both approaches are detailed in Appendix~\ref{app:purity} and lead to
\begin{equation}
\label{eq:purity_exact}
p_k^{-1} = 1 + 2 \mathcal{L}_{k} + \mathcal{L}_{k}^2 - \left| \mathcal{M}_{k} \right|^2 \, ,
\end{equation}
where
\begin{subequations}
\begin{eqnarray}
\label{def:Lk}
\mathcal{L}_{k} & = & 2 k \int_{\eta_{\mathrm{in}}}^{\eta} \mathfrak{F}_{\bm{k}}(\eta') \left| u_{k}(\eta') \right|^2 \dd \eta' \, , \\
\label{def:Mk}
\mathcal{M}_{k} & = & 2 k \int_{\eta_{\mathrm{in}}}^{\eta} \mathfrak{F}_{\bm{k}}(\eta') \, u_{k}^2(\eta') \dd \eta' \, .
\end{eqnarray}
\end{subequations}
Using $\mathcal{L}_{k} \geq 0$ and the Cauchy--Schwarz inequality $\mathcal{L}_{k}^2 \geq \left| \mathcal{M}_{k} \right|^2$, one can check that the expression~\eqref{eq:purity_exact} always respects the constraint $0 \leq p_k \leq 1$ on the purity. Finally, we can invert Eqs.~\eqref{def:squeezing_para} to obtain expressions for the squeezing parameter $r_k$ and the squeezing angle $\varphi_k$ in terms of the covariance-matrix elements. The squeezing parameter $r_k$ is given by
\begin{equation}
\label{eq:rk_exact}
r_k = \frac{1}{2} \cosh^{-1} \left( \frac{\gamma_{11} + \gamma_{22}}{2 p_k^{-1/2}} \right) .
\end{equation}
For the squeezing angle $\varphi_k$, we can derive expressions for both $\sin (2\varphi_k)$ and $\cos (2\varphi_k)$, which read
\begin{subequations}
\label{eq:sin_cos_varphik_exact}
\begin{align}
\sin \left( 2 \varphi_k \right)
& = - \frac{\gamma_{12}}{\sqrt{\left( \displaystyle\frac{\gamma_{22} - \gamma_{11}}{2} \right)^2 + \gamma_{12}^2}} \, , \\ \notag \\
\cos \left( 2 \varphi_k \right)
& = \frac{\displaystyle\frac{\gamma_{22} - \gamma_{11}}{2}}{\sqrt{\left( \displaystyle\frac{\gamma_{22} - \gamma_{11}}{2} \right)^2 + \gamma_{12}^2}} \, , \\ \notag \\
\tan \left( 2 \varphi_k \right)
& = - \frac{2 \gamma_{12}}{\gamma_{22} - \gamma_{11}} \, .
\end{align}
\end{subequations}
The non-linearity of the relation between the squeezing parameters and the covariance-matrix elements means that, for these parameters, there is no compact exact formula analogous to Eq.~\eqref{eq:purity_exact} for the purity. In the next section, we derive approximations for these quantities in the late-time limit, which is the regime relevant for inflation.

\subsection{Application: Decoherence in a de Sitter phase}

To numerically evaluate the effective squeezing parameters of the cosmological perturbations, we now focus on the de Sitter case, where $H = \mathrm{constant}$ and $a(\eta) = -1 / (H \eta)$. Then $\omega$ is given by
\begin{equation}
\label{def:omega_dS}
\frac{\omega^2 \left( k , \eta \right)}{k^2} = 1 - \frac{2}{x^2} \, ,
\end{equation}
where we have defined a dimensionless time variable $x = - k \eta \,(>0)$, which implicitly depends on our choice of scale~$k$. In addition, given the form of the kernel~\eqref{def:coupling}, the perturbations freely evolve until $x_{E} = ( H \ell_E )^{-1}$, such that $a(x_{E}) = k \ell_{E}$. The Heaviside function in Eq.~\eqref{def:coupling} therefore reads $\Theta ( 1 - \frac{k \ell_E}{a} ) = \Theta ( 1 - \frac{x}{x_{E}} )$. We pick $\eta_{\mathrm{in}}$ to be earlier than this time so that the starting time of the integral is always set by $x_E$. Furthermore, since the perturbations are initially freely evolving, we can work with the standard assumption that they start in the Bunch--Davies vacuum, so that $u_{k} \underset{x \to \infty}{\sim} e^{-i x}$. Imposing this initial condition and the Wronskian normalisation condition given below Eq.~\eqref{eq:MS} on the solution of the Mukhanov--Sasaki equation, we obtain
\begin{equation}
\label{eq:mode_fn_dS}
u_{k} = \left( 1 + \frac{i}{x} \right) e^{i x} \, .
\end{equation}
Finally, we also set the reference time in the kernel to $a_{\star} = a_{k} = k H^{-1}$, the value of the scale factor when the modes of interest $\pm \bm{k}$ cross the Hubble radius. This can always be achieved by a $k$-dependent redefinition of $k_{\Gamma}$.

\subsubsection{Late-time approximation without decoherence}

\begin{figure}[H]
  \centering
  \includegraphics[width=0.7\textwidth]{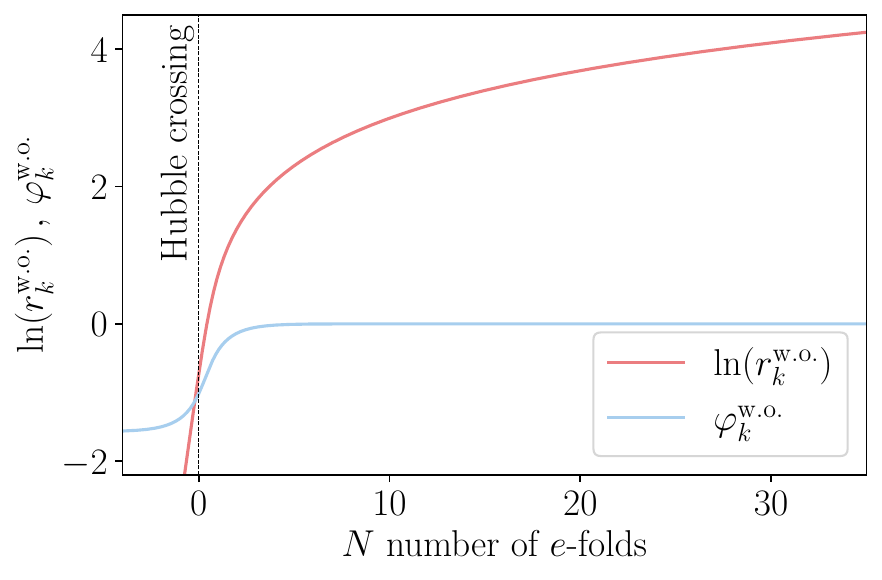}
  \caption{ Evolution of the squeezing parameters as a function of the number of $e$-folds in the absence of interaction. The red line shows the values of $ \ln (r_k^{\rm w.o.})$ computed using Eq.~\eqref{eq:no_deco_rk}. The blue line shows the values of $\varphi_k^{\rm w.o.}$ computed using Eq.~\eqref{eq:no_deco_phik}.
  \label{fig:squeezing_pure_state}
  }
\end{figure}

For reference, let us state the expressions of the covariance-matrix elements and the squeezing parameters in the absence of decoherence. Our goal is to eventually evaluate quantities at the end of inflation, for which a customary time variable is the number of $e$-folds $N = \ln [a(\eta)/a_k]$, normalised by $a_k$. For de Sitter space,  we then have $x = e^{-N}$, and cosmological modes are expected to have undergone at least $N = 60$ $e$-folds of inflation. Thus, we are required to evaluate quantities for exponentially small $x$, such that $x = e^{-N} \leq e^{-60} \ll 1$. Therefore it useful to give late-time approximations for the pure-state solutions. We have
\begin{subequations}
\begin{alignat}{1}
\label{eq:cov_dS}
\gamma_{11}(\eta) &= 1+\frac{1}{x^2} \underset{x \to 0}{=} \frac{1}{x^2} + \mathcal{O} \left( 1 \right) , \\ 
\gamma_{12}(\eta) &= \frac{1}{x^3} \underset{x \to 0}{=} \frac{1}{x^3} , \\
\gamma_{22}(\eta) &= 1-\frac{1}{x^2}+\frac{1}{x^4} \underset{x \to 0}{=} \frac{1}{x^4} + \mathcal{O} \left( x^{-2} \right) ,
\end{alignat}
\end{subequations}
and the associated squeezing parameters are
\begin{align}
\label{eq:no_deco_rk}
r^{\mathrm{w.o.}}_k\left(\eta\right) &= \frac{1}{2} \mathrm{arccosh} 
\left[ 1+\frac{1}{2 x^4} \right] \underset{x \to 0}{=} - 2 \log \left( x \right) + \mathcal{O} \left( x^4 \right) \, ,\\ 
\label{eq:no_deco_phik}
\varphi_k^{\mathrm{w.o.}} \left(\eta\right) & = - \frac{1}{2}
\arctan \left(\frac{2 x }{1-2 x^2} \right)   
-\frac{\pi}{2} \Theta\left(-k\eta-\frac{1}{\sqrt{2}}\right) \underset{x \to 0}{=} - x + \mathcal{O} \left( x^3 \right) \, ,
\end{align}
where a superscript ``w.o.'' indicates that the quantity is evaluated in the non-interacting case. Their evolution is shown in Fig.~\ref{fig:squeezing_pure_state}.

\subsubsection{Late-time approximations with decoherence}

We now evaluate the effective squeezing parameters $r_k$, $\varphi_k$ and the purity $p_k$ in the presence of interactions modelled by Eqs.~\eqref{eq:lindblad_Caldeira_Leggett}--\eqref{def:coupling}. This is done by first evaluating the integrals appearing in the exact expressions for the covariance-matrix elements, Eqs.~\eqref{eq:gij:exact}, and inserting the result into Eqs.~\eqref{eq:purity_exact}, \eqref{eq:rk_exact} and \eqref{eq:sin_cos_varphik_exact}. Evaluating these quantities at late times, when $x \sim e^{-50}$, is numerically demanding for at least two reasons. First, computing the $\gamma_{ij}$ themselves requires integrating oscillatory functions. Second, computing the squeezing parameters leads to near-perfect cancellations of exponentially large terms. These problems were already encountered and partially solved for the purity in~\cite{Martin:2021znx}. Here we extend their treatment to the case of relatively large $p$ and $k_{\Gamma}/k$. We give detailed computations and reproduce the results of~\cite{Martin:2021znx} in Appendix~\ref{app:purity_dS}.

First, let us recall the late-time approximations of the covariance-matrix elements derived in~\cite{Martin:2021znx}:
\begin{subequations}
\label{eq:latetime_gammaij}
\begin{align}
\label{eq:latetime_gamma11}
\gamma_{11} & = 
   \frac{1}{x^2}\left[ 1 - 2 \left(\frac{k_\Gamma}{k}\right)^2 B_{11} + \mathcal{O} \left(x^2 \right) - 2 \left(\frac{k_\Gamma}{k}\right)^2 A_{11} \, x^{8-p} + \mathcal{O} \left( x^{10-p} \right) \right] \, ,  \\
\label{eq:latetime_gamma12}
\gamma_{12} & =  \frac{1}{x^3}\left[ 1 - 2 \left(\frac{k_\Gamma}{k}\right)^2 B_{12} + \mathcal{O} \left(x^2 \right) - 2 \left(\frac{k_\Gamma}{k}\right)^2 A_{12} \, x^{8-p} + \mathcal{O} \left( x^{10-p} \right) \right]  \, , \\ 
\label{eq:latetime_gamma22}
\gamma_{22} & =  \frac{1}{x^4}\left[ 1 - 2 \left(\frac{k_\Gamma}{k}\right)^2 B_{22} + \mathcal{O} \left(x^2 \right) -
     2 \left(\frac{k_\Gamma}{k}\right)^2 
      A_{22}\, x^{8-p} + \mathcal{O} \left( x^{10-p} \right) \right] \, ,   
\end{align}
\end{subequations}
with expressions for the coefficients $A_{ij}$ and $B_{ij}$ ($i,j=1,2$), which depend on $p$ and $x_E$, given in Eqs.~\eqref{eq:latetime_gammaij_coef_A11}--\eqref{eq:latetime_gammaij_coef_B22}. Note that, for small $x$, we have the ordering $\gamma_{22} \gg \lvert \gamma_{12} \rvert \gg \gamma_{11} \gg 1$. This proves useful when deriving expansions.

We first derive a late-time approximation for $p_k$, which will then be used to derive approximations for $r_k$ and $\varphi_k$. One way to derive such an approximation is to insert the expansions Eqs.~\eqref{eq:latetime_gammaij} into Eq.~\eqref{eq:purity_determinant}. While one can obtain the leading-order term in this way, it does not provide an optimal estimate of the order of the next contribution. To estimate this contribution, we use the exact expression~\eqref{eq:purity_exact} and expand the integrals $\mathcal{L}_k$ and $\mathcal{M}_k$ appearing there\footnote{
Another approach would be to use Eq.~\eqref{eq:transport_purity} and integrate the right-hand side containing $\gamma_{11}$ from $x$ to $x_{\mathrm{in}} = - k \eta_{\mathrm{in}}$. However, $x_{\mathrm{in}}$ needs not be small, and thus the leading-order approximation Eq.~\eqref{eq:latetime_gamma11} may not be valid over the whole integration domain. We therefore have to use the full expression Eq.~\eqref{eq:g11:exact}.
}. This approach yields
\begin{align}
\begin{split}
\label{eq:latetime_purity}
p_k^{-1} & = 1 + \left( \frac{k_{\Gamma}}{k} \right)^2 \left[ A_{\sigma}^{(1)} + B_{\sigma}^{(1)} x^{2-p} + \mathcal{O}\left( x^{4-p} \right) \right] \\
& \qquad + \left( \frac{k_{\Gamma}}{k} \right)^4 \left[ A_{\sigma}^{(2)} + B_{\sigma}^{(2)} x^{2-p} + \mathcal{O}\left( x^{4-p} \right) + D_{\sigma}^{(2)} \left(p,x_E\right) x^{10-2p} + \mathcal{O}\left( x^{12-2p} \right) \right] \, ,
\end{split}
\end{align}
with the expressions of the coefficients given in Eqs.~\eqref{eq:latetime_pk_coef_A}--\eqref{eq:latetime_pk_coef_D}. This formula~\eqref{eq:latetime_purity} matches Eq.~(5.17) of~\cite{Martin:2021znx} for $p<8$ in the small-$x_E$ and small-$k_{\Gamma}/k$ limits, but differs for $p>8$ and large $x_E$ and $k_{\Gamma}/k$. Let us pause to explain the reason. Since the coupling is linear in the master equation~\eqref{eq:lindblad_Caldeira_Leggett}, the covariance-matrix elements receive only contributions linear in $\mathfrak{F}_k$, which, in the parametrisation~\eqref{def:coupling}, means that they are strictly of order $(k_{\Gamma}/k)^2$ and can only contain powers of $a$ containing $p$, but not $2 p$ for instance. On the other hand, the effective squeezing parameters are non-linearly related to the covariance-matrix elements. For instance, $p_k^{-1}$ is quadratic in the $\gamma_{ij}$, and thus it can receive contributions of different orders in $k_{\Gamma}/k$ and powers of $a$ that contain larger multiples of $p$. Even if $k_{\Gamma}/k \ll 1$, these terms may become dominant at late times if they are multiplied by large powers of the scale factor. This is precisely what happens in Eq.~\eqref{eq:latetime_purity} for $p>8$, where the dominant term is $x^{10-2p}$ and of order $(k_{\Gamma}/k)^4$, as seen in the expression of $D_{\sigma}$ in Eq.~\eqref{eq:latetime_pk_coef_D}. In~\cite{Martin:2021znx}, the authors dropped these terms by implicitly assuming $k_{\Gamma}/k \ll 1$ and evaluating $\gamma_{11}$ in the free theory before performing the integral. If we assume that the master equation~\eqref{eq:lindblad_Caldeira_Leggett} is valid at late times, then there is no reason to make this assumption, and a more accurate formula is given by Eq.~\eqref{eq:latetime_purity}.

\begin{figure}[H]
  \centering
  \begin{subfigure}[b]{0.48\textwidth}
    \includegraphics[width=\textwidth]{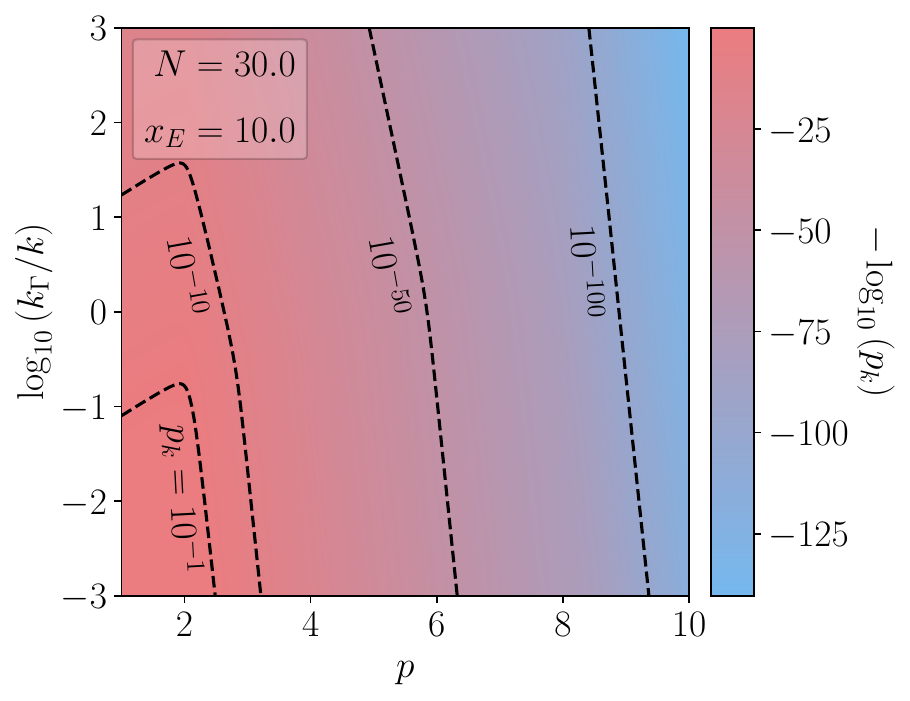}
    \caption{\label{fig:purity_r_N30} $N=30$}
  \end{subfigure}
  \hfill
  \begin{subfigure}[b]{0.48\textwidth}
    \includegraphics[width=\textwidth]{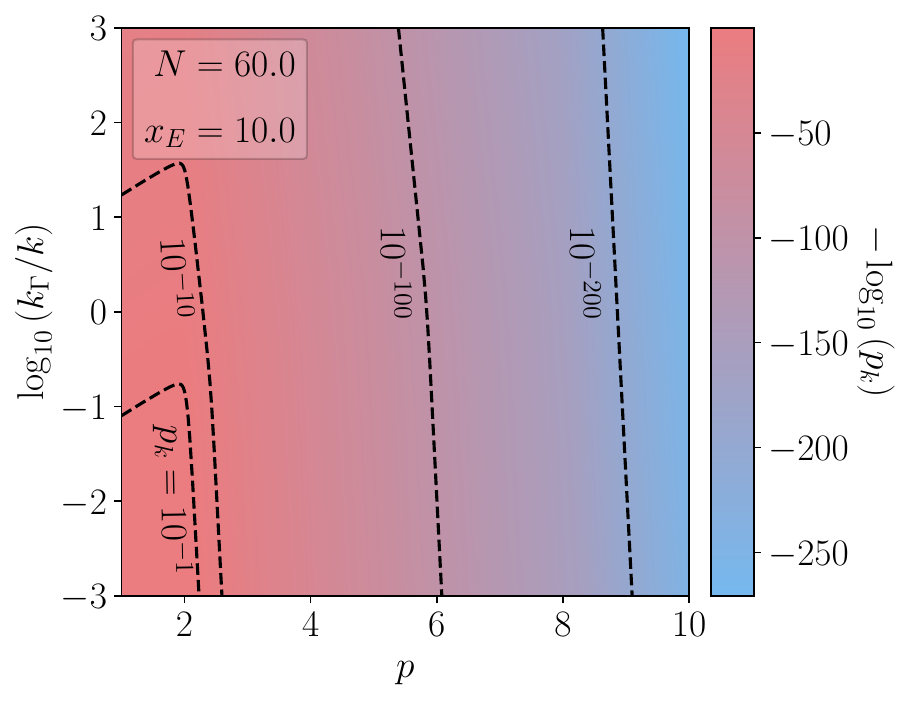}
    \caption{\label{fig:purity_N_60} $N=60$}
  \end{subfigure}
  \caption{ Values of the purity $p_k$ as a function of $p$ and $\log_{10}{(k_{\Gamma}/k)}$ after $N=30.0$ (left) and $N=60$ (right) $e$-folds of inflation.
  \label{fig:purity}
  }
\end{figure}

In Fig.~\ref{fig:purity}, we plot the value of the purity $p_k$ as a function of $p$ and $k_{\Gamma}/k$ after $N=30$ and $N=60$ $e$-folds from the horizon exit of the pivot scale, i.e. $x=e^{-30}$ and $x=e^{-60}$. We picked an arbitrary value $x_{E}=10$ for the coherence length, but changing its value is straightforward as it only fixes the starting point of decoherence. Fig.~\ref{fig:purity} shows that even for very small values of $k_{\Gamma}/k$, when $p>2$, the purity will eventually be strongly suppressed. The smallness of the purity at the end of inflation is generally interpreted as the cosmological perturbations having “classicalised”. Yet, as we discuss in Sec.~\ref{sec:quantum_corr}, the purity alone is not sufficient to correctly assess the level of quantum correlations in the state. One also needs to compute the values of the other squeezing parameters in the presence of decoherence, which we now proceed to do, starting with the squeezing parameter $r_k$.

To evaluate $r_k$ in the presence of decoherence, we start from Eq.~\eqref{eq:rk_exact} and rewrite the inverse hyperbolic function as a logarithm. Details are given in Appendix~\ref{app:squeezing_parameter}. By rearranging terms, we can isolate the dominant contribution and obtain
\begin{equation}
\label{eq:latetime_rk}
r_k = \frac{1}{2} \ln \left( \frac{\gamma_{22}}{p_k^{-1/2}} \right) + \left( \frac{\gamma_{12}}{\gamma_{22}} \right)^2 + \mathcal{O} \left( x^{4} \right) \, ,
\end{equation}
where $\gamma_{12}/\gamma_{22}=\mathcal{O}(x)$ always gives a subdominant contribution. We can then use the expansion Eq.~\eqref{eq:latetime_gamma22} for $\gamma_{22}$ and Eq.~\eqref{eq:latetime_purity} for $p_k$ to expand the first term and obtain a more explicit formula. The dominant term then depends on $p$ and the expansion reads
\begin{equation}
\label{eq:cases_rk}
\begin{alignedat}{3}
r_k & = - 2 \ln \left( x \right) + \frac{1}{2} \ln \left[ \frac{ 1
  - 2 \left(\frac{k_\Gamma}{k}\right)^2
  B_{22}\left( p , x_{E} \right) }{ \sqrt{1 + A_{\sigma} \left( p , x_{E} \right) }} \right] + \mathcal{O}\left(x^2\right)
  ~~ & \mathrm{for} & \quad p<2 \, , \\
    & = - \frac{1}{4} \left( 10 - p \right) \ln \left( x \right) + \frac{1}{2} \ln \left[ \frac{ 1
  - 2 \left(\frac{k_\Gamma}{k}\right)^2
  B_{22} \left( p , x_{E} \right) }{ \sqrt{B_{\sigma}\left( p , x_{E} \right) }} \right] + \mathcal{O}\left(x^2\right)
  ~~ & \mathrm{for} & \quad 2<p<8 \, , \\
  & = - \frac{1}{2} \ln \left( x \right) + \frac{1}{2} \ln \left[ \frac{
  - 2 \left(\frac{k_\Gamma}{k}\right)^2
  A_{22} \left( p \right) }{ \sqrt{D_{\sigma} \left( p , x_{E} \right) }} \right] + \mathcal{O}\left(x^2\right)
  ~~ & \mathrm{for} & \quad p>8 \, .
\end{alignedat}
\end{equation}

\begin{figure}[H]
  \centering
  \includegraphics[width=0.49\textwidth]{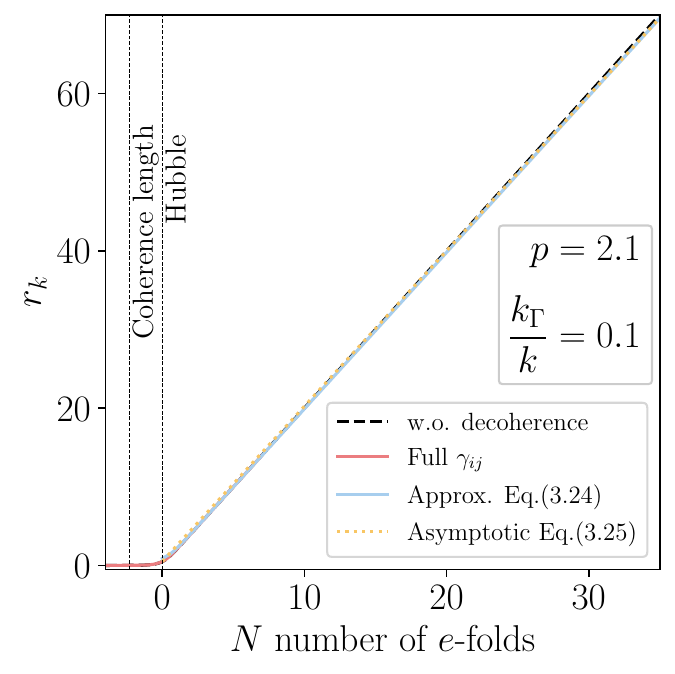}
  \includegraphics[width=0.49\textwidth]{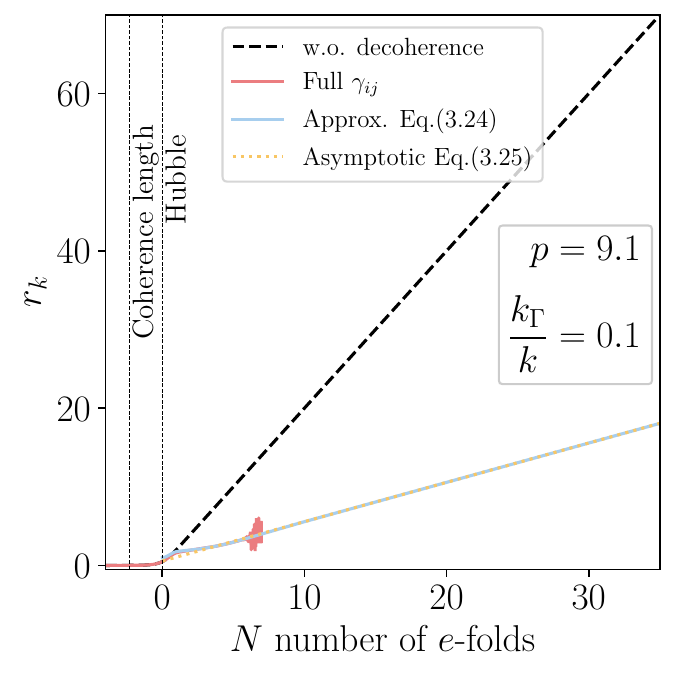}
  \caption{ Evolution of the squeezing parameter $r_k$ as a function of the number of $e$-folds in the presence of interaction for $k_{\Gamma}/k = 0.1$, $p=2.1$ (left panel) and $p=9.1$ (right panel). The full red line is obtained using Eq.~\eqref{eq:rk_exact} with the covariance-matrix elements $\gamma_{ij}$ computed with Eqs.~\eqref{eq:gij:exact} where the integrals of Eq.~\eqref{def:integrals_gammaij} are numerically evaluated. We stop plotting this line around $N=7$ $e$-folds when it starts to show numerical error for $p=9.1$. The full light blue line is computed using the late-time approximation~\eqref{eq:latetime_rk} of $r_k$ with the covariance-matrix elements evaluated from Eqs.~\eqref{eq:latetime_gammaij}. The yellow dotted line is computed using Eq.~\eqref{eq:cases_rk}.
  \label{fig:squeezing_deco}
  }
\end{figure}

We checked numerically that all terms appearing inside the logarithms and square roots are positive in the parameter ranges where they are evaluated. Note that for $p>8$, the change in the leading-order power of the $\gamma_{ij}$ compensates that in the leading power of $p_k$ in such a way that the leading power in $r_k$ no longer depends on $p$. Importantly, this asymptotic expression in $x$ does not take into account the order in $k_{\Gamma}/k$ appearing in the different terms. Thus, it can be inaccurate at late times $x \ll 1$ if, simultaneously, the interaction is very suppressed, $k_{\Gamma}/k \ll 1$, so that the terms coming from the interaction, although asymptotically dominant, are still subdominant. As an illustration, the approximation would fail for $k_{\Gamma}/k = 10^{-2}$ and $p=8.1$ even for $N=60$. We detail this point in Appendix~\ref{app:validity_cases}. Since in practice we want to consider $N=60$ and small coupling values, when performing numerical evaluations we use Eq.~\eqref{eq:latetime_rk}, where all terms are kept. In Fig.~\ref{fig:squeezing_deco} we compare the asymptotic form Eq.~\eqref{eq:cases_rk} to other expressions for the squeezing parameter $r_k$.

To isolate the effect of the interaction on the squeezing parameter $r_k$, we compute the difference with respect to the pure-state case (i.e. without decoherence), $\delta r_k = r_k - r^{\mathrm{w.o.}}_k$, where the pure-state expression $r_k^{\mathrm{w.o.}}$ is given in Eq.~\eqref{eq:no_deco_rk}. We find
\begin{equation}
\label{eq:cases_delta_rk}
\begin{alignedat}{3}
\delta r_k & = \frac{1}{2} \ln \left[ \frac{ 1
  - 2 \left(\frac{k_\Gamma}{k}\right)^2
  B_{22}\left( p , x_{E} \right) }{ \sqrt{1 + A_{\sigma} \left( p , x_{E} \right) }} \right] + \mathcal{O}\left(x^2\right)
  \quad & \mathrm{for} & \quad p<2 \, , \\
    & = \frac{1}{4} \left( p - 2 \right) \ln x + \frac{1}{2} \ln \left[ \frac{ 1
  - 2 \left(\frac{k_\Gamma}{k}\right)^2
  B_{22} \left( p , x_{E} \right) }{ \sqrt{B_{\sigma}\left( p , x_{E} \right) }} \right] + \mathcal{O}\left(x^2\right)
  \quad & \mathrm{for} & \quad 2<p<8 \, , \\
  & = \frac{3}{2} \ln x + \frac{1}{2} \ln \left[ \frac{
  - 2 \left(\frac{k_\Gamma}{k}\right)^2
  A_{22} \left( p \right) }{ \sqrt{D_{\sigma} \left( p , x_{E} \right) }} \right] + \mathcal{O}\left(x^2\right)
  \quad & \mathrm{for} & \quad p>8 \, .
\end{alignedat}
\end{equation}

\begin{figure}[H]
  \centering
  \begin{subfigure}[b]{0.48\textwidth}
    \includegraphics[width=\textwidth]{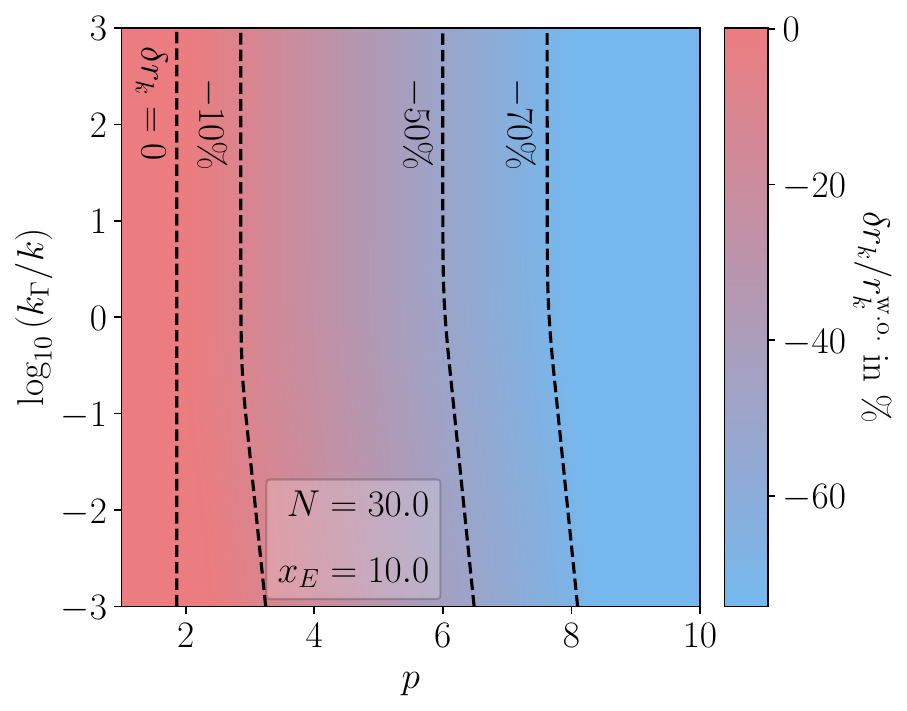}
    \caption{\label{fig:delta_r_N30} $N=30$}
  \end{subfigure}
  \hfill
  \begin{subfigure}[b]{0.48\textwidth}
    \includegraphics[width=\textwidth]{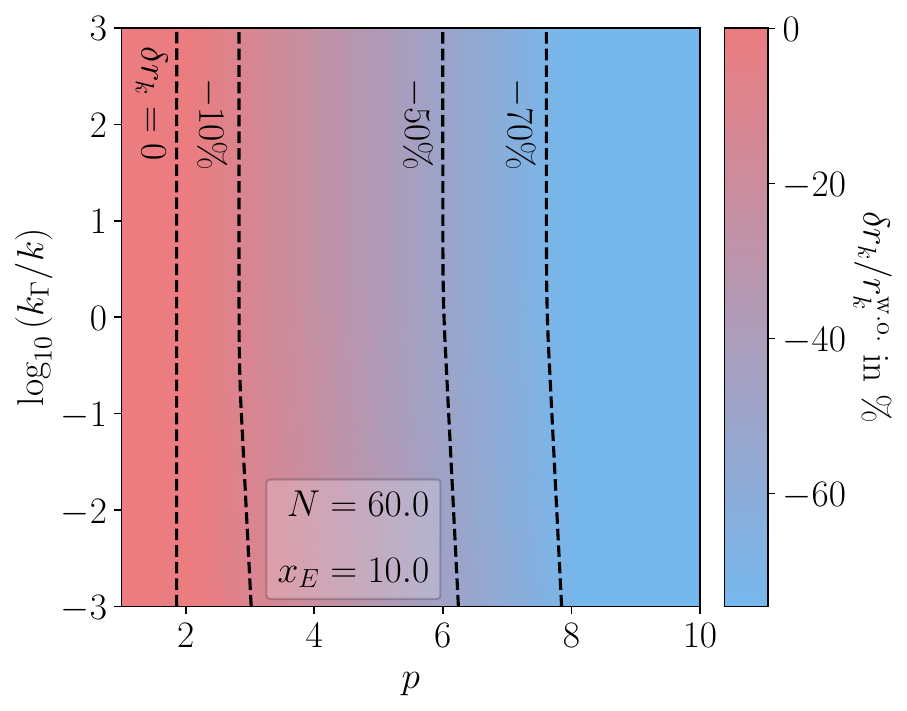}
    \caption{\label{fig:delta_r_N60} $N=60$}
  \end{subfigure}
  \caption{\label{fig:delta_rk}
  Values of $\delta r_k/ r_k^{\rm w.o.}=r_k/r_k^{\rm w.o.}-1$ in percentage as a function of $p$ and $\log_{10}{(k_{\Gamma}/k)}$ for $N=30.0$ (left) and $N=60$ (right) $e$-folds.
  }
\end{figure}

Let us comment on these expressions. First, Eq.~\eqref{eq:cases_rk} shows that, irrespective of the value of $p$, the squeezing parameter is asymptotically positive and diverging; at late times, the state of the perturbations is squeezed even in the presence of decoherence. Secondly, however, in this model decoherence typically reduces the squeezing, since for sufficiently small $x$, $\delta r_k < 0$ when $p>2$. When $p<2$, the squeezing can be increased, but only by a bounded, time-independent amount which vanishes as $k_{\Gamma}/k \to 0$ and tends to a constant in the large $k_{\Gamma}/k$ limit. The smallness of this enhancement of squeezing is confirmed by Fig.~\ref{fig:delta_rk}, where it is found to be at most of order $10^{-1}$ in the region $1<p<2$. Thirdly, the expressions Eqs.~\eqref{eq:cases_rk}--\eqref{eq:cases_delta_rk} are continuous in $p$. Indeed, first, it is straightforward to check that the coefficient of $\ln x$ is continuous, and, in addition, the next-order term is also continuous by taking the limits $p \to 2$ and $p \to 8$ on both sides of the equality. We illustrate the general de-squeezing behaviour by plotting $\delta r_k$, normalised by the pure-state value of $r_k$, in Fig.~\ref{fig:delta_rk} as a function of $p$ and $k_{\Gamma}/k$.

\begin{figure}
\centering
\begin{subfigure}{0.495\linewidth}
\centering
\includegraphics[width=\linewidth]{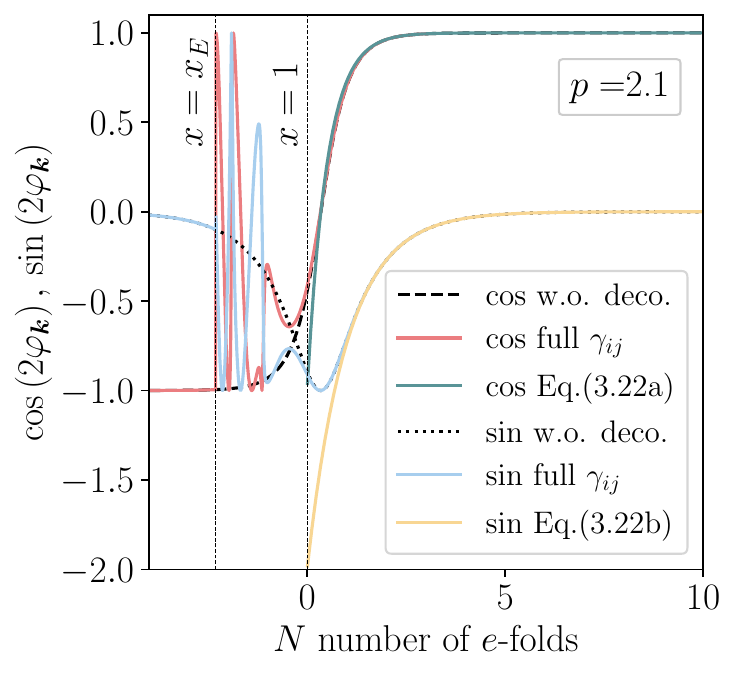}
\caption{$p=2.1$}
\end{subfigure}
\hfill
\begin{subfigure}{0.495\linewidth}
\centering
\includegraphics[width=\linewidth]{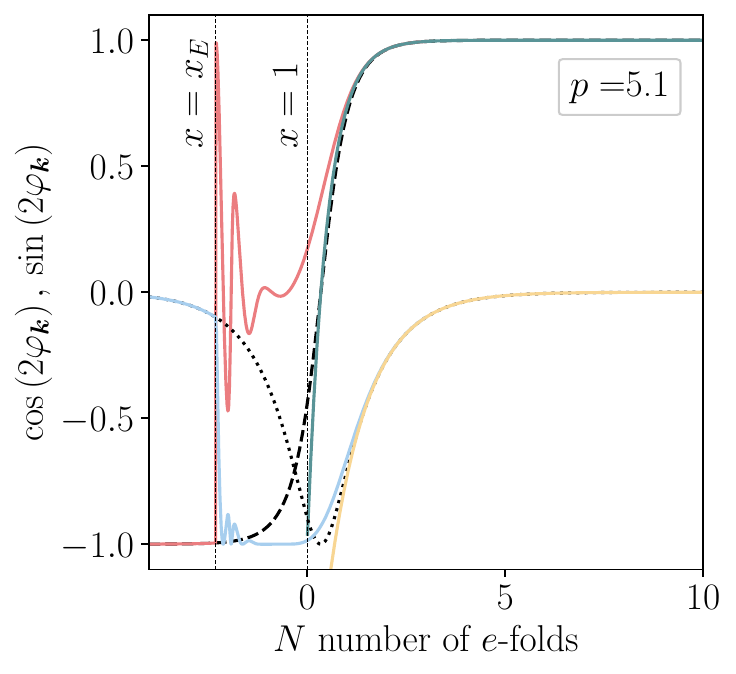} 
\caption{$p=5.1$}
\end{subfigure}
\caption{\label{fig:cos_sin_varying_p}
Evolution of $\cos(2 \varphi_k)$ and $\sin (2 \varphi_k)$ as functions of $N$, the number of $e$-folds, in the presence of decoherence, for fixed $k_\Gamma/k = 10$ and varying $p$. The dashed (respectively dotted) black line shows the values of $\cos$ (resp. $\sin$) without decoherence. The red (respectively blue) line represents the values of $\cos$ (respectively $\sin$) with decoherence, evaluated using Eqs.~\eqref{eq:sin_cos_varphik_exact}, with the covariance-matrix elements computed using Eqs.~\eqref{eq:gij:exact} by numerically evaluating the integrals~\eqref{def:integrals_gammaij}. The green (respectively yellow) line shows the values of cos  (respectively $\sin$) with decoherence, evaluated using the asymptotic expressions Eqs.~\eqref{eq:cases_sin_cos_2phik}. These last expressions are only plotted after the Hubble-crossing time $x = 1$, shown by a vertical dotted line. For reference, we also show the coherence-length crossing time $x = x_E$ with a second vertical dotted line.
}
\end{figure}

\begin{figure}
\centering
\begin{subfigure}{0.495\linewidth}
\includegraphics[width=\linewidth]{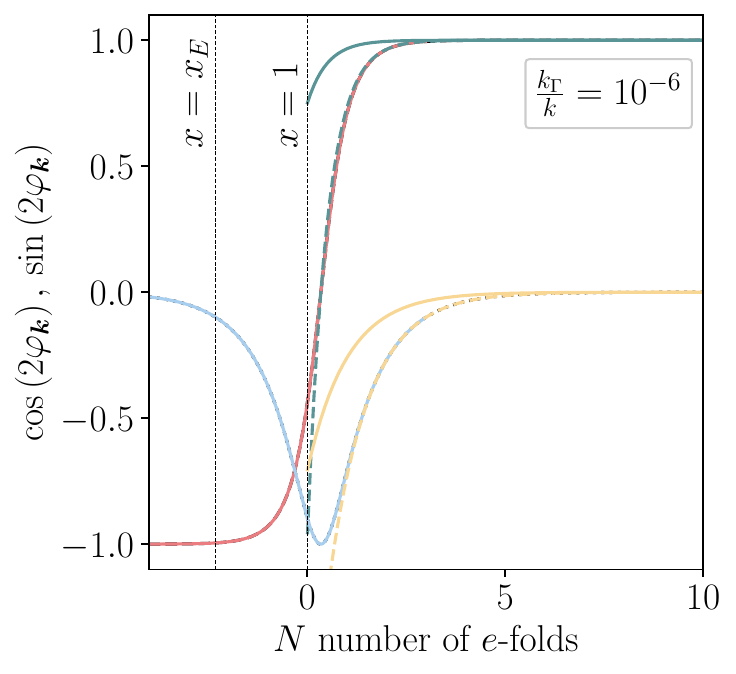}
\caption{\label{fig:sin_cos_kgam1e-06} $k_\Gamma/k =10^{-6}$}
\end{subfigure}
\hfill
\begin{subfigure}{0.495\linewidth}
\includegraphics[width=\linewidth]{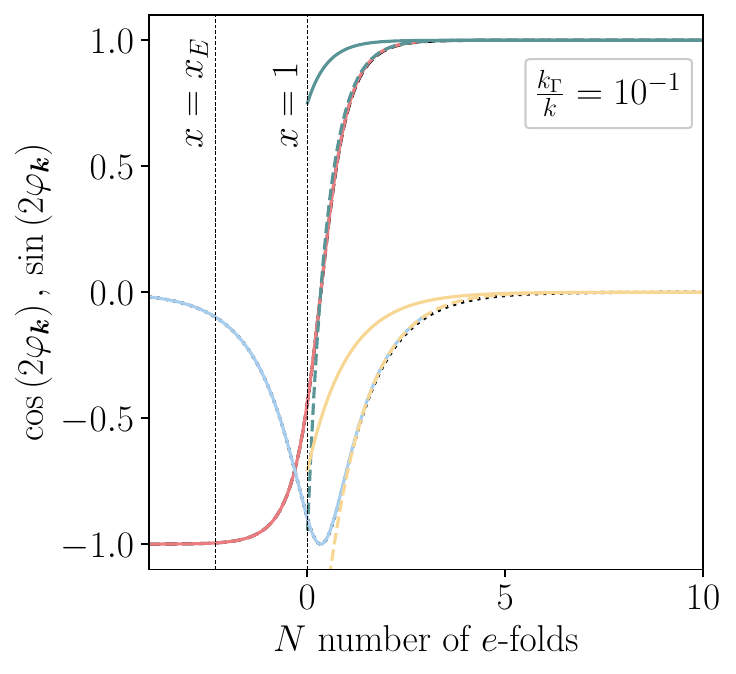} 
\caption{$k_\Gamma/k =10^{-1}$}
\end{subfigure}
\hfill
\begin{subfigure}{0.495\linewidth}
\includegraphics[width=\linewidth]{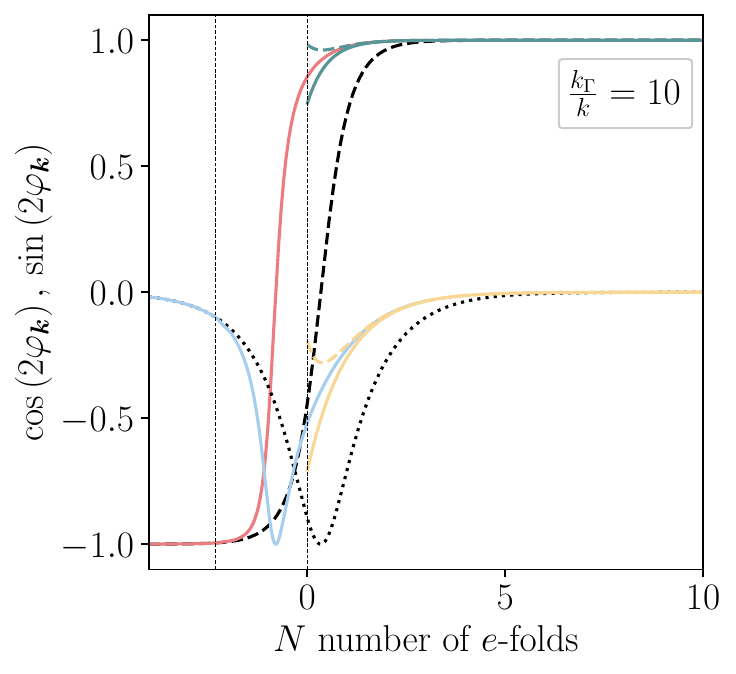} 
\caption{$k_\Gamma/k =10$}
\end{subfigure}
\hfill
\begin{subfigure}{0.495\linewidth}
\includegraphics[width=\linewidth]{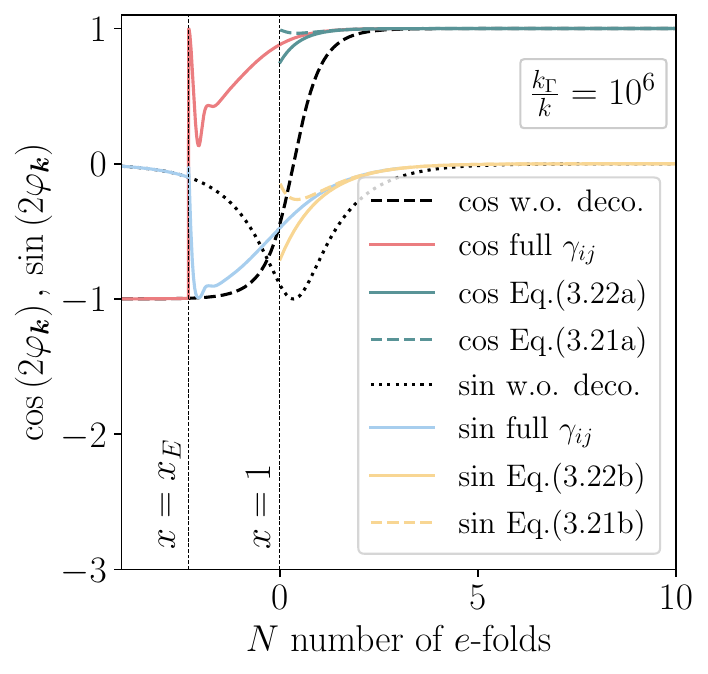} 
\caption{$k_\Gamma/k =10^{6}$}
\end{subfigure}
\caption{\label{fig:cos_sin_varying_kgam}
Evolution of $\cos(2 \varphi_k)$ and $\sin (2 \varphi_k)$ as functions of $N$, the number of $e$-folds, in the presence of decoherence, for fixed $p = 9.1$ and varying $k_\Gamma/k$. The same legend as Fig.~\ref{fig:cos_sin_varying_p} is used, where in addition we plot, as dashed green (respectively yellow) lines, the values of $\cos$ (respectively $\sin$) computed using the late-time expansions in Eqs.~\eqref{eq:latetime_sin_cos_2phik}, where the covariance-matrix elements $\gamma_{ij}$ are evaluated using their late-time expansions given in Eqs.~\eqref{eq:latetime_gammaij}.
}
\end{figure}

Finally, we compute the effect of decoherence on the squeezing angle $\varphi_k$. We start by deriving the expansions of both $\sin (2\varphi_k)$ and $\cos (2\varphi_k)$. The details are given in Appendix~\ref{app:squeezing_angle}. First, we obtain expressions useful for numerical computations by expanding Eq.~\eqref{eq:sin_cos_varphik_exact}. We have
\begin{subequations}
\label{eq:latetime_sin_cos_2phik}
\begin{align}
\cos \left( 2 \varphi_k \right)  & =  1 - 2 \left( \frac{\gamma_{12}}{\gamma_{22}} \right)^2 + \mathcal{O} \left( x^4 \right) \, , \\
\sin \left( 2 \varphi_k \right)  & = - 2 \frac{\gamma_{12}}{\gamma_{22}} + \mathcal{O} \left( x^3 \right) \, .
\end{align}
\end{subequations}
Then, inserting the expansions of $\gamma_{ij}$ and distinguishing cases for $p$, we obtain
\begin{subequations}
\label{eq:cases_sin_cos_2phik}
\begin{align}
\label{eq:cases_cos_2phik}
\cos \left( 2 \varphi_k \right) & = 1 - 2 x^2 \left[ \Theta \left(8-p \right) + \frac{A_{12}^2}{A_{22}^2} \Theta \left(p- 8\right) \right]  + \mathcal{O}\left(x^4\right) \, , \\
\label{eq:cases_sin_2phik}
\sin \left( 2 \varphi_k \right) &  = -  2 x \left[ \Theta \left( 8-p \right) + \frac{A_{12}}{A_{22}} \Theta \left(p - 8\right) \right] + \mathcal{O}\left(x^3\right) \, .
\end{align} 
\end{subequations}
One can check numerically that these expressions are also continuous at $p=8$. We have $A_{12}/A_{22}>0$ for $p>8$, so that, asymptotically, $\sin ( 2 \varphi_k ) < 0$ and $\cos ( 2 \varphi_k ) > 0$. Thus, $2 \varphi_k \in [-\pi/2 , \pi/2]$, and so $2 \varphi_k = \mathrm{arctan} [ \tan (2 \varphi_k )]$ without additional $\pi$ terms. We obtain
\begin{align}
\begin{split}
\label{eq:cases_phik}
\varphi_k  &  = - x \left[ \Theta \left(8-p\right) + \frac{A_{12}}{A_{22}} \Theta \left(p- 8\right) \right] + \mathcal{O}\left(x^3\right) \, .
\end{split}    
\end{align}
We can also define the difference with the de Sitter expression, $\delta \varphi_k = \varphi_k -  \varphi_k^{\mathrm{w.o.}}$, to see how decoherence affects the squeezing angle. Indeed, as can be seen from the above expression, we always obtain $\delta \varphi_k \to 0$: decoherence, when modelled by Eq.~\eqref{eq:lindblad_Caldeira_Leggett}, does not affect the asymptotic behaviour of the squeezing angle $\varphi_k$. Therefore, a colormap similar to Fig.~\ref{fig:delta_rk} would not be very informative, and we instead plot the values of $\sin (2\varphi_k)$ and $\cos (2\varphi_k)$ as functions of time, labelled by the number of $e$-folds. In Fig.~\ref{fig:cos_sin_varying_p} we show their evolution for different values of $p$ at fixed $k_{\Gamma}/k$, and in Fig.~\ref{fig:cos_sin_varying_kgam} we show their evolution for different values of $k_{\Gamma}/k$ at fixed $p$. In particular, Fig.~\ref{fig:sin_cos_kgam1e-06}, corresponding to $k_{\Gamma}/k = 10^{-6}$ and $p=9.1$, demonstrates, as detailed in Appendix~\ref{app:validity_cases}, that the asymptotic forms Eqs.~\eqref{eq:cases_sin_cos_2phik} perform poorly for $p>8$ and small $k_{\Gamma}/k$, while Eqs.~\eqref{eq:latetime_sin_cos_2phik} remain accurate.

\section{Effect of decoherence on quantum correlations \label{sec:quantum_corr}}

While most studies of decoherence focus on computing how interactions degrade the purity of the state, e.g.~\cite{Burgess:2022nwu,Lopez:2025arw,Burgess:2025dwm}, it is in principle necessary to know the full quantum state in order to properly diagnose the presence or absence of quantum features. In particular, for the class of states and partitions considered here, quantifiers of quantum correlations depend on the relative magnitudes of the purity $p_k$ and the squeezing parameter $r_k$~\cite{Martin:2022kph}. Since we have shown in the previous section that the squeezing parameter $r_k$ can also be strongly affected by decoherence, we analyse here how this modification translates into the evaluation of three quantum-correlation criteria: logarithmic negativity, quantum discord, and a Bell-inequality violation. First, we briefly present these criteria, and then we use them to evaluate the quantumness of correlations in the presence of interactions using the results of Sec.~\ref{sec:deco_squeezing}.

\subsection{Quantifiers of quantum correlations}
\label{sec:quantif_quantum}

\subsubsection{Non-separability and Logarithmic Negativity}

First, we consider the separability of the state, which is the natural generalisation of the notion of entanglement to mixed, i.e. non-pure, states. The $\pm \bm{k}$ modes are said to be in a non-separable state if their density matrix cannot be written as a sum of products of density matrices, i.e. we cannot find coefficients $\{ \lambda_i \}$ and density matrices $\{ \hat{\varrho}_{+\bm{k}}^i \}$ and $\{ \hat{\varrho}_{-\bm{k}}^i \}$ acting respectively on $H_{+\bm{k}}$ and $H_{-\bm{k}}$ such that $\hat{\varrho}_{\pm \bm{k}} = \sum_i \lambda_i \hat{\varrho}_{+\bm{k}}^i \otimes \hat{\varrho}_{-\bm{k}}^i$. In the rest of this work, when referring to entangled states, we mean non-separable states. For bipartite Gaussian states it has been shown~\cite{Simon:1999lfr} that the Peres–Horodecki criterion~\cite{Peres:1996dw,Horodecki:1996nc} provides a necessary and sufficient condition for separability, which can be simply expressed in terms of covariance-matrix elements. In the $\pm \bm{k}$ partition, the condition takes a particularly elegant form. One can show~\cite{Martin:2022kph} that the state is non-separable if and only if the semi-minor axis of the squeezing ellipse is shorter than in the vacuum, i.e.
\begin{equation}
\label{eq:bk_nonsep}
    b_k = p_k^{-1/4} e^{- r_k} < 1 \, ,
\end{equation}
where we used the expression~\eqref{def:axes_ellipse} for $b_k$. To obtain a quantifier of the degree of this entanglement, we can use the logarithmic negativity~\cite{Vidal:2002zz}, which for this class of states is simply given by~\cite{Brady:2022ffk}
\begin{equation}
\label{def:LN}
    \mathrm{LN} = \mathrm{max} \left[ 0 , - \log_2 \left( b_k^2 \right) \right] = \mathrm{max} \left[ 0 , - \frac{2}{\ln 2} \ln \left( b_k \right) \right] \, .
\end{equation}

\subsubsection{Quantum discord}

Secondly, we discuss the quantum discord introduced in~\cite{Ollivier:2001fdq,Henderson:2001wrr}, which allows one to distinguish classical from quantum correlations between two systems. We consider here the systems $+\bm{k}$ and $-\bm{k}$ defined by the partition $\pm \bm{k}$. Briefly, the quantum discord is built from two expressions, $\mathcal{I}_{\pm \bm{k}}$ and $\mathcal{J}_{\pm \bm{k}}$, for the mutual information, an entropy-based quantifier of correlations, between the systems $+\bm{k}$ and $-\bm{k}$. The expression for $\mathcal{I}_{\pm \bm{k}}$ only involves the entropies of $+\bm{k}$, $-\bm{k}$ and of the joint system $\pm \bm{k}$, whereas that of $\mathcal{J}_{\pm \bm{k}}$ involves conditional entropies. These two expressions coincide in their classical definitions thanks to Bayes’ theorem, but can differ in the quantum setting where, intuitively, conditional entropies require specifying a set of measurement operators that disturb the state. The quantum discord $\mathcal{D}_{\pm \bm{k}}$ is defined as their difference, $\mathcal{D}_{\pm \bm{k}} = \mathcal{I}_{\pm \bm{k}} - \mathcal{J}_{\pm \bm{k}}$, which can be shown to be positive. Since its classical counterpart vanishes, it quantifies the strictly quantum correlations between $+\bm{k}$ and $-\bm{k}$. Properties and detailed expressions of the discord can be found in~\cite{Ollivier:2001fdq,Henderson:2001wrr,Bera:2017lmd}. The quantum discord is in general difficult to evaluate; however, exact expressions are known for Gaussian states in terms of the covariance-matrix elements~\cite{Adesso:2014npz}. Its expression for scalar cosmological perturbations in the presence of the decoherence channel~\eqref{eq:lindblad_Caldeira_Leggett} was derived in~\cite{Martin:2021znx}. For the partition $\pm \bm{k}$, we have
\begin{equation}
\label{eq:discord_p_r}
\mathcal{D}_{\pm \bm{k}} = f\!\left[\sigma\left(p_k,r_k\right)\right] - 2 f\!\left(p_k^{-1/2}\right) + f\!\left[\frac{\sigma\left(p_k,r_k\right)+p_k^{-1}}{\sigma\left(p_k,r_k\right)+1}\right] \, ,
\end{equation}
where
\begin{equation}
\label{eq:defsigma}
\sigma\left(p_k,r_k\right)= p_k^{-1/2} \cosh\!\left(2 r_k\right) \, .
\end{equation}
The function $f(x)$ is defined for $x \geq 1$ by
\begin{equation}
\label{def:function_f}
    f(x)=\left(\frac{x+1}{2}\right)\log\!\left(\frac{x+1}{2}\right)-\left(\frac{x-1}{2}\right)\log\!\left(\frac{x-1}{2}\right) .
\end{equation}
Note that since $p_k \leq 1$, we have $\sigma\!\left(p_k,r_k\right) \geq 1$, and all arguments of $f$ in Eq.~\eqref{eq:discord_p_r} are therefore larger than $1$.

\subsubsection{An asymptotic form of the discord}

Before moving to the specific case of cosmological perturbations in de Sitter space, we derive an asymptotic expression for the discord of a generic two-mode thermal squeezed state in the regime of large squeezing, $e^{-r_k} \ll 1$, and small purity, $p_k \ll 1$. This approximation therefore applies to the late-time state of cosmological perturbations evolving via Eq.~\eqref{eq:lindblad_Caldeira_Leggett} for $p>2$ in our setup (see Sec.~\ref{sec:deco_squeezing}). The first step is to identify the relevant contributions to the discord by singling out the terms that are negligible because they depend only on $e^{-4 r_k}$ and $p_k$.
For this purpose, we follow~\cite{Martin:2022kph} and rewrite the discord in terms of the lengths of the semi-minor axis $b_k$ and semi-major axis $a_k$. Note that, while large squeezing, $e^{-r_k} \ll 1$, and small purity, $p_k \ll 1$, imply a large semi-major axis, $a_k \gg 1$, they do not fix the length of the semi-minor axis $b_k$. This suggests that this quantity may control the magnitude of the discord in this regime. We then rewrite the discord in terms of $b_k$, $p_k$ and $e^{-r_k}$ (see Appendix~\ref{app:approx_q_measures} for the derivation) as
\begin{align}
\begin{split}
\label{eq:discord_expansion_ready}
\mathcal{D}_{\pm \bm{k}}
& = \log_2 \left[ 1 + \frac{1}{2 b_k^2} - p_k + \frac{p_k^{1/2} e^{-2 r_k}}{2} - 2 p_k^{1/2} e^{-2 r_k} \frac{1 - p_k}{1 + e^{-4 r_k} + 2 p_k^{1/2} e^{-2 r_k}} \right] \\
& \qquad\qquad
+ g\left( \frac{1 + e^{-4 r_k}}{2 p_k^{1/2} e^{-2 r_k}} \right) - 2 g\left(p_k^{-1/2}\right)
+ g\left( 1 + 2 b_k^2 \frac{1 - p_k}{1 + e^{-4 r_k} + 2 p_k^{1/2} e^{-2 r_k}} \right) \, ,
\end{split}
\end{align}
where
\begin{equation}
\label{def:function_g}
  g(x) = f(x) - \log_{2} \left( \frac{x}{2} \right) - \frac{1}{\ln 2} \, ,
\end{equation}
is a negative, bounded and strictly increasing function.
This equation is still exact, but, similarly to Eq.~\eqref{eq:latetime_rk}, it is useful for numerical evaluations since the dominant contributions are clearly identified in the first term of each evaluation of $f$ and $g$, and are only corrected by adding small contributions. We then expand the different functions in the limits $e^{-r_k} \ll 1$ and $p_k \ll 1$, and obtain
\begin{equation}
\label{eq:discord_semi_minor_squeezing}
\begin{alignedat}{2}
\mathcal{D}_{\pm \bm{k}}
& = \log_2 \Bigl[ 1 + \frac{1}{2 b_k^2} - p_k + \frac{3}{2} p_k^{1/2} e^{-2 r_k}
+ \mathcal{O} \left( p_k^{3/2} e^{-2 r_k} \right)
+ \mathcal{O} \left( p_k^{1/2} e^{-6 r_k} \right)
+ \mathcal{O} \left( p_k e^{-4 r_k} \right) \Bigr] \\[0.3ex]
& \qquad\qquad + \frac{p_k}{3 \ln 2} + \mathcal{O} \left( p_k^{2} \right)
+ \mathcal{O} \left( p_k e^{-4 r_k} \right) + g \Bigl\{ 1 + 2 b_k^2 \bigl[ 1 - p_k - e^{-4 r_k} - 2 a_k^{-2} \\[0.3ex]
& \hspace{5cm}
+ \mathcal{O} \left( p_k^{3/2} e^{-2 r_k} \right)
+ \mathcal{O} \left( p_k^{1/2} e^{-6 r_k} \right)
+ \mathcal{O} \left( p_k e^{-4 r_k} \right) \bigr] \Bigr\} \, .
\end{alignedat}
\end{equation}
Equation~\eqref{eq:discord_semi_minor_squeezing} is a refinement of Eq.~(31) of~\cite{Martin:2022kph}, where we explicitly give the subdominant contributions for small but finite $p_k$ and $e^{-r_k}$. It shows that, for a generic two-mode squeezed thermal state with small purity and large squeezing, the value of the discord is controlled by the size of the semi-minor axis $b_k$. Since $b_k$ at late times for cosmological perturbations is either very large or very small (see Sec.~\ref{sec:quantumness_dS_deco}), we further expand Eq.~\eqref{eq:discord_semi_minor_squeezing} in these two limits.
We obtain
\begin{equation}
\label{eq:discord_approx_semi_minor}
\begin{alignedat}{3}
\mathcal{D}_{\pm \bm{k}}
& = - 2 \log_2 b_k - \frac{1}{\ln 2}
+ \mathcal{O} \left( b_k^2 \ln b_k \right)
+ \mathcal{O} \left( p_k \right)
\quad & \mathrm{for} & \quad b_k \ll 1 \, , \\
& = \frac{b_k^{-2}}{2 \ln 2}
+ \mathcal{O} \left( b_k^{-4} \right)
+ \mathcal{O} \left( p_k \right)
\quad & \mathrm{for} & \quad b_k \gg 1 \, .
\end{alignedat}
\end{equation}
It is natural that the value of the discord is controlled by the size of the semi-minor axis~$b_k$. First, since it is a ratio between the squeezing and the purity, $b_k^2 = e^{-2 r_k} / \sqrt{p_k}$, it characterises the competition between squeezing and decoherence. Moreover, as shown in Eq.~\eqref{def:LN}, it controls the separability of the state and the value of the logarithmic negativity. The coincidence of these different correlation measures is illustrated in Fig.~\ref{fig:quantum_correlations} below.

\subsubsection{GKMR Bell inequality}

Finally, we discuss the violation of a Bell inequality as a quantifier of quantum correlations. In its most standard CHSH form~\cite{Clauser:1969ny}, the inequality involves a combination of expectation values for spin degrees of freedom. While we are dealing here with bosonic modes, we can define pseudo-spin operators, obeying a spin algebra, for which we can write a Bell inequality in the usual form. Several such pseudo-spin operators can be defined; here we adopt those proposed in~\cite{Gour:2003wsm}. For a bosonic pair $\hat{q}, \hat{\pi}$, they are defined using the eigenstates $\ket{q}$ of the position operator $\hat{q}$. We have
\begin{subequations}
\begin{alignat}{2}
    \hat{\sigma}_x 
    &=& \int_{-\infty}^{\infty} \mathrm{sign}(q) \ket{q} \bra{q} \dd q \,, \\
  \hat{\sigma}_y
    &=& - i \int_{-\infty}^{\infty} \mathrm{sign}(q) \ket{q} \bra{q} \dd q \,, \\
  \hat{\sigma}_z
    &=& - \int_{-\infty}^{\infty} \ket{q} \bra{-q} \dd q \, ,
\end{alignat}
\end{subequations}
and one can check that these operators satisfy the spin commutation relations
\begin{equation}
    [\hat{\sigma}_\mu , \hat{\sigma}_\nu] = 2 i \epsilon^{\mu\nu\lambda} \hat{\sigma}_\lambda \, .
\end{equation}
We denote $\hat{\sigma}_i^{\bm{k}}$ the operators constructed from the position operator $\hat{q}_{\bm{k}}$ of the pair $\hat{q}_{\bm{k}}, \hat{\pi}_{\bm{k}}$ associated to the mode $\bm{k}$. Then, using the operators for the pair of modes $\pm \bm{k}$, we can define a Bell operator $\hat{B}_{\pm \bm{k}}$ such that its expectation value is given by~\cite{Gour:2003wsm}
\begin{equation}
    \ev{\hat{B}_{\pm \bm{k}}}
    = 2 \sqrt{
        \ev{\hat{\sigma}_z^{\bm{k}} \hat{\sigma}_z^{-\bm{k}}}^2
        + \ev{\hat{\sigma}_x^{\bm{k}} \hat{\sigma}_x^{-\bm{k}}}^2
    } \,.
\end{equation}
One can show that, if the measured values of the spins are described by a joint probability distribution, then $\ev{\hat{B}_{\pm \bm{k}}} \leq 2$. Thus, a quantum state for which
\begin{equation}
\label{eq:bell_inequality}
\ev{\hat{B}_{\pm \bm{k}}} > 2 \,,
\end{equation}
that is, for which the Bell inequality is violated, exhibits features that cannot be explained by a classical probability distribution.
One can show that, for a two-mode squeezed thermal state, such as that obtained for the cosmological perturbations considered here, $\ev{\hat{\sigma}_z^{\bm{k}} \hat{\sigma}_z^{-\bm{k}}}$ and $\ev{\hat{\sigma}_x^{\bm{k}} \hat{\sigma}_x^{-\bm{k}}}$ can be expressed in terms of the squeezing parameter $r_k$, the squeezing angle $\varphi_k$ and the purity $p_k$ as~\cite{Martin:2022kph}
\begin{eqnarray}
&& \ev{\hat{\sigma}_z^1 \hat{\sigma}_z^2} = p_k \,, \\
&& \ev{\hat{\sigma}_x^1 \hat{\sigma}_x^2}
= - \frac{2}{\pi}
\arcsin \left[ \lvert \cos (2 \varphi_k) \rvert \tanh (2 r_k) \right] \, .
\end{eqnarray}
The expectation value of the Bell operator then reads
\begin{equation}
\label{eq:Bell_ev}
    \ev{\hat{B}_{\pm \bm{k}}}
    = 2 \sqrt{
        p_k^2
        + \frac{4}{\pi^2}
        \arcsin^2 \left[ \lvert \cos (2 \varphi_k) \rvert \tanh (2 r_k) \right]
    } \,.
\end{equation}

\subsection{Decoherence in de Sitter}
\label{sec:quantumness_dS_deco}

\subsubsection{Late-time expansions}
\label{sec:latetime_quantumness_deco}

We now discuss how the open dynamics of Eq.~\eqref{eq:lindblad_Caldeira_Leggett} affect the measures of quantum correlations as functions of the phenomenological parameters $k_\Gamma$ and $p$. We are particularly interested in the expressions of these measures at late times, $x \to 0^{+}$. First, we consider the logarithmic negativity, which depends only on the parameter $b_k = p_k^{-1/4} e^{-r_k}$, see Eq.~\eqref{def:LN}. In the absence of decoherence, we have $p_k = 1$, and thus
\begin{equation}
\label{eq:latetime_bk_purestate}
b_k^{\mathrm{w.o.}} = e^{-r_k^{\mathrm{w.o.}}} = x^2 + \mathcal{O}\left(x^6\right) \, .
\end{equation}
In the presence of decoherence, we can adopt Eq.~\eqref{eq:latetime_rk}, which is useful for numerical computations, to obtain
\begin{equation}
\label{eq:latetime_bk}
\ln b_k = \frac{1}{2} \ln \left( \frac{p_k^{-1}}{\gamma_{22}} \right) - \left( \frac{\gamma_{12}}{\gamma_{22}} \right)^2 + \mathcal{O} \left( x^{4} \right) \, .
\end{equation}
We can then expand $p_k$ using Eq.~\eqref{eq:latetime_purity}, and $\gamma_{22}$ using Eq.~\eqref{eq:latetime_gamma22}, and distinguish cases in $p$ to obtain
\begin{equation}
\label{eq:cases_bk}
\begin{alignedat}{3}
\ln b_k & = \frac{1}{2} \ln \left[ \frac{x^4 \left( 1 + A_{\sigma} \right)}{1 - 2 \left( \frac{k_\Gamma}{k} \right)^2 B_{22}} \right] + \mathcal{O}\left(x^{2-p}\right) \quad & \mathrm{for} & \quad p<2 \, , \\
& = \frac{1}{2} \ln \left[ \frac{x^{6-p} B_{\sigma}}{1 - 2 \left( \frac{k_\Gamma}{k} \right)^2 B_{22}} \right] + \mathcal{O}\left(x^{p-2}\right) + \mathcal{O}\left(x^2\right) \quad & \mathrm{for} & \quad 2<p<8 \, , \\
& = \frac{1}{2} \ln \left[ \frac{x^{6-p} D_{\sigma}}{- 2 \left( \frac{k_\Gamma}{k} \right)^2 A_{22}} \right] + \mathcal{O}\left(x^2\right) \quad & \mathrm{for} & \quad 8<p \, .
\end{alignedat}
\end{equation}
Here we keep the dependence of the coefficients $A_{ij}$, etc., on $p$, $x_E$ and $k_\Gamma$ implicit for ease of display.
Again, note that the domain of validity of the approximation for $p>8$ depends implicitly on $k_\Gamma$, see Appendix~\ref{app:validity_cases}. One can check that this expression is continuous across the transitions. Note also that, for $p<2$, the first-order term matches the pure-state case~\eqref{eq:latetime_bk_purestate}, so that the interactions do not affect the logarithmic negativity at first order in this case. One can then obtain an approximation for the separability threshold as a function of $x$, $k_\Gamma$ and $p$ by solving $b_k = 1$ in the above expression. Working at first non-trivial order in $x$, we find that $p = 6$ is a critical value: at late times, for $p<6$ the state is entangled, while it is separable for $p>6$.

Similarly, we derive a late-time expansion for the quantum discord. In the absence of decoherence, we have $p_k = 1$, and thus
\begin{equation}
\label{eq:latetime_discord_purestate}
\mathcal{D}_{\pm \bm{k}}^{\mathrm{w.o.}} = f \left[ 1 + \frac{1}{2 x^4} \right]
= - 4 \log_2 x + \frac{1}{\ln 2} - 2 + \mathcal{O}\left(x^4\right) \, .
\end{equation}
In the presence of decoherence, we start from Eq.~\eqref{eq:discord_expansion_ready} which can be numerically evaluated using the asymptotic expressions for $b_k$, $p_k$ and $e^{-4 r_k}$. To obtain an explicit form, we again have to distinguish cases in $p$ to identify the dominant term. First, at late times, for $p>2$ the purity $p_k$ is small and the squeezing is large $e^{-r_k} \ll 1$. We can thus directly use Eq.~\eqref{eq:discord_approx_semi_minor} together with the late-time approximation Eq.~\eqref{eq:latetime_bk}. Second, for $p<2$ we need to expand other terms more carefully; this is done in Appendix~\ref{app:approx_q_measures}, and the resulting expression is given in Eq.~\eqref{eq:cases_discord_bis}. Combining the two, we obtain
\begin{equation}
\label{eq:cases_discord}
\begin{alignedat}{3}
\mathcal{D}_{\pm \bm{k}} & = - 4 \log_2 x
+ \log_2 \left[ 1 - 2 \left( \frac{k_\Gamma}{k} \right)^2 B_{22} \right]
+ \frac{1}{\ln 2} - 2 \\
& \qquad - 2 f \left[ \sqrt{1 + A_{\sigma}} + \mathcal{O}\left(x^{2-p}\right) \right]
+ \mathcal{O}\left(x^{2-p}\right) \quad & \mathrm{for} & \quad p<2 \, , \\
& = - (6-p) \log_2 x
+ \log_2 \left[ \frac{1 - 2 \left( \frac{k_\Gamma}{k} \right)^2 B_{22}}{B_{\sigma}} \right]
- \frac{1}{\ln 2} \\
& \qquad + \mathcal{O}\left(x^{p-2}\right)
+ \mathcal{O}\left(x^2\right)
+ \mathcal{O}\left(x^{6-p} \ln x\right)
\quad & \mathrm{for} & \quad 2<p<6 \, , \\
& = x^{p-6} \, \frac{1 - 2 \left( \frac{k_\Gamma}{k} \right)^2 B_{22}}{2 B_{\sigma} \ln 2}
+ \mathcal{O}\left(x^{p-4}\right)
\quad & \mathrm{for} & \quad 6<p<8 \, , \\
& = x^{p-6} \, \frac{- \left( \frac{k_\Gamma}{k} \right)^2 A_{22}}{D_{\sigma} \ln 2}
+ \mathcal{O}\left(x^{p-4}\right)
\quad & \mathrm{for} & \quad 8<p \, .
\end{alignedat}    
\end{equation}
One can straightforwardly check that this expression is continuous in $p$ except at $p = 2$, which requires more careful treatment. Note that, for $p<2$, as for the logarithmic negativity, the first-order term matches the pure-state case~\eqref{eq:latetime_discord_purestate}.

Finally, we turn to the GMKR Bell-operator expectation value. In the absence of decoherence, we have
\begin{equation}
\label{eq:latetime_Bell_purestate}
\frac{\ev{\hat{B}_{\pm \bm{k}}}_{\mathrm{w.o.}}^{2}}{4}
= 2 - \frac{8 x}{\pi} + \mathcal{O}\left(x^{2}\right) \, .
\end{equation}
In the presence of decoherence, we approximate this expectation value using the late-time expansions of the squeezing parameter~\eqref{eq:cases_rk} and of $\cos (2 \varphi_k)$, as detailed in Appendix~\ref{app:approx_q_measures}. We then obtain
\begin{equation}
\label{eq:latetime_Bell}
\frac{\ev{\hat{B}_{\pm \bm{k}}}^{2}}{4}
= 1 + p_k^2 - \frac{8}{\pi} \sqrt{\frac{\gamma_{11}}{\gamma_{22}}}
+ \mathcal{O}\left(x^2\right) \, .
\end{equation}
Finally, we distinguish cases to write a more explicit expression using the expansion of the purity~\eqref{eq:latetime_purity} and of the covariance-matrix elements Eqs.~\eqref{eq:latetime_gammaij}. We obtain
\begin{equation}
\label{eq:cases_Bell}
\begin{alignedat}{3}
\frac{\ev{\hat{B}_{\pm \bm{k}}}^{2}}{4}
& = 1 + \frac{1}{\left[ 1 + A_{\sigma} \left( p , x_{E} \right) \right]^2}
- \frac{8 x}{\pi}
+ \mathcal{O}\left(x^{2-p}\right)
\quad & \mathrm{for} & \quad p<2 \, , \\
& = 1 + \frac{x^{2p-4}}{B_{\sigma} \left( p , x_{E} \right)^2}
- \frac{8 x}{\pi}
+ \mathcal{O}\left(x^{2p-2}\right)
+ \mathcal{O}\left(x^3\right)
\quad & \mathrm{for} & \quad 2<p<8 \, , \\
& = 1 - \frac{8 x}{\pi} \sqrt{\frac{A_{11}}{A_{22}}}
+ \mathcal{O}\left(x^3\right)
\quad & \mathrm{for} & \quad 8<p \, .
\end{alignedat}
\end{equation}

At first order in $x$, for $p<2$, the Bell inequality is violated asymptotically since $\ev{\hat{B}_{\pm \bm{k}}} \to 2 + 2/(1 + A_{\sigma})^2$. For $2<p<5/2$, the first term in the second line of Eq.~\eqref{eq:cases_Bell}, which is positive, dominates, and thus the expectation value approaches $2$ from above, $\ev{\hat{B}_{\pm \bm{k}}} \to 2^{+}$. Although the inequality is not violated asymptotically in this case, it is always slightly violated for any finite $x$. On the other hand, for $p>5/2$, the second term, which is negative, dominates and we have $\ev{\hat{B}_{\pm \bm{k}}} \to 2^{-}$. The Bell inequality is therefore not violated asymptotically. Finally, for $p>8$, the term coming from the purity is of order $\mathcal{O}\left(x^{4p-20}\right)$ and is thus negligible even compared with the first-order term $\mathcal{O}\left(x^3\right)$. As a result, the approximation always gives $\ev{\hat{B}_{\pm \bm{k}}} < 2$, and we cannot identify the critical point where $\ev{\hat{B}_{\pm \bm{k}}} = 2$. Such a point exists, but would correspond either to an extremely small number of $e$-folds after coherence-length crossing, or to an extremely small value of $k_\Gamma/k$, in order to suppress the effect of decoherence. In summary, asymptotically $p = 5/2$ is a critical value: for $p<5/2$ the Bell inequality is violated as $x \to 0^{+}$, and it is not violated for $p>5/2$.

\subsubsection{Discussion}

\begin{figure}[H]
  \begin{subfigure}[b]{\textwidth}
    \includegraphics[width=0.49\textwidth]{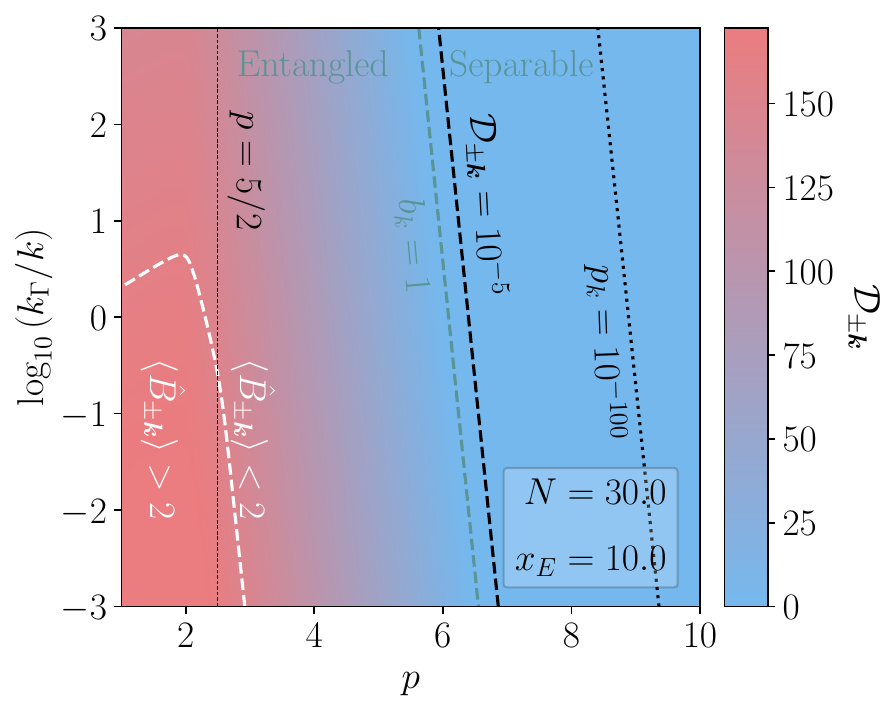}
    \includegraphics[width=0.49\textwidth]{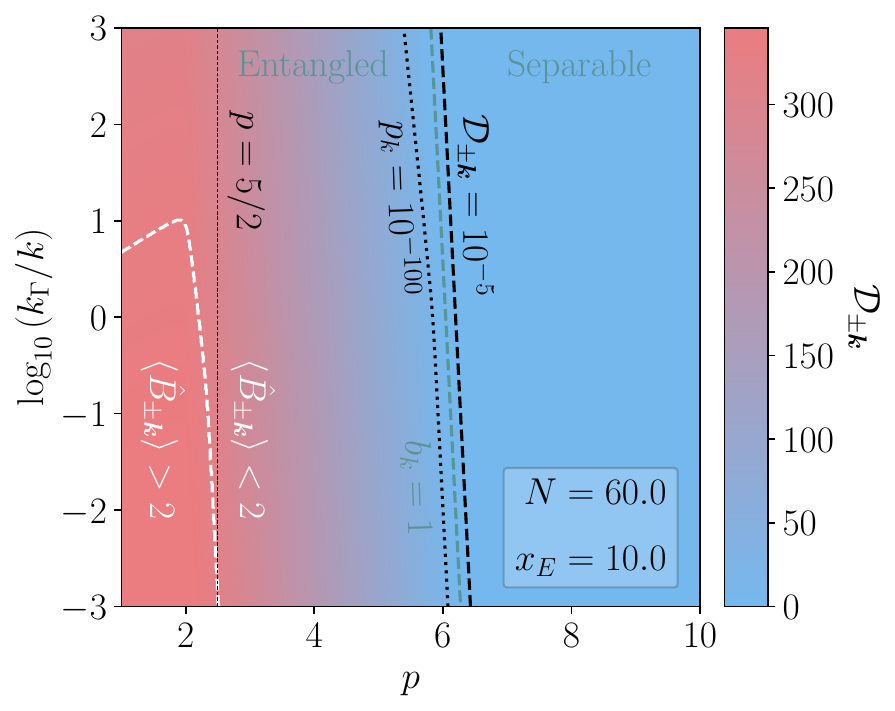}
    \caption{\label{fig:quantum_corr} Effect of interactions taken into account for all parameters $r_k$, $\varphi_k$ and $p_k$. }
  \end{subfigure}
  \hfill
  \begin{subfigure}[b]{\textwidth}
    \includegraphics[width=0.49\textwidth]{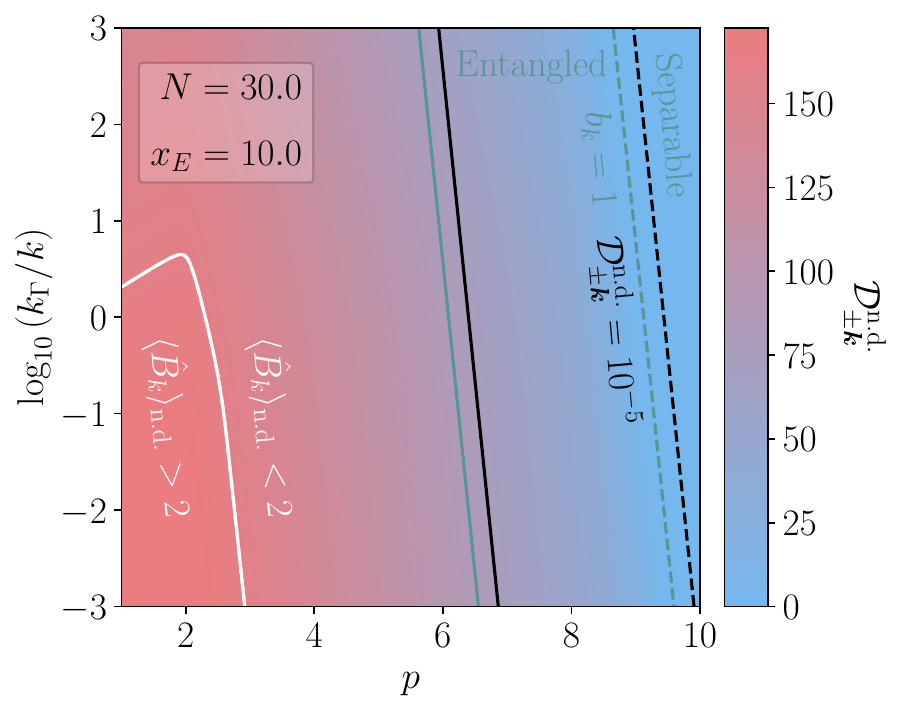}
    \includegraphics[width=0.49\textwidth]{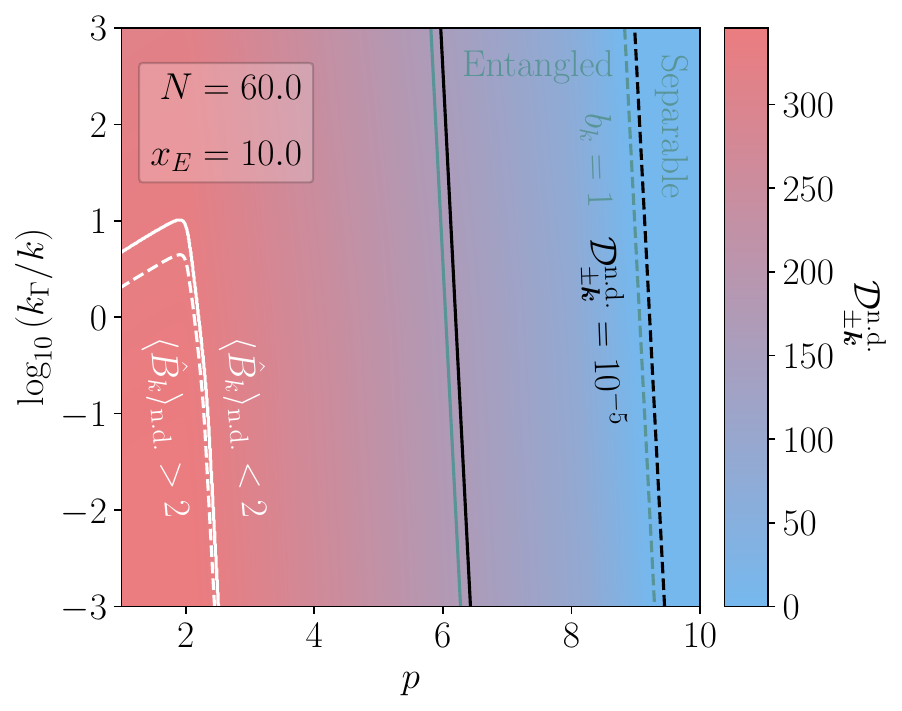}
    \caption{\label{fig:quantum_corr_no_desqueezing} Effect of interactions only taken into account for $p_k$. }
  \end{subfigure}
  \caption{\label{fig:quantum_correlations}
  Values of the quantum discord $\mathcal{D}_{\pm \bm{k}}$ as a function of $p$ and $k_\Gamma/k$ after $N=30$ (left) and $N=60$ (right) $e$-folds of inflation. The top panels show values computed using the effective squeezing parameters in the presence of interaction. More precisely, we use Eqs.~\eqref{eq:latetime_purity}, \eqref{eq:latetime_ak} and \eqref{eq:latetime_bk}, inserted into the discord formula Eq.~\eqref{eq:discord_expansion_ready}. The dashed green lines show the entanglement threshold, corresponding to $b_k = 1$, computed using Eq.~\eqref{eq:latetime_bk}. The dashed white lines show $\ev{\hat{B}_{\pm \bm{k}}} = 2$, i.e. the violation threshold for the Bell inequality~\eqref{eq:bell_inequality}. The expectation value of the Bell operator is computed using Eq.~\eqref{eq:latetime_Bell}. The lower panels show the values of the quantum discord computed using the purity in the presence of interaction, as given by Eq.~\eqref{eq:latetime_purity}, and the squeezing parameters in the absence of interaction, Eqs.~\eqref{eq:no_deco_rk}--\eqref{eq:no_deco_phik}. These three parameters are then used to compute $a_k$ and $b_k$ via Eq.~\eqref{def:axes_ellipse}, and the discord is obtained using Eq.~\eqref{eq:discord_expansion_ready}. The entanglement and Bell-inequality violation thresholds are computed in the same way as in the top panels. In addition, for comparison, we show as solid lines the contours from the top panels for the same values of $N=30$ (left) and $N=60$ (right).
}
\end{figure}

We now summarise our results on these quantum-correlation quantifiers by plotting in Fig.~\ref{fig:quantum_correlations} their values in the $p$--$\log (k_\Gamma/k)$ plane after $N=30$ (left panel) and $N=60$ (right panel) $e$-folds of inflation. Let us comment on their content.

First, Fig.~\ref{fig:quantum_corr} reproduces the known hierarchy between the different quantum-correlation quantifiers. While the Bell inequality is only violated for entangled states and entangled states have a non-vanishing quantum discord, a non-vanishing quantum discord does not imply that the state is entangled and entangled state does not necessarily violate a Bell inequality. Second, Fig.~\ref{fig:quantum_corr} shows that quantum correlations become weaker, at fixed time $x$, when the interaction is stronger, either because it is more time-dependent, i.e. $p$ increases, or, to a lesser extent, because its overall magnitude $k_{\Gamma}/k$ is larger. Third, by comparing the left and right panels, we see that as time increases the dependence on $k_{\Gamma}/k$ becomes less relevant. This is in agreement with the findings of Sec.~\ref{sec:latetime_quantumness_deco}, where we identified that the asymptotic behaviour of the quantum correlations mostly depends on the value of $p$. Turning to the logarithmic negativity and the quantum discord, the figures confirm that $p=6$ is a critical value: for $p<6$ the state has a large negativity, is entangled, and has a large quantum discord at late times, while it is separable and has a small quantum discord for $p>6$. Going from the left to the right panel of Fig.~\ref{fig:quantum_corr}, we observe the emergence of this asymptotic behaviour as the contours for the value of the discord and for $b_k=1$ become more vertical. Moreover, although not plotted here, the contour level for $\mathcal{D}_{\pm \bm{k}} = 1$ lies on top of that for $b_k=1$. This confirms that the behaviour of the discord is asymptotically controlled by the value of $b_k$, and is in agreement with the results of~\cite{Martin:2022kph}. To make quantitative sense of this overlap, we use Eq.~\eqref{eq:discord_expansion_ready} to evaluate the discord by taking $b_k=1$ and setting to zero all terms proportional to powers of $p_k$ and $e^{-r_k}$. We find that the quantum discord is indeed of order unity, $\mathcal{D}_{\pm \bm{k}} \approx f(3) - 1/\ln 2 \approx 0.55$. Similarly, for the Bell inequality, we observe that the contour level $\ev{\hat{B}_{\pm \bm{k}}}=2$ becomes more vertical at later times, and closer to the critical value $p=5/2$ identified in Sec.~\ref{sec:latetime_quantumness_deco}. However, the convergence to this asymptotic behaviour is slower than for the quantum discord and the logarithmic negativity, and even for $N=60$ (right panel) the contour still strongly depends on the value of $k_{\Gamma}/k$. A fourth important observation is that there is no direct relation between the value of the purity $p_k$ and the level of quantum correlations. In particular, we see that for $N=60$, even states with a purity as small as $10^{-100}$ can still be entangled. Thus, it is incorrect to infer that a state has only classical properties solely based on the smallness of its purity.

Since most studies on the decoherence of cosmological perturbations only evaluate their purity, e.g.~\cite{Burgess:2022nwu,Lopez:2025arw,Burgess:2025dwm}, a natural approach is to estimate the level of quantum correlations by assuming that the squeezing parameters $r_k$ and $\varphi_k$ are not significantly affected by the interactions leading to decoherence. One then represents the state by the parameters $(p_k,\, r_k^{\mathrm{w.o.}},\, \varphi_k^{\mathrm{w.o.}})$ and uses them to evaluate the correlation quantifiers of Sec.~\ref{sec:quantif_quantum}. This was the approach adopted in the conclusion of~\cite{Micheli:2023qnc}. However, we showed in Sec.~\ref{sec:deco_squeezing} that interactions modelled by Eq.~\eqref{eq:lindblad_Caldeira_Leggett} typically induce a strong reduction of $r_k$, i.e. a de-squeezing of the perturbation. As a result, an estimate based on $(p_k,\, r_k^{\mathrm{w.o.}},\, \varphi_k^{\mathrm{w.o.}})$ leads to an overestimation of the quantum correlations of the state. To assess the magnitude of this error, we plot in Fig.~\ref{fig:quantum_corr_no_desqueezing} the same quantum-correlation quantifiers as in Fig.~\ref{fig:quantum_corr}, using a quantum state with effective squeezing parameters $(p_k,\, r_k^{\mathrm{w.o.}},\, \varphi_k^{\mathrm{w.o.}})$, i.e. we only take into account the effect of decoherence on the purity $p_k$ and keep the pure-state values $r_k^{\mathrm{w.o.}}$ and $\varphi_k^{\mathrm{w.o.}}$ for the squeezing parameters. In Fig.~\ref{fig:quantum_corr_no_desqueezing}, we also reproduce (as solid lines) some of the contour levels from Fig.~\ref{fig:quantum_corr} to facilitate the comparison.

We first notice that the contour of the expectation value of the Bell operator is not strongly affected by this incorrect estimate of $r_k$ and $\varphi_k$. This can be understood by noting that in Eq.~\eqref{eq:cases_Bell}, in the region $p<5/2$ where the contour lies, the dominant contribution always comes from the purity $p_k$, while the sub-dominant term coming from the squeezing parameters $r_k$ and $\varphi_k$ matches the contribution obtained from the pure-state values $r_k^{\mathrm{w.o.}}$ and $\varphi_k^{\mathrm{w.o.}}$. As a result, the contours estimated from states with parameters $(p_k,\, r_k^{\mathrm{w.o.}},\, \varphi_k^{\mathrm{w.o.}})$ and $(p_k,\, r_k,\, \varphi_k)$ agree at leading order, as observed in Fig.~\ref{fig:quantum_corr_no_desqueezing}. On the other hand, we find that the values of the quantum discord and the separability threshold are strongly affected by an incorrect estimate of the squeezing parameter $r_k$. For instance, for any values of $p$ and $k_{\Gamma}/k$ lying between the solid and dashed green lines in Fig.~\ref{fig:quantum_corr_no_desqueezing}, the state is found to be entangled when using the squeezing $r_k^{\mathrm{w.o.}}$, while it is in fact separable once the de-squeezing is properly taken into account.

\section{Conclusion and Perspectives} \label{sec:conclusion}

In this paper, we have analysed the effects of interactions of primordial inhomogeneities on their quantum state, parametrised by effective squeezing parameters, and on their degree of quantum correlations as measured by several quantifiers. We have adopted a simple interaction model in which the dynamics of the perturbations is described by a Lindblad equation with a non-unitary term quadratic in the Mukhanov--Sasaki variable, parametrised by two parameters $k_\Gamma$ and $p$. First, we have computed for the first time the variation of the squeezing parameters $r_k$ and $\varphi_k$ due to interactions over a large region of the phenomenological parameter space spanned by $k_\Gamma$ and $p$, extending the computations of~\cite{Martin:2021znx} for the purity $p_k$. This constitutes the first set of results of this paper. While it is expected and well known that interactions suppress the purity $p_k$ of the state, we have also demonstrated that our interaction model typically reduces $r_k$ as well, and can do so significantly; see Fig.~\ref{fig:delta_rk}. On the other hand, although some deviations from the case without decoherence can be observed at early times, the squeezing angle $\varphi_k$ is essentially independent of the values of $p$ and $k_\Gamma$.

Using these values of the effective squeezing parameters, we then computed how interactions modify the degree of quantum correlations between opposite-momentum pairs $\pm \bm{k}$, as measured by three quantum-correlation quantifiers: quantum discord, logarithmic negativity, and Bell-inequality violation. This constitutes the second set of results of this paper. First, these results extend the computations of~\cite{Martin:2021znx}, which were limited to the quantum discord. Second, we showed that the asymptotic late-time behaviour of these quantifiers is universal in the sense that it depends primarily, at leading order, on the scale-factor dependence of the interaction, $a^{p-3}$, and only weakly on its overall strength $k_{\Gamma}/k$. This behaviour is illustrated in Fig.~\ref{fig:quantum_corr}. Furthermore, we demonstrated that estimating the degree of quantum correlations in the presence of interactions based solely on a computation of the purity $p_k$, supplemented by the pure-state values $r_k^{\mathrm{w.o.}}$ and $\varphi_k^{\mathrm{w.o.}}$, leads to a gross overestimation of these correlations when using the quantum discord or the logarithmic negativity. This is evident from a comparison between Fig.~\ref{fig:quantum_corr} and Fig.~\ref{fig:quantum_corr_no_desqueezing}, where the same quantifiers are plotted for a state in which only the effect of interactions on $p_k$ is taken into account, while the squeezing parameters are fixed to their pure-state values $r_k^{\mathrm{w.o.}}$ and $\varphi_k^{\mathrm{w.o.}}$. This result highlights that determining whether a state has been classicalised by its interactions, i.e. whether it no longer exhibits quantum correlations, requires knowledge of its full quantum state, and not only of its purity.

Our treatment of decoherence was intentionally broad, encompassing a wide class of possible models. Once a specific framework or set of assumptions is chosen, however, one can explicitly formulate the interaction between the system and the environment, as has been done, for example, in~\cite{Burgess:2022nwu,Burgess:2025dwm,Sou:2022nsd}. Applying our analysis to these cases, which consider a minimal level of interaction induced by the second-order dynamics of cosmological perturbations, would allow one to derive an upper bound on the degree of quantum correlations at the end of inflation. While the master equation of~\cite{Burgess:2022nwu} has a form similar to our Eq.~\eqref{eq:lindblad_Caldeira_Leggett}, the more complete equation of~\cite{Burgess:2025dwm} contains additional non-unitary terms that would need to be included. However, the main difficulty in applying our results to these interactions is that the ``Lamb shift'' terms appearing in~\cite{Burgess:2022nwu,Burgess:2025dwm} contain divergences; see Eq.~(3.36) of~\cite{Burgess:2025dwm}. The authors argue that these divergences can be renormalised by introducing counterterms generated by $1/M_{\mathrm{Pl}}^2$ corrections to the Einstein--Hilbert action. Such procedure, beyond the scope of the present paper, must be carried out before the present analysis can be applied to this interaction.

Finally, we may ask about the generality of our investigation. First, it is not obvious whether the de-squeezing we have observed is universal or specific to an interaction proportional to the Mukhanov--Sasaki field $\hat{v}$. Second, the validity of our formulation may warrant careful scrutiny, particularly in regions of parameter space corresponding to large values of $p$ and $k_{\Gamma}/k$. Although we have briefly addressed this issue in Appendix~\ref{sec:validity}, further clarification is required. A comprehensive treatment of these questions is deferred to future work.

\section*{Acknowledgements}
This work was supported by JSPS KAKENHI Grant Numbers 25K01004 (TT) and MEXT KAKENHI 23H04515 (TT), 25H01543 (TT). AM thanks J\'{e}r\^{o}me Martin, Fumiya Sanno and Vincent Vennin for discussions on topics related to the present article.

\pagebreak
\appendix
\noindent
{\LARGE \bf Appendix} 

\section{Applicability domain of our Lindblad equation \label{app:scope} }

In this Appendix we sketch how one can arrive at Eq.~\eqref{eq:lindblad_Caldeira_Leggett} from a specific field interaction model. This will allow us to discuss its domain of applicability and explain the origin of the different terms appearing in it. A fuller and more rigorous derivation is given in~\cite{Martin:2018zbe} in general, and in~\cite{Burgess:2022nwu} for the case of decoherence induced by the second-order dynamics of the cosmological perturbations. 

\subsection{Derivation}

First, consider the following interaction Hamiltonian with a linear coupling:
\begin{equation}
\label{def:Hint}
\hat{H}_{\mathrm{int}}(\eta) =  G(\eta) \int 
\hat{v} (\bm{x}) \otimes \hat{E}(\eta,\bm{x}) \dd^3 \bm{x} = G(\eta) \int_{\setR^{3+}} \sum_{s=\mathrm{R,I}} \hat{v}_{\bm k}^{\mathrm{s}} \otimes \hat{E}_{\bm k}^{\mathrm{s}} \left( \eta \right) \dd ^3\bm k  \, ,
\end{equation}
where $G(\eta)$ is a time-dependent coupling constant and $\hat{E}(\eta,\bm{x})$ is an operator acting only on the environment Hilbert space; their concrete expressions define the chosen interaction model. In the second equality, we transformed the Hamiltonian into Fourier space and defined the Hermitian and anti-Hermitian parts of the operators. Our goal is to characterise how the $\pm \bm{k}$ correlations generated by the free evolution are modified when the interaction~\eqref{def:Hint} is taken into account. This is, in general, a difficult problem, as it requires first solving the coupled Liouville equation Eq.~\eqref{eq:Liouville_free_RI} for the system and the environment, and then tracing out the environmental degrees of freedom to obtain the reduced density matrix of the system $\hat{\varrho}(\eta)$. Since we are only concerned with the dynamics of the system, we can instead trace out the environmental degrees of freedom from the start and effectively treat the evolution of the system as open, rather than solving the coupled closed-system dynamics of the system and the environment. Under a set of assumptions on the coupling and the state of the environment, we can perform this trace without knowing the state of the environment exactly and obtain an approximate dynamics for $\hat{\varrho}(\eta)$ in the form of a non-unitary equation called a master equation. The first step is to go to the interaction picture and iteratively integrate the full dynamics to cast the equation in an integro-differential form:
\begin{align}
\begin{split}
\hat{\rho}_{\mathrm{S/E}} \left(\eta + \delta \eta \right) - \hat{\rho}_{\mathrm{S/E}} \left( \eta \right) & = - i \int_{\eta}^{\eta+\delta \eta} \left[ \tilde{H}_{\mathrm{int}} \left( \eta^{\prime} \right) , \tilde{\rho}_{\mathrm{S/E}} \left( \eta^{\prime} \right)  \right] \dd \eta^{\prime}  \, , \\
& = - i \int_{\eta}^{\eta+\delta \eta} \left[ \tilde{H}_{\mathrm{int}} \left( \eta^{\prime} \right) , \tilde{\rho}_{\mathrm{S/E}} \left( \eta \right)  \right] \dd \eta^{\prime}   \\
& \qquad 
- \int_{\eta}^{\eta+\delta \eta} \int_{\eta}^{\eta^{\prime}} \left[ \tilde{H}_{\mathrm{int}} \left( \eta^{\prime} \right) , \left[ \tilde{H}_{\mathrm{int}} \left( \eta^{\prime \prime} \right) , \tilde{\rho}_{\mathrm{S/E}} \left( \eta^{\prime \prime} \right)  \right]  \right] \dd \eta^{\prime} \dd \eta^{\prime \prime}  \, ,
\end{split}
\end{align}
where $\tilde{\rho}_{\mathrm{S/E}}$ is the density matrix of the full system and $\tilde{O}(\eta)$ refers to operators in the interaction picture.
The derivation then proceeds by tracing out the environment on both sides:
\begin{align}
\begin{split}
\label{eq:step1_Lindlad}
\hat{\rho} \left(\eta + \delta \eta \right) - \hat{\rho} \left( \eta \right) & = - i \int_{\eta}^{\eta+\delta \eta} \mathrm{Tr}_{\mathrm{E}}  \left\{  \left[ \tilde{H}_{\mathrm{int}} \left( \eta^{\prime} \right) , \tilde{\rho}_{\mathrm{S/E}} \left( \eta \right)  \right] \right\} \dd \eta^{\prime} \\
&  \qquad - \int_{\eta}^{\eta+\delta \eta} \int_{\eta}^{\eta^{\prime}} \mathrm{Tr}_{\mathrm{E}}  \left\{ \left[ \tilde{H}_{\mathrm{int}} \left( \eta^{\prime} \right) , \left[ \tilde{H}_{\mathrm{int}} \left( \eta^{\prime \prime} \right) , \tilde{\rho}_{\mathrm{S/E}} \left( \eta^{\prime \prime} \right)  \right]  \right] \right\} \dd \eta^{\prime} \dd \eta^{\prime \prime}  \, .
\end{split}
\end{align}
By definition, we have $\mathrm{Tr}_{\mathrm{E}} \left[ \tilde{\rho}_{\mathrm{S/E}} (\eta)  \right] = \tilde{\rho} (\eta)$, and we define the reduced density matrix of the environment as $\tilde{\rho}_{\mathrm{E}} (\eta) = \mathrm{Tr}_{\mathrm{S}} \left[ \tilde{\rho}_{\mathrm{S/E}} (\eta)  \right]$. The interaction correlates the system and the environment, so that $\tilde{\rho}_{\mathrm{S/E}} (\eta) \neq \tilde{\rho} (\eta) \otimes \tilde{\rho}_{\mathrm{E}} (\eta)$. Yet, since in the absence of interactions they evolve independently, $\tilde{\rho}_{\mathrm{S/E}} (\eta) = \tilde{\rho} (\eta) \otimes \tilde{\rho}_{\mathrm{E}} (\eta)$, the non-factorisable part of the density matrix must arise from terms in $\tilde{H}_{\mathrm{int}}$.
We assume that the interaction is perturbative, i.e.\ $\tilde{H}_{\mathrm{int}} \ll 1$ in some sense, so that this contribution is small, and, for simplicity, that the first term is of order $\mathcal{O} (\tilde{H}_{\mathrm{int}}^2)$, so that
\begin{equation}
\tilde{\rho}_{\mathrm{S/E}} (\eta) = \tilde{\rho} (\eta) \otimes \tilde{\rho}_{\mathrm{E}} (\eta) +  \mathcal{O} (\tilde{H}_{\mathrm{int}}^2) \, .
\end{equation}
The first term in Eq.~\eqref{eq:step1_Lindlad} is \emph{a priori} of order $\mathcal{O} (\tilde{H}_{\mathrm{int}})$, but it can be removed by an appropriate redefinition of the operator $\hat{E}$. To see how, first consider the trace appearing in this term, which reads
\begin{align}
\begin{split}
\mathrm{Tr}_{\mathrm{E}}  \left\{  \left[ \tilde{H}_{\mathrm{int}} \left( \eta^{\prime} \right) , \tilde{\rho}_{\mathrm{S/E}} \left( \eta \right)  \right] \right\} = G(\eta) \int  \mathrm{Tr}_{\mathrm{E}} \left[ \tilde{E}(\eta,\bm{x}) \tilde{\rho}_{\mathrm{E}} (\eta) \right]   \left[ \hat{v} (\bm{x}) , \tilde{\rho} (\eta)  \right] \dd^3 \bm{x} +  \mathcal{O} (\tilde{H}_{\mathrm{int}}^3) \, .
\end{split}
\end{align}
Assuming that the environment is left unaffected by the interaction with the system and remains stationary, its reduced state is time-independent and identical to that of the free theory: we write $\tilde{\rho}_{\mathrm{E}} (\eta) \approx \hat{\rho}_{\mathrm{E}}^{\mathrm{free}}$. We then have
\begin{equation}
\mathrm{Tr}_{\mathrm{E}} \left[ \tilde{E}(\eta,\bm{x}) \tilde{\rho}_{\mathrm{E}} (\eta) \right]  =   \mathrm{Tr}_{\mathrm{E}} \left[ \hat{E}(\eta,\bm{x}) \hat{\rho}_{\mathrm{E}}^{\mathrm{free}}  \right]  = \langle \hat{E}(\eta,\bm{x}) \rangle \, ,
\end{equation}
where in the last equality we identified the expectation value of the operator in the free theory. By redefining the operator appearing in the interaction Hamiltonian as $\hat{E} \to \hat{E} - \langle \hat{E} \rangle \hat{\mathds{1}}$, we can remove this term. As a result, the remainder of the first term in Eq.~\eqref{eq:step1_Lindlad} is of order $\mathcal{O} (\tilde{H}_{\mathrm{int}}^3)$ and is thus subdominant compared to the second term in the equation, which is of order $\mathcal{O} (\tilde{H}_{\mathrm{int}}^2)$. Let us now analyse this term. We have
\begin{equation}
\hat{\rho} \left(\eta + \delta \eta \right) - \hat{\rho} \left(\eta \right) \approx  - \int_{\eta}^{\eta+\delta \eta} \int_{\eta}^{\eta^{\prime}} \mathrm{Tr}_{E} \left\{ \left[ \tilde{H}_{\mathrm{int}} \left( \eta^{\prime} \right) , \left[ \tilde{H}_{\mathrm{int}} \left( \eta^{\prime \prime} \right) , \tilde{\rho} \left(\eta^{\prime \prime} \right) \otimes \hat{\rho}_{\mathrm{E}}^{\mathrm{free}}  \right]  \right] \right\} \dd \eta^{\prime} \dd \eta^{\prime \prime} \, .
\end{equation}
Using the form of the interaction Hamiltonian~\eqref{def:Hint}, the contribution of the environment in the second term is encoded in the autocorrelation function
\begin{equation}
    C_{B} \left( \eta , \eta^{\prime} ; \bm{x} , \bm{x}^{\prime} \right) = \mathrm{Tr}_{\mathrm{E}} \left[ \hat{\rho}_{\mathrm{E}}^{\mathrm{free}} \hat{E}(\eta,\bm{x}) \hat{E}(\eta^{\prime},\bm{x}^{\prime})  \right] \, .
\end{equation}
Note that, by cyclicity of the trace and because $\hat{E}$ is Hermitian, we have $C_{B} ( \eta , \eta^{\prime} ; \bm{x} , \bm{x}^{\prime} )^{\star} = C_{B} ( \eta^{\prime} , \eta ; \bm{x}^{\prime} , \bm{x} )$.
The equation then reads
\begin{align}
\begin{split}
\label{eq:step2_Lindlad}
& \hat{\rho} \left(\eta + \delta \eta \right) - \hat{\rho} \left(\eta \right)  \approx  \\
& - \int_{\eta}^{\eta+\delta \eta} \int_{\eta}^{\eta^{\prime}} G\left( \eta^{\prime} \right) G\left( \eta^{\prime \prime} \right) \int_{\setR^3} \int_{\setR^3}  \big\{ C_{B} ( \eta^{\prime} , \eta^{\prime \prime} ; \bm{x}^{\prime} , \bm{x}^{\prime \prime} ) \left[ \tilde{v} \left( \bm{x}^{\prime} , \eta^{\prime} \right) ,  \tilde{v} \left( \bm{x}^{\prime \prime}, \eta^{\prime \prime} \right) \tilde{\rho} \left(\eta^{\prime \prime} \right) \right]  \\
& 
\qquad\qquad
+ C_{B} ( \eta^{\prime} , \eta^{\prime \prime} ; \bm{x}^{\prime} , \bm{x}^{\prime \prime} )^{\star} \left[ \tilde{\rho} \left(\eta^{\prime \prime} \right) \tilde{v} \left( \bm{x}^{\prime \prime}, \eta^{\prime \prime} \right) , \tilde{v} \left( \bm{x}^{\prime} , \eta^{\prime} \right) \right] \big\} \dd \eta^{\prime} \dd \eta^{\prime \prime} \dd \bm{x}^{\prime} \dd \bm{x}^{\prime \prime} \, .   
\end{split}
\end{align}
The key final step is to assume that $C_{B} ( \eta , \eta^{\prime} ; \bm{x} , \bm{x}^{\prime} )$ is sharply peaked and supported around $\eta = \eta^{\prime}$. This allows us to simplify the time integrals and obtain an equation whose right-hand side depends only on $\tilde{\varrho}$ evaluated at $\eta$, rather than on its entire history. Such an equation is dubbed Markovian. Assuming that the system evolves on a timescale much longer than the typical autocorrelation time $t_c$ over which $C_{B}$ is supported, we have
\begin{equation}
    \int_{\eta}^{\eta^{\prime}} \dd \tau \, C_{B} ( \eta^{ \prime} , \eta^{\prime \prime} ) f\left( \eta^{ \prime} , \eta^{\prime \prime}  \right) \approx f\left( \eta^{\prime} , \eta^{\prime}  \right) \int_{- \infty}^{\eta^{\prime}} \dd \tau \, C_{B} ( \eta^{ \prime} , \eta^{\prime \prime} ) \, ,
\end{equation}
i.e.\ the autocorrelation selects only the contribution with $\eta^{\prime} = \eta^{\prime \prime}$. In addition, since $C_{B}$ is supported around $\eta^{\prime} = \eta^{\prime \prime}$, we can safely extend the integration domain on the right-hand side to $- \infty$ without changing the value of the integral. This allows us to suppress the explicit dependence on $\eta$. Using this approximation in Eq.~\eqref{eq:step2_Lindlad} and taking $\delta \eta \to 0$, we obtain
\begin{align}
\begin{split}
\label{eq:step3_Lindblad}
\partial_{\eta} \hat{\rho}  &  \approx  -  \int_{\setR^3} \int_{\setR^3} \dd \bm{x}^{\prime} \dd \bm{x}^{\prime \prime}  \big\{ \mathrm{Re} \left[ \mathfrak{F}\left( \eta ; \bm{x}^{\prime}, \bm{x}^{\prime\prime} \right) \right]  \left[ \tilde{v} \left( \bm{x}^{\prime} , \eta \right) , \left[   \tilde{v} \left( \bm{x}^{\prime \prime}, \eta \right) , \tilde{\rho} \left(\eta \right) \right] \right]  \\
& - i \, \mathrm{Im} \left[ \mathfrak{F}\left( \eta ; \bm{x}^{\prime}, \bm{x}^{\prime\prime} \right) \right] \left[ \tilde{v} \left( \bm{x}^{\prime} , \eta \right)  \tilde{v} \left( \bm{x}^{\prime \prime}, \eta \right) , \tilde{\rho} \left(\eta \right)  \right]   \\
& - i \,  \mathrm{Im} \left[ \mathfrak{F}\left( \eta ; \bm{x}^{\prime}, \bm{x}^{\prime\prime} \right) - \mathfrak{F}\left( \eta ; \bm{x}^{\prime\prime} ,  \bm{x}^{\prime} \right) \right]  \tilde{v} \left( \bm{x}^{\prime} , \eta \right) \tilde{\rho} \left(\eta \right)  \tilde{v} \left( \bm{x}^{\prime \prime}, \eta \right) \big\} \, ,
\end{split}
\end{align}
with
\begin{equation}
 \mathfrak{F}\left( \eta ; \bm{x}^{\prime}, \bm{x}^{\prime\prime} \right)  = \int^{\eta}_{-\infty}     G \left( \eta \right) G \left( \eta^{\prime} \right)  C_{B} \left( \eta , \eta^{\prime} ; \bm{x}^{\prime}, \bm{x}^{\prime\prime}  \right) \dd \eta^{\prime}  \, .
\end{equation}
Finally, let us assume that the environment is homogeneous and isotropic, so that the correlation function reads
\begin{equation}
    C_{B} \left( \eta , \eta^{\prime} ; \bm{x}, \bm{x}^{\prime}  \right) = \int \frac{\dd \bm{k}}{\left( 2 \pi \right)^{3/2}} e^{i \bm{k} \cdot \left( \bm{x} - \bm{x}^{\prime} \right)} \mathcal{C}_{B} \left( \eta , \eta^{\prime} ; k \right) \, ,
\end{equation}
where $\mathcal{C}_{B} \left( \eta , \eta^{\prime} ; k \right)$ depends only on the modulus of $k$. Thus, in particular, $C_{B} ( \eta , \eta^{\prime} ; \bm{x} , \bm{x}^{\prime} ) = C_{B} ( \eta^{\prime} , \eta ; \bm{x}^{\prime} , \bm{x} )$. This symmetry cancels the last term in Eq.~\eqref{eq:step3_Lindblad}. Then, going to Fourier space and separating into $\mathrm{R}$ and $\mathrm{I}$ sectors, we obtain
\begin{align}
\begin{split}
\label{eq:Markov_master}
\frac{\mathcal{V}}{(2\pi)^3}
\frac{\partial \tilde{\varrho}_{ \bm{k},\mathrm{s}}}{\partial \eta}
& =  - i \mathrm{Im} \left[ \mathfrak{F}_{\bm{k}}(\eta) \right] \left[ \left( \tilde{v}_{\bm k}^{\mathrm{s}} \right)^2,
\tilde{\varrho}_{ \bm{k},\mathrm{s}} \right] - \mathrm{Re} \left[ \mathfrak{F}_{\bm{k}}(\eta)  \right] \left[ \tilde{v}_{\bm k}^{\mathrm{s}} , \left[ \tilde{v}_{\bm k}^{\mathrm{s}}, \tilde{\varrho}_{ \bm{k},\mathrm{s}} \right] \right] \, ,
\end{split}
\end{align}
where
\begin{equation}
\mathfrak{F}_{\bm{k}}(\eta) = (2 \pi)^{3/2} \int^{\eta}_{-\infty} G \left( \eta \right) G \left( \eta^{\prime} \right) \mathcal{C}_{B} \left( \eta , \eta^{\prime} ; \bm{k}  \right) \dd \eta^{\prime} \, .
\end{equation}
Let us make some comments on this final equation.
First, note that Eq.~\eqref{eq:Markov_master} contains two additions with respect to the free dynamics given by Eq.~\eqref{eq:Liouville_free_RI}. The first term is a ``Lamb shift''-type term arising from the imaginary part of the kernel, which contributes to the unitary dynamics of the $\pm \bm{k}$ modes and can be interpreted as a modification of the bare Hamiltonian. The second term comes from the real part of the kernel and gives a non-unitary contribution that generates dissipation and decoherence for the $\pm \bm{k}$ modes. This is the general form of a Lindblad equation.

Yet, Eq.~\eqref{eq:Markov_master} still does not match Eq.~\eqref{eq:lindblad_Caldeira_Leggett}. To obtain this equation, we follow the simple, but restrictive, derivation of~\cite{Martin:2018zbe}, where it is assumed that the environment is in a state where $C_{B}$ decays exponentially in time with a rate $t_{\mathrm{c}} (\eta)$, and is real, so that no Lamb shift term appears These assumptions transform Eq.~\eqref{eq:Markov_master} to an equation of the simplified form~\eqref{eq:lindblad_Caldeira_Leggett}, which we use in this work. In addition, we can write the resulting kernel as
\begin{equation}
\label{eq:kernel_simple}
  \mathfrak{F}_{\bm{k}}(\eta) \propto  G ( \eta )^2 t_{\mathrm{c}} (\eta) \mathcal{C}_{B} \left( \eta, 0 ; \bm{k}  \right) \, .
\end{equation}
This form shows that the magnitude and time dependence of the decoherence kernel, which we parametrised using $k_{\Gamma}/k$ and $a^{p-3}(\eta)$, depend on the interaction constant $G (\eta)$, but also on the environmental autocorrelation function $\mathcal{C}_{B}$ and the timescale $t_{\mathrm{c}}$.

\subsection{Validity \label{sec:validity}}

The derivations of Eqs.~\eqref{eq:Markov_master} and~\eqref{eq:lindblad_Caldeira_Leggett} rest crucially on the assumptions that the interaction is perturbative and that the environment leads to an approximately Markovian dynamics. These assumptions must be carefully checked when considering a concrete interaction model. In general, it is tricky to estimate for which values of the phenomenological parameters $k_{\Gamma}/k$ and $p$ the equation~\eqref{eq:lindblad_Caldeira_Leggett} is valid. For instance, the perturbative assumption requires $\tilde{H}_{\mathrm{int}}$ to be small, so that $p$ and $k_{\Gamma}$ cannot be arbitrarily large. However, there is no one-to-one relation between the magnitude of $\tilde{H}_{\mathrm{int}}$ and that of $\mathfrak{F}_{\bm{k}}$. For example, even in the simple model of~\cite{Martin:2018zbe}, $\mathfrak{F}_{\bm{k}}$ also contains the autocorrelation time $t_{c}$ and autocorrelation function $\mathcal{C}_{B}$, which can be time-dependent, see Eq.~\eqref{eq:kernel_simple}. Therefore, in this paper we remain completely agnostic about the range of validity of the equation and solve it in full generality.

To close this Appendix, we mention two specific situations in which the assumptions required to derive the master equation were shown to be valid. First, in~\cite{Martin:2018zbe} the authors demonstrated that an environment made up of a heavy scalar field interacting with the cosmological perturbations during slow-roll inflation satisfies these assumptions, provided the coupling is small enough. They derived Eq.~\eqref{eq:lindblad_Caldeira_Leggett} with a time dependence $p \approx 5$. Second, in~\cite{Burgess:2022nwu} the authors considered the self-coupling of inflationary perturbations in slow-roll inflation, obtained at second order in perturbation theory~\cite{Maldacena:2002vr}. These interactions thus set the minimal amount of decoherence expected in any single-field slow-roll inflation. Considering the interaction of a pair of modes $\pm \bm{k}$ corresponding to CMB scales, interacting with shorter, but still super-Hubble, scales, they identified a dominant interaction channel among the different cubic terms. This channel gives a Hamiltonian of the form~\eqref{def:Hint} and leads to a master equation of the form Eq.~\eqref{eq:Markov_master}. The time dependence of the imaginary part also gives $p \approx 5$. This computation was refined and corrected in~\cite{Burgess:2025dwm}, where the authors took into account all leading-order terms and derived a generalised equation in which one must consider additional kernels in $\hat{p}_{\bm k}^2$ and cross-terms $\hat{v}_{\bm k} \hat{p}_{-\bm k}$. We further comment on the application of our results to those of~\cite{Burgess:2022nwu,Burgess:2025dwm} in the conclusion; see Sec.~\ref{sec:conclusion}.

\section{Late time approximations in de Sitter in presence of decoherence}
\label{app:cov_mat}

\subsection{Covariance matrix elements}

Exact expressions can be obtained for the covariance-matrix elements in de Sitter, where the mode function is given by Eq.~\eqref{eq:mode_fn_dS}, and in the presence of interactions modelled by Eq.~\eqref{eq:lindblad_Caldeira_Leggett}. Their form was derived in~\cite{Martin:2021znx} and depend on the evaluation of the following integral:
\begin{align}
\begin{split}
\label{eq:defAalpha}
  A_\alpha \left( x ; x_E\right) & = A^{R}_\alpha + i A^{I}_\alpha  \, \\
  & = \int _{x_{E}}^{x} e^{2i y} y^{\alpha}{\dd} y \, ,\\
  &=-2^{-1-\alpha}
  (-i)^{-1-\alpha}\left[\Gamma\left(1+\alpha,-2ix\right)
    -\Gamma\left(1+\alpha,-2ix_E\right)\right],
\end{split}
\end{align}
where we first split $A_\alpha$ into its real and imaginary parts, and used $\Gamma(a,z)=\displaystyle\int_z^{+\infty} t^{a-1} e^{-t}{\dd}{t}$, the incomplete Gamma
function~\cite{olverNISTHandbookMathematical2010}. We are interested in obtaining expressions valid at late times $x \to 0^{+}$, where the integral is convergent. We thus define
\begin{equation}
\mathcal{A}^{R}_\alpha  = A^{R}_\alpha \left( 0 ; x_E\right) \, , \quad \mathcal{A}^{I}_\alpha   = A^{I}_\alpha \left( 0 ; x_E\right) \, ,
\end{equation}
making the $x_E$ dependence implicit. To obtain the leading-order terms in the expressions of the effective squeezing parameters and correlation measures, we need to expand $A_\alpha$ to sufficiently high order as follows:
\begin{alignat}{1}
\begin{split}
\label{eq:expansion_Aalpha}
A_\alpha \left( x ; x_E\right) & = \mathcal{A}^{R}_{\alpha} + \frac{x^{1+\alpha}}{1 + \alpha} - \frac{2 x^{3+\alpha} }{3 + \alpha} - \frac{2 x^{5+\alpha} }{3 (5 + \alpha)}  - \frac{4 x^{7+\alpha} }{45 (7 + \alpha)} + \mathcal{O} \left( x^{9+\alpha} \right) \\
+ & i \left[ \mathcal{A}^{I}_{\alpha} + \frac{2 x^{2+\alpha}}{2 + \alpha} - \frac{4 x^{3+\alpha} }{ 3(4 + \alpha)}  + \frac{4 x^{6+\alpha} }{15 (6 + \alpha)}  - \frac{8 x^{8+\alpha} }{315 (8 + \alpha)} + \mathcal{O} \left( x^{10+\alpha} \right) \right] \, .
\end{split}
\end{alignat}
Using this expansion, we can obtain corresponding expansions for the covariance-matrix elements.
Details are given in Appendix D of~\cite{Martin:2021znx} (the main text only quotes the small-$x_E$ expansions). The above discussion gives Eqs.~\eqref{eq:latetime_gammaij}, which we reproduce here:
\begin{align*}
\begin{split}
\gamma_{11} & = 
   \frac{1}{x^2}\left[ 1 - 2  \left(\frac{k_\Gamma}{k}\right)^2 B_{11}\left(p,x_E\right) +  \mathcal{O}  \left(x^2 \right) - 2  \left(\frac{k_\Gamma}{k}\right)^2  A_{11}\left(p\right) \, x^{8-p} +  \mathcal{O}  \left( x^{10-p} \right) \right] \, ,  \\ \notag \\
\gamma_{12} & =  \frac{1}{x^3}\left[ 1 - 2  \left(\frac{k_\Gamma}{k}\right)^2 B_{12}\left(p,x_E\right) +  \mathcal{O}  \left(x^2 \right)  - 2  \left(\frac{k_\Gamma}{k}\right)^2 A_{12}\left(p\right) \, x^{8-p} +  \mathcal{O} \left( x^{10-p} \right) \right]  \, , \\  \notag \\
\gamma_{22} & =  \frac{1}{x^4}\left[ 1 - 2  \left(\frac{k_\Gamma}{k}\right)^2 B_{22}\left(p,x_E\right) +  \mathcal{O}  \left(x^2 \right) -
     2  \left(\frac{k_\Gamma}{k}\right)^2 
      A_{22}\left(p\right) \, x^{8-p} +  \mathcal{O} \left( x^{10-p} \right) \right] \, ,
\end{split}
\end{align*}
where 
\begin{align}
\label{eq:latetime_gammaij_coef_A11}
A_{11} \left( p \right) & = -\frac{2}{(p-8)(p-5)(p-2)} \, , \\ \notag \\
B_{11} \left( p , x_{E} \right) & = \frac{1}{2}\left[\frac{(\ell_{E}H)^{p-4}}{p-4}
      +\frac{(\ell_{E}H)^{p-2}}{p-2}-{\cal A}_{1-p}^\mathrm{R}
      -2{\cal A}_{2-p}^\mathrm{I}+{\cal A}_{3-p}^\mathrm{R}\right] \, , \\ \notag \\
A_{12} \left( p \right) & = -\frac{(p-6)}{(p-8)(p-5)(p-2)} \, , \\ \notag \\
B_{12} \left( p , x_{E} \right) & = B_{11} \, , \\ \notag \\
A_{22}\left( p \right)  & = -\frac{26+p(p-11)}{(p-8)(p-5)(p-2)} \, , \\ \notag \\
\label{eq:latetime_gammaij_coef_B22}
B_{22} \left( p , x_{E} \right) & = B_{11} \, .
\end{align}
Two remarks are in order. First, we checked that for $p>8$, the term neglected in $\gamma_{11}$ is of order $\mathcal{O} \left( x^{8-p} \right)$, despite being indicated as $\mathcal{O} \left( x^{7-p} \right)$ in~\cite{Martin:2021znx}. Second, by definition $\gamma_{11} > 0$, irrespective of the value of $k_{\Gamma}$ and $p$. Thus, for $p<8$, where the constant term dominates, it must always be positive. Taking $k_{\Gamma}/k \to \infty$, this implies that $B_{11}(p,x_E) < 0$ for $p<8$ and arbitrary $x_E$.

\subsection{Exact expression of purity in presence of decoherence}
\label{app:purity}

A first method to obtain an exact expression for the purity is to integrate Eq.~\eqref{eq:transport_purity} from $\eta_{\mathrm{in}}$ to $\eta$. We have
\begin{align}
\begin{split}
p^{-1}_k = 1 & + 4 k \int_{\eta_{\mathrm{in}}}^{\eta} \mathfrak{F}_{\bm{k}} \left( \eta^{\prime} \right) \gamma_{11} \left( \eta^{\prime} \right) \dd \eta^{\prime}  \, , \\ 
 = 1 & + 4 k \int_{\eta_{\mathrm{in}}}^{\eta} \mathfrak{F}_{\bm{k}} \left( \eta^{\prime} \right) \lvert u_{k} \rvert^2 \left( \eta^{\prime} \right) \dd \eta^{\prime} \\
& + 16 k^2 \int_{\eta_{\mathrm{in}}}^{\eta} \mathfrak{F}_{\bm{k}} \left( \eta^{\prime} \right) \left\{ \int_{\eta_{\mathrm{in}}}^{\eta^{\prime}} \mathfrak{F}_{\bm{k}}\left( \eta^{\prime \prime} \right) \mathrm{Im}^2\left[u_{k}( \eta^{\prime } )
  u_{k}^{\star}(\eta^{\prime \prime})\right] \dd \eta^{\prime \prime} \right\}  \dd \eta^{\prime} \, , \\
= 1 & + 4 k \int_{\eta_{\mathrm{in}}}^{\eta} \mathfrak{F}_{\bm{k}} \left( \eta^{\prime} \right) \lvert u_{k} \rvert^2 \left( \eta^{\prime} \right) \dd \eta^{\prime} \\
& + 8 k^2 \underset{\substack{\eta_{\mathrm{in}} \leq \eta^{\prime}  \leq \eta \\ \eta_{\mathrm{in}} \leq  \eta^{\prime \prime} \leq \eta}}{\int} \mathfrak{F}_{\bm{k}}\left( \eta^{\prime } \right) \mathfrak{F}_{\bm{k}}\left( \eta^{\prime \prime} \right) \mathrm{Im}^2\left[u_{k}( \eta^{\prime } )
  u_{k}^{\star}(\eta^{\prime \prime})\right] \dd \eta^{\prime \prime}  \dd \eta^{\prime} \, , 
\end{split}
\end{align}
where, from the second to the third line, we used the symmetry of the integrand to double the integration domain from $\eta_{\mathrm{in}} \leq \eta^{\prime} \leq \eta^{\prime \prime} \leq \eta$ to $(\eta_{\mathrm{in}} \leq \eta^{\prime} \leq \eta^{\prime \prime} \leq \eta) \cup ( \eta_{\mathrm{in}} \leq \eta^{\prime \prime} \leq \eta^{\prime}  \leq \eta ) = \{ (\eta^{\prime} , \eta^{\prime \prime}) \in [\eta_{\mathrm{in}} , \eta]^2 \}$. Expanding the imaginary part, we obtain
\begin{equation}
\mathrm{Im}^2\left[u_{k}( \eta^{\prime } )
  u_{k}^{\star}(\eta^{\prime \prime})\right] = \frac{1}{4} \left\{ 2 \lvert u_{k} \rvert^2 \left( \eta^{\prime} \right) \lvert u_{k} \rvert^2 \left( \eta^{\prime \prime} \right)   - \left[ u_{k}^2 \left( \eta^{\prime} \right) \left(u_{k}^{\star} \right)^2 \left( \eta^{\prime \prime} \right) + \mathrm{c.c.}  \right] \right\} \, .
\end{equation}
Inserting this expression under the integral and switching the dummy indices $\eta^{\prime \prime}  \leftrightarrow \eta^{\prime }$, the two terms on the right-hand side give the same contribution and we get
\begin{align}
\begin{split}
\underset{\substack{\eta_{\mathrm{in}} \leq \eta^{\prime}  \leq \eta \\ \eta_{\mathrm{in}} \leq  \eta^{\prime \prime} \leq \eta}}{\int} \mathfrak{F}_{\bm{k}}\left( \eta^{\prime } \right) \mathfrak{F}_{\bm{k}}\left( \eta^{\prime \prime} \right) \mathrm{Im}^2\left[u_{k}( \eta^{\prime } )
  u_{k}^{\star}(\eta^{\prime \prime})\right] \dd \eta^{\prime \prime}  \dd \eta^{\prime} & = \frac{1}{2} \left( \mathcal{L}_{k}^2 - \lvert \mathcal{M}_{k} \rvert^2 \right) \, ,
\end{split}
\end{align}
where $\mathcal{L}_{k}$ and $\mathcal{M}_{k}$ are given by Eqs.~\eqref{def:Lk}--\eqref{def:Mk}, which yields Eq.~\eqref{eq:purity_exact}.

A second method to obtain Eq.~\eqref{eq:purity_exact} is to directly insert the exact expressions for $\gamma_{ij}$ into Eq.~\eqref{eq:purity_determinant}. It is instructive to see how the computation proceeds. First, we have
\begin{align}
\begin{split}
\label{eq:purity_exact_direct}
p_k^{-1} & = \frac{1}{k^2} \left( \lvert u_{k} \rvert^2  \lvert u_{k}^{\prime} \rvert^2 - \mathrm{Re}^2 \left[  u_{k} \left( u_{k}^{\prime} \right)^{\star}\right] \right) \\
 & \qquad+ \mathcal{I}_{k} \frac{\lvert u_{k}^{\prime} \rvert^2}{k^2} + \mathcal{K}_{k} \lvert u_{k} \rvert^2 - 2 \mathcal{J}_{k} \frac{\mathrm{Re} \left[  u_{k} \left( u_{k}^{\prime} \right)^{\star}\right]}{k} 
 + \mathcal{I}_{k} \mathcal{K}_{k} - \mathcal{J}_{k}^2 \, ,
\end{split}
\end{align} 
and we can compute each of these terms directly. An important relation is
\begin{equation}
 \lvert u_{k} \rvert^2  \lvert u_{k}^{\prime} \rvert^2 - \mathrm{Re}^2 \left[  u_{k} \left( u_{k}^{\prime} \right)^{\star}\right] = - \frac{W}{4} = k^2 \, ,
\end{equation}
where we identified the Wronskian $W$ and used its normalisation
$W  = 2 i k$ for the solution considered here, as explained below Eq.~\eqref{eq:MS}. By simple but tedious algebra, expanding the real and imaginary parts in the integrals, factoring out the functions depending on $\eta^{\prime}$ and $\eta^{\prime \prime}$, and rearranging the terms in $u_{k}$ using the Wronskian, we obtain
\begin{align}
\mathcal{I}_{k} \frac{\lvert u_{k}^{\prime} \rvert^2}{k^2} + \mathcal{K}_{k} \lvert u_{k} \rvert^2 - 2 \mathcal{J}_{k} \frac{\mathrm{Re} \left[  u_{k} \left( u_{k}^{\prime} \right)^{\star}\right]}{k} & = 2 \mathcal{L}_{k} \, , \\ \notag \\ 
\mathcal{I}_{k} \mathcal{K}_{k} - \mathcal{J}_{k}^2 & =  \mathcal{L}_{k}^2 -  \lvert \mathcal{M}_{k} \rvert^2 \, .
\end{align}
One can check that inserting these expressions into Eq.~\eqref{eq:purity_exact_direct} indeed reproduces Eq.~\eqref{eq:purity_exact}.

\subsection{Late-time approximation of purity}
\label{app:purity_dS}

We now consider the case of de Sitter space, where the mode function $u_k$ is given by Eq.~\eqref{eq:mode_fn_dS}, and take the coupling to be of the form given in Eq.~\eqref{def:coupling}. We can then explicitly evaluate the integrals entering Eq.~\eqref{eq:purity_exact} as follows:
\begin{align}
\mathcal{L}_{k} & =  \left( \frac{k_{\Gamma}}{k} \right)^2  \left(  \frac{x_{\mathrm{E}}^{2-p}}{2-p} + \frac{x_{\mathrm{E}}^{4-p}}{4-p} - \frac{x^{2-p}}{2-p} - \frac{x^{4-p}}{4-p} \right) \, , \\ \notag \\
\mathcal{M}_{k} & = - \left( \frac{k_{\Gamma}}{k} \right)^2  \left( - A_{1-p} + A_{3-p} + 2 i  A_{2-p} \right) \, .
\end{align}

In the main text, we computed the purity for exponentially small values of $x$. To perform this computation numerically, we need an analytic approximation of the above expressions. As explained in the main text above Eq.~\eqref{eq:latetime_purity}, this is most easily achieved by starting from Eq.~\eqref{eq:purity_exact} and expanding the integrals $\mathcal{L}_{k}$ and $\mathcal{M}_{k}$ in powers of $x$, while keeping all orders in $k_{\Gamma}$ and $x_E$. In practice, this only requires expanding the terms in $A_{\alpha}$ using Eq.~\eqref{eq:expansion_Aalpha}. Doing so leads to
\begin{equation}
\label{eq:latetime_Lk}
\mathcal{L}_{k} = \left( \frac{k_{\Gamma}}{k} \right)^2  \left( A_{\mathcal{L}} + B_{\mathcal{L}} x^{2-p} + C_{\mathcal{L}} x^{4-p} \right) \, ,
\end{equation}
with
\begin{equation}
A_{\mathcal{L}} = \frac{x_{\mathrm{E}}^{4-p}}{4-p} + \frac{x_{\mathrm{E}}^{2-p}}{2-p} \, , \quad B_{\mathcal{L}}  = - \frac{1}{2-p}  \, , \quad 
C_{\mathcal{L}}  =  - \frac{1}{4-p}  \, ,
\end{equation}
and
\begin{align}
\begin{split}
\label{eq:latetime_Mk}
& \mathcal{M}_{k} = - i \left( \frac{k_{\Gamma}}{k} \right)^2  \left[ A_{\mathcal{M},\mathrm{I}} + B_{\mathcal{M},\mathrm{I}} x^{5-p} + C_{\mathcal{M},\mathrm{I}} x^{7-p} + \mathcal{O}\left( x^{9-p} \right)   \right] \\
& \hspace{1.5em}- \left( \frac{k_{\Gamma}}{k} \right)^2  \left[ A_{\mathcal{M},\mathrm{R}} + B_{\mathcal{M},\mathrm{R}} x^{2-p} + C_{\mathcal{M},\mathrm{R}} x^{4-p} + D_{\mathcal{M},\mathrm{R}} x^{6-p} + E_{\mathcal{M},\mathrm{R}} x^{8-p} + \mathcal{O}\left( x^{10-p} \right)  \right] \, ,
\end{split}
\end{align}
where
\begin{eqnarray}
A_{\mathcal{M},\mathrm{R}} & = & - \mathcal{A}^{R}_{1-p} + \mathcal{A}^{R}_{3-p} - 2 \mathcal{A}^{I}_{2-p} \, , \\ [6pt]
B_{\mathcal{M},\mathrm{R}}  & = & - \frac{1}{2-p}  \, , \\ [6pt]
C_{\mathcal{M},\mathrm{R}}  & = &  - \frac{1}{4-p}  \, , \\ [6pt]
D_{\mathcal{M},\mathrm{R}}  & = & 0 \, , \\ [6pt]
E_{\mathcal{M},\mathrm{R}}  & = & \frac{2}{9 (8-p)} \, , \\ [6pt]
A_{\mathcal{M},\mathrm{I}} & = & - \mathcal{A}^{I}_{1-p} + \mathcal{A}^{I}_{3-p} + 2 \mathcal{A}^{R}_{2-p}  \, , \\ [6pt]
B_{\mathcal{M},\mathrm{I}}  & = & - \frac{2}{3(5-p)}  \, , \\ [6pt]
C_{\mathcal{M},\mathrm{I}}  & = &  - \frac{4}{15(7-p)} \, .
\end{eqnarray}
Note that $D_{\mathcal{M},\mathrm{R}}$ vanishes due to cancellations between terms. Inserting the expressions in Eqs.~\eqref{eq:latetime_Lk} and~\eqref{eq:latetime_Mk} into Eq.~\eqref{eq:purity_exact}, we obtain
\begin{align}
\begin{split}
p_k^{-1} & = 1 + \left( \frac{k_{\Gamma}}{k} \right)^2 \left[  A_{\sigma}^{(1)} + B_{\sigma}^{(1)} x^{2-p} + \mathcal{O}\left(  x^{4-p} \right)  \right]    \\
& + \left( \frac{k_{\Gamma}}{k} \right)^4 \left[  A_{\sigma}^{(2)} + B_{\sigma}^{(2)} x^{2-p} + \mathcal{O}\left(  x^{4-p} \right)  + D_{\sigma}^{(2)} \left(p,x_E\right) x^{10-2p} + \mathcal{O}\left(  x^{12-2p} \right)  \right]   \, .
\end{split}
\end{align}
The coefficients of this expansion are
\begin{eqnarray}
\label{eq:latetime_pk_coef_A}
&& A_{\sigma} \left( p , x_{E} \right)   =  \left( \frac{k_{\Gamma}}{k} \right)^2 A_{\sigma}^{(1)} + \left( \frac{k_{\Gamma}}{k} \right)^4 A_{\sigma}^{(2)}  \, , \\ [6pt]
&& \qquad\qquad
A_{\sigma}^{(1)}   =  2 A_{\mathcal{L}} \, , \qquad
A_{\sigma}^{(2)}   =  A_{\mathcal{L}}^2 - A_{\mathcal{M},\mathrm{R}}^2 - A_{\mathcal{M},\mathrm{I}}^2 \, , \\ [6pt]
\label{eq:latetime_pk_coef_B}
&& B_{\sigma} \left( p , x_{E} \right)    =  \left( \frac{k_{\Gamma}}{k} \right)^2 B_{\sigma}^{(1)} + \left( \frac{k_{\Gamma}}{k} \right)^4 B_{\sigma}^{(2)}  \, ,  \\ [6pt]
\qquad\qquad
&& \qquad\qquad
B_{\sigma}^{(1)}  =  2 B_{\mathcal{L}}  \, , \qquad
B_{\sigma}^{(2)}  =  2 \left( 
A_{\mathcal{L}} B_{\mathcal{L}} - A_{\mathcal{M},\mathrm{R}} B_{\mathcal{M},\mathrm{R}} 
\right)\, ,  \\ [6pt]
\label{eq:latetime_pk_coef_C}
&& C_{\sigma} \left( p , x_{E} \right)   =   C_{\mathcal{L}}^2 - C_{\mathcal{M},\mathrm{R}}^2 + 2 B_{\mathcal{M},\mathrm{R}} \, , \\ [6pt] 
\label{eq:latetime_pk_coef_D}
&& D_{\sigma} \left( p , x_{E} \right)   =   \left( \frac{k_{\Gamma}}{k} \right)^2 D_{\sigma}^{(1)} + \left( \frac{k_{\Gamma}}{k} \right)^4 D_{\sigma}^{(2)}  \, , \\ [6pt] 
&& \qquad\qquad D_{\sigma}^{(1)}   =   0  \, , \qquad
D_{\sigma}^{(2)}  =  - \left( 2 C_{\mathcal{M},\mathrm{R}}  D_{\mathcal{M},\mathrm{R}} + 2 B_{\mathcal{M},\mathrm{R}}  E_{\mathcal{M},\mathrm{R}} + B_{\mathcal{M},\mathrm{I}}^2 \right) \, .
\qquad\qquad
\end{eqnarray}
Although it does not appear in the final expression, we have defined the term $C_{\sigma}$, which would be associated with $x^{8-2p}$. This term vanishes, but this is due to a non-trivial combination of cancellations between $\mathcal{L}_k$ and $\mathcal{M}_k$, as well as the accidental cancellation of $D_{\mathcal{M},\mathrm{R}}$. In addition, by inserting the expansions for the covariance-matrix elements~\eqref{eq:latetime_gammaij} into the expression for the purity, Eq.~\eqref{eq:purity_determinant}, and identifying the different terms in the series, one can check that the following relations hold:
\begin{align}
\begin{split}
B_{\sigma} & =  2 \left( \frac{k_{\Gamma}}{k} \right)^2 \Big\{  2 A_{12} \left[1 - 2  \left(\frac{k_\Gamma}{k}\right)^2 
     B_{12}\left(p,x_E\right) \right]  \\
    & \hspace{2em}- A_{11} A_{22} \left[ 1 -
     2  \left(\frac{k_\Gamma}{k}\right)^2 
     B_{22}\left(p,x_E\right)\right] \left[ 1 -
     2  \left(\frac{k_\Gamma}{k}\right)^2 
     B_{11}\left(p,x_E\right) \right] \Big\} \, , 
\end{split} \\
D_{\sigma} & = 4 \left( \frac{k_{\Gamma}}{k} \right)^4 \left[ A_{11} A_{22} - A_{12}^2 \right] \,.
\label{eq:relation_coeff_purity}
\end{align}

\subsection{Squeezing parameter $r_k$}
\label{app:squeezing_parameter}

To derive an expansion for $r_k$, we start from Eq.~\eqref{eq:rk_exact} and express the inverse hyperbolic cosine as a logarithm. We have
\begin{align}
\begin{split}
\label{eq:rk_comp_eff}
r_k & = \frac{1}{2} \ln \left[ \frac{\gamma_{22}+ \gamma_{11}}{ 2 p_k^{-1/2}} + \sqrt{\left( \frac{\gamma_{22}+ \gamma_{11}}{ 2 p_k^{-1/2}} \right)^2 -1 }\right] \, , \\
& =  \frac{1}{2} \ln \left( \frac{ \gamma_{22}}{p_k^{-1/2}} \right)  +  \ln \left[ \frac{1}{2} + \frac{\gamma_{11}}{2 \gamma_{22}}   + \frac{1}{2} \left( 1 - \frac{\gamma_{11}}{\gamma_{22}} \right) \sqrt{1 + \frac{4 \gamma_{12}^2}{\gamma_{22}^2} \frac{1}{\left( 1 - \frac{\gamma_{11}}{\gamma_{22}} \right)^2}} \right] \, ,
\end{split}
\end{align}
where we used the expression of the purity in terms of the covariance-matrix elements, Eq.~\eqref{eq:purity_determinant}. Rewriting $r_k$ in the form of Eq.~\eqref{eq:rk_comp_eff} isolates the large contribution in the first term, while the second term only depends on small ratios. Using that $\gamma_{12}/\gamma_{22} = \mathcal{O}(x)$ and $\gamma_{11}/\gamma_{22} = \mathcal{O}(x^{2})$, we can expand the argument of the second term and obtain
\begin{equation}
    \frac{1}{2} + \frac{\gamma_{11}}{2 \gamma_{22}}   + \frac{1}{2} \left( 1 - \frac{\gamma_{11}}{\gamma_{22}} \right) \sqrt{1 + \frac{4 \gamma_{12}^2}{\gamma_{22}^2} \frac{1}{\left( 1 - \frac{\gamma_{11}}{\gamma_{22}} \right)^2}} = 1 + \left( \frac{\gamma_{12}}{\gamma_{22}} \right)^2 + \mathcal{O} \left( x^{4} \right) \, .
\end{equation}
Inserting this expansion into Eq.~\eqref{eq:rk_comp_eff}, we recover Eq.~\eqref{eq:latetime_rk}. This expression is the most useful for numerical approximation but is unfortunately not very explicit, since we still have to compare the behaviours of $\gamma_{22}^2$ and $p_k^{-1}$. To obtain a more explicit formula, we need to use the expansions of $\gamma_{22}$ and $p_k^{-1}$ for small $x$ and expand the first term, which gives
\begin{align}
\begin{split}
r_k & = - 2  \ln \left( x \right)  \\
& \qquad + \frac{1}{2} \ln \left\{ \frac{  1
  - 2  \left(\frac{k_\Gamma}{k}\right)^2
  B_{22}  + \mathcal{O} \left(x^2\right) - 2 \left(\frac{k_\Gamma}{k}\right)^2
  A_{22} x^{8-p} + \mathcal{O} \left(x^{10-p}\right) }{ \sqrt{1 + A_{\sigma} + B_{\sigma} x^{2-p} + \mathcal{O} \left(x^{4-p}\right) + D_{\sigma} x^{10-2p } + \mathcal{O} \left(x^{12-2p}\right) } } \right\} \, ,
\end{split}
\end{align}
where the dominant term inside the brackets depends on the value of $p$. Note that the $\mathcal{O} \left(x^2\right)$ term arising from the expansion of the second term has been reabsorbed into the logarithm. First, for $p<2$, the denominator corresponding to $p_k^{-1}$ tends to a non-vanishing constant, and so does the numerator. Second, for $2<p<8$, the numerator still tends to a constant, while the denominator diverges and the term in $x^{2-p}$ dominates. Third, for $p>8$, both the numerator and the denominator diverge: the term in $x^{8-p}$ dominates in the numerator, and that in $x^{10-2p}$ dominates in the denominator. The resulting expressions are given in Eq.~\eqref{eq:cases_rk}.

\subsection{Squeezing angle $\varphi_k$}
\label{app:squeezing_angle}

Similarly to our approach for $r_k$, we use the hierarchy between the values of the $\gamma_{ij}$ to isolate the small contributions to the asymptotic values of the cosine and sine. In this case, we only need to note that for $x \ll 1$, $\gamma_{22} > \gamma_{11}$, so that we can factor out terms under the square root in Eq.~\eqref{eq:sin_cos_varphik_exact} to obtain
\begin{subequations}
\begin{align}
\sin \left( 2 \varphi_k \right) & = \frac{\tan \left( 2 \varphi_k \right)}{\sqrt{1+ \tan^2 \left( 2 \varphi_k \right)}} \, , \\
\cos \left( 2 \varphi_k \right)  & =  \frac{1}{\sqrt{1+ \tan^2 \left( 2 \varphi_k \right)}} \, ,
\end{align}
\end{subequations}
with the tangent expansion
\begin{align}
\begin{split}
\tan \left( 2 \varphi_k \right) & = - 2 \frac{\gamma_{12}}{\gamma_{22}} \left[ 1 - \frac{\gamma_{11}}{\gamma_{22}} \right]^{-1}  \, , \\
& =  - 2 \frac{\gamma_{12}}{\gamma_{22}} + \mathcal{O} \left( x^3 \right) \, ,
\end{split}
\end{align}
where we have only used $\gamma_{12}/\gamma_{22} = \mathcal{O}(x)$ and $\gamma_{11}/\gamma_{22} = \mathcal{O}(x^2)$. Inserting this into the expressions for the cosine and sine, we recover Eq.~\eqref{eq:latetime_sin_cos_2phik}. Finally, taking the $\arctan$ of the expression for the tangent, we obtain
\begin{equation}
 \varphi_k  = - x \frac{\gamma_{12}}{\gamma_{22}} + \mathcal{O} \left( x^3 \right) \, .
\end{equation}
We now have to use Eqs.~\eqref{eq:latetime_gamma12} and \eqref{eq:latetime_gamma22}, and distinguish cases in $p$ in order to obtain an explicit expression. If $p<8$, the constant terms in $\gamma_{12}$ and $\gamma_{22}$ dominate in the numerator and denominator; otherwise, the term in $x^{8-p}$ dominates. In both cases, the fraction tends to a constant, and we recover Eq.~\eqref{eq:cases_phik}. The same procedure leads to Eqs.~\eqref{eq:cases_cos_2phik} and \eqref{eq:cases_sin_2phik}. Note that once the expression for the sine is known, that for the cosine can be derived directly using $\cos^2 ( 2 \varphi_k ) + \sin^2 ( 2 \varphi_k ) = 1$.

\subsection{Domain of validity case-distinguishing formulas}
\label{app:validity_cases}

Finally, we want to comment on the range of validity of the approximations in which we distinguish different values of $p$, such as in Eq.~\eqref{eq:cases_rk}. These approximations involve selecting the dominant term in the expansion solely based on its scaling with $x$, without checking the value of the coefficient in front. However, the coefficient of the dominant term in $x$ can turn out to be so suppressed that the term only becomes dominant after a very large number of $e$-folds, larger than the $N=60$ $e$-folds considered here. 

For instance, let us consider the expansion of the purity in Eq.~\eqref{eq:latetime_purity}. We have to compare the magnitude of the term in $x^{2-p}$ with that in $x^{10-2p}$. Their ratio is
\begin{equation}
\label{def:ratio_asymptot_validity}
\frac{B_{\sigma} x^{2-p}}{D_{\sigma} x^{10-2p }} = \left[ \frac{B_{\sigma}^{(1)}}{D_{\sigma}^{(2)}} \left( \frac{k_{\Gamma}}{k}\right)^{-2} + \frac{B_{\sigma}^{(2)}}{D_{\sigma}^{(2)}} \right] x^{8-p } \, .
\end{equation}
Assuming the coefficients $B_{\sigma}$ and $D_{\sigma}$ to be of order unity, the term in $x^{10-2p}$ thus dominates only when both $x^{8-p} \ll 1$ and $\left( k_{\Gamma}/k \right)^{-2} x^{8-p} \ll 1$. Hence, for very small values of the interaction with the environment, $k_{\Gamma}/k\ll 1$, the quantity $\left( k_{\Gamma}/k \right)^{-2}x^{8-p}$ can still be large, and the term in $x^{2-p}$ can remain dominant. As an illustration, consider $p=8.1$ and $k_{\Gamma}/k=10^{-2}$. Then, even after $N=60$ $e$-folds of inflation, the ratio in Eq.~\eqref{def:ratio_asymptot_validity} is of order unity, since $\left( k_{\Gamma}/k \right)^{-2} x^{8-p} \approx 5$ and the approximation in Eq.~\eqref{eq:cases_rk} is thus still inaccurate for $N=60$. To obtain a better approximation, we should keep the full expression given in Eq.~\eqref{eq:latetime_rk}. The same type of reasoning applies to the other formulae where we distinguished different values of $p$, such as Eqs.~\eqref{eq:cases_sin_cos_2phik} for the squeezing angle, or Eq.~\eqref{eq:cases_discord} for the quantum discord.

\section{Useful approximations of quantum correlations measures in presence of decoherence}
\label{app:approx_q_measures}

In this Appendix,  we derive useful approximations of the quantum discord given in Eq.~\eqref{eq:discord_p_r} and the Bell operator expectation value given in Eq.~\eqref{eq:Bell_ev}. 

\subsection{Discord for small purity and large squeezing}

We start by deriving an expression for the discord in the large-squeezing and small-purity regime in which cosmological perturbations are typically found. Following~\cite{Martin:2022kph}, we first rewrite the discord in terms of $a_k = p_k^{-1/4} e^{r_k}$, the semi-major axis, and $b_k = p_k^{-1/4} e^{-r_k}$, the semi-minor axis lengths:
\begin{equation}
\label{eq:discord_semi_axes}
\mathcal{D}_{\pm \bm{k} } = f\left[ \frac{a_k^2+b_k^2}{2}\right] - 2 f\left(a_k b_k\right) + f\left(1 + 2 b_k^2 \frac{1 - \frac{1}{a_k^2 b_k^2}}{1 + \frac{b_{\bm k}^2}{a_{\bm k}^2} + \frac{2}{a_k^2}}\right) \, .
\end{equation}
Next, we want to expand the discord in the small-squeezing $b_k/a_k = e^{-2 r_k} \ll 1$ and small-purity $(a_k b_k)^{-1} = \sqrt{p_k} \ll 1$ limits. Note that, while in this case $a_k \gg 1$, the asymptotic behaviour of $b_k$ is not fixed. This is a hint that $b_k$ will be the relevant parameter controlling the value of the discord. Since its value is a ratio involving the squeezing parameter $r_k$ and the purity $p_k$, whether it is large or small depends on the relative magnitude of squeezing and decoherence.  Before expanding, we rearrange the terms in the discord to single out contributions that can be large, and write the remainder as combinations of small quantities. This gives Eq.~\eqref{eq:discord_expansion_ready}, which involves both the functions $f$, defined in Eq.~\eqref{def:function_f}, and $g$, defined in Eq.~\eqref{def:function_g}. Their asymptotic behaviours are given by
\begin{subequations}
\label{eq:asymptotic_g}
\begin{alignat}{1}
g(x) & \underset{x \to 1}{=} 1 - \frac{1}{\ln 2} + \mathcal{O} \left[(x-1) \ln(x-1)\right] \, , \\
g(x) & \underset{x \to + \infty}{=} - \frac{1}{6 \ln 2} \frac{1}{x^2} + \mathcal{O} \left( \frac{1}{x^4} \right) \, ,
\end{alignat}
\end{subequations}
and
\begin{subequations}
\label{eq:asymptotic_f}
\begin{alignat}{1}
f(x) & \underset{x \to 1}{=} - \frac{1}{2} (x-1) \log_2(x-1) + \mathcal{O} \left(x-1\right) \, , \\
f(x) & \underset{x \to + \infty}{=} \log_{2}(x) + \frac{1}{\ln 2} - 1 + \mathcal{O} \left( \frac{1}{x^2} \right) \, .
\end{alignat}
\end{subequations}
Using these approximations, at first order in $p_k$ and $e^{-2 r_k}$, we obtain Eq.~\eqref{eq:discord_semi_minor_squeezing}. Finally, we expand the function $g$ in the small- and large-$b_k$ cases. First, for $b_k \ll 1$, the largest small contribution is given by the terms in $p_k$, since $e^{-4 r_k} = b_k^2 a_k^{-2} \ll a_k^{-2} = p_k b_k^{2} \ll p_k$. We thus have
\begin{equation}
\mathcal{D}_{\pm \bm{k} } = - 2 \log_2 b_k - \frac{1}{\ln 2} + \mathcal{O} \left( b_k^2 \ln b_k \right) + \mathcal{O} \left( p_k \right) \, , \qquad \mathrm{for} \, \,\,b_k \ll 1 \, .
\end{equation}
The case $b_k \gg 1$ is more subtle, because the logarithmic term then gives a contribution proportional to $b_k^{-2}$ that is small and need not be dominant compared to the other small terms in $p_k$ and $e^{-4 r_k}$. First, reversing the inequalities of the previous case, we find the ordering of the small contributions
\begin{equation}
e^{-4 r_k} = b_k^2 a_k^{-2} \gg a_k^{-2} = p_k b_k^{-2} \gg p_k \, .
\end{equation}
Second, we can check that $b_k^{-2} = a_k^{2} p_k \gg p_k$. Third, if the purity is small enough compared to the squeezing, $p_k < e^{-12 r_k}$, then the terms related to the squeezing parameter dominant over that in $b_k$, $b_k^{-2} < e^{-4 r_k}$. However, in this expansion the terms proportional to $e^{-2 r_k}$ are always further suppressed by powers of $p_k$, which is smaller than $b_k^{-2}$. Thus, the term proportional to $b_k^{-2}$ is indeed dominant. Therefore, we have
\begin{equation}
\mathcal{D}_{\pm \bm{k} } = \frac{b_k^{-2}}{2 \ln 2} + \mathcal{O} \left( b_k^{-4} \right) \, , \qquad \mathrm{for} \,\,\, b_k \gg 1 \, .
\end{equation}
Combining both cases gives Eq.~\eqref{eq:discord_approx_semi_minor}.

\subsection{Discord at late-time}

We derive a late-time approximation of the quantum discord. For $p>2$, the purity goes to zero and we can directly use Eqs.~\eqref{eq:discord_semi_minor_squeezing} and \eqref{eq:latetime_bk}. However, for $p<2$, we must re-expand all three evaluations of the function $f$ in Eq.~\eqref{eq:discord_p_r}. The first term depends on the asymptotic behaviour of $\sigma(p_k,r_k)$. The second term depends on that of $p_k$, which we already computed in Eq.~\eqref{eq:latetime_purity}. The third depends on $p_k\,\sigma(p_k,r_k)$. We start with $\sigma(p_k,r_k)$, which can be rewritten as
\begin{equation}
\sigma(p_k,r_k) = \frac{a_k^2}{2} \left( 1 + e^{-4 r_k} \right) \, .
\end{equation}
Since the squeezing is always positive asymptotically, the asymptotic behaviour of $\sigma(p_k,r_k)$ is exponentially close to that of $a_k$. First, using Eq.~\eqref{eq:latetime_rk}, we obtain a very simple expansion for $a_k$ that does not involve the value of the purity $p_k$:
\begin{align}
\begin{split}
\label{eq:latetime_ak}
\ln a_k & = \frac{1}{2} \ln \left( \gamma_{22} \right) + \left( \frac{\gamma_{12}}{\gamma_{22}} \right)^2 + \mathcal{O} \left(x^4\right) \, .
\end{split}
\end{align}
One can check that the asymptotic expressions for $a_k$, Eq.~\eqref{eq:latetime_ak}, and $b_k$, Eq.~\eqref{eq:latetime_bk}, satisfy $a_k b_k = \sqrt{p_k}$ and $\ln b_k - \ln a_k = -2 r_k$. Next we obtain an estimate for the small parameter $e^{-4 r_k}$. First, by comparing the late-time expansions of $\gamma_{22}$ in Eq.~\eqref{eq:latetime_gamma22} and of $p_k^{-1}$ in Eq.~\eqref{eq:latetime_purity}, we find that $\gamma_{22}^2 \gg p_k^{-1}$ always holds, and that at most their ratio is of order $x^2$. We thus take the upper bound $p_k^{-1}/\gamma_{22}^2 = \mathcal{O}(x^2)$, which is convenient as it avoids the need to distinguish cases. Taking the exponential of Eq.~\eqref{eq:latetime_rk}, we obtain
\begin{equation}
\label{eq:latetime_expminus4rk}
e^{-4 r_k} = \frac{p_k^{-1}}{\gamma_{22}^2} \left[ 1 + \mathcal{O} \left(x^2\right) \right] = \frac{p_k^{-1}}{\gamma_{22}^2} + \mathcal{O} \left(x^4\right) \, ,
\end{equation}
so that $e^{-4 r_k}$ is at most of order $x^2$. Inserting the two expansions, Eqs.~\eqref{eq:latetime_ak} and \eqref{eq:latetime_expminus4rk}, into the expression of $\sigma(p_k,r_k)$ in terms of $a_k$ and $e^{-4 r_k}$, we obtain
\begin{equation}
\label{eq:latetime_sigmak}
\begin{alignedat}{1}
\ln \sigma(p_k,r_k) & = 2 \ln a_k - \ln 2 + e^{-4 r_k} + \mathcal{O} \left( e^{-8 r_k} \right) \, , \\
& = \ln \gamma_{22} - \ln 2 + 2 \left( \frac{\gamma_{12}}{\gamma_{22}} \right)^2 + e^{-4 r_k} + \mathcal{O} \left(x^4\right) \, ,
\end{alignedat}
\end{equation}
where we used Eq.~\eqref{eq:latetime_rk} and the fact that $e^{-4 r_k} = \mathcal{O}(x^2)$. We then have
\begin{equation}
\begin{alignedat}{1}
f \left[ \sigma(p_k,r_k) \right] & = \log_2 \sigma(p_k,r_k) + \frac{1}{\ln 2} - 1 + \mathcal{O} \left[ \sigma(p_k,r_k)^{-2} \right] \, , \\
& = \log_2 \gamma_{22} + \frac{1}{\ln 2} - 2 + \mathcal{O} \left(x^2\right) \, .
\end{alignedat}
\end{equation}
The term proportional to $\sigma^{-2}$ is of order $\gamma_{22}^{-2}$, which is always at most $\mathcal{O}(x^8)$ and thus negligible. We now move to the third term. The argument of $f$ in the third term reads
\begin{equation}
\begin{alignedat}{1}
\frac{\sigma\left(p_k,r_k\right)+p_k^{-1}}{\sigma\left(p_k,r_k\right)+1} & = 1 + \frac{p_k^{-1} - 1}{\sigma\left(p_k,r_k\right)} \frac{1}{1+\sigma\left(p_k,r_k\right)^{-1}} \, , \\
& = 1 + \frac{2 \left( p_k^{-1} - 1 \right)}{\gamma_{22}} \left[ 1 + \mathcal{O} \left(x^2\right) \right] \, ,
\end{alignedat}
\end{equation}
where we used the fact that $\sigma\left(p_k,r_k\right)^{-1}$ is at least of order $x^4$ to neglect the dominant contribution of $1/[1+\sigma\left(p_k,r_k\right)^{-1}]$ compared to the sub-dominant ones coming from $( p_k^{-1} - 1)/\sigma\left(p_k,r_k\right)$. Since the argument tends to $1$ for $p<6$ and to infinity for $p>6$, we must now distinguish cases in order to expand $f$. We find
\begin{equation}
\begin{alignedat}{3}
f \left[ \frac{\sigma\left(p_k,r_k\right)+p_k^{-1}}{\sigma\left(p_k,r_k\right)+1} \right] & = - \frac{p_k^{-1} - 1}{\gamma_{22}} \log_2 \left[ \frac{p_k^{-1} - 1}{\gamma_{22}} \right] + \mathcal{O} \left( \frac{p_k^{-1} - 1}{\gamma_{22}} \right) \quad & \mathrm{for} & \quad p<6 \, , \\
& = \log_2 \left( \frac{p_k^{-1} - 1}{\gamma_{22}} \right) + \frac{1}{\ln 2} + \mathcal{O} \left(x^2\right) + \mathcal{O} \left[ \left( \frac{p_k^{-1}}{\gamma_{22}} \right)^{-1} \right] \quad & \mathrm{for} & \quad p>6 \, ,
\end{alignedat}
\end{equation}
where we used that for $p>6$, $p_k$ is at least of order $\mathcal{O}(x^4)$. Finally, combining all terms, we obtain the following expression for the discord:
\begin{equation}
\label{eq:cases_discord_bis}
\begin{alignedat}{3}
\mathcal{D}_{\pm \bm{k} } & = - \log_2 \left( \gamma_{22}^{-1} \right) + \frac{1}{\ln 2} - 2 - 2 f \left( p_k^{-1/2} \right) + \mathcal{O} \left(x^2\right) \quad & \mathrm{for} & \quad p<2 \, , \\
& = - \log_2 \left( \frac{p_k^{-1}}{\gamma_{22}} \right) - \frac{1}{\ln 2} + \mathcal{O} \left(p_k\right) + \mathcal{O} \left(x^2\right) + \mathcal{O} \left[ \frac{p_k^{-1}}{\gamma_{22}} \log_2 \left( \frac{p_k^{-1}}{\gamma_{22}} \right) \right] \quad & \mathrm{for} & \quad 2<p<6 \, , \\
& = \mathcal{O} \left(p_k\right) + \mathcal{O} \left(x^2\right) + \mathcal{O} \left[ \left( \frac{p_k^{-1}}{\gamma_{22}} \right)^{-1} \right] \quad & \mathrm{for} & \quad p>6 \, .
\end{alignedat}
\end{equation}
Here we used the fact that for $p<2$ we have $(p_k^{-1}-1)/\gamma_{22} = \mathcal{O}(x^4)$. Equation~\eqref{eq:cases_discord_bis} gives the expression of the discord for $p<2$, where we cannot use Eq.~\eqref{eq:discord_approx_semi_minor}, which requires the purity to be small. For $2<p<6$, one can check using Eq.~\eqref{eq:latetime_bk} that this expression matches Eq.~\eqref{eq:discord_approx_semi_minor}. On the other hand, for $p>6$ the leading terms cancel and no estimate is obtained. Fortunately, for $p>6$ we can use Eq.~\eqref{eq:discord_approx_semi_minor}. We summarise the late-time approximation of the discord for arbitrary $p$ in Eq.~\eqref{eq:cases_discord}.

\subsection{Bell operator at late-time}

We expand the expectation value of the Bell operator given in Eq.~\eqref{eq:Bell_ev}. The first term is given by $p_k$, and the second contains $\tanh \left( 2 r_k \right)\lvert \cos \left( 2 \varphi_k \right) \rvert$, which we must expand. First, we expand the hyperbolic tangent using Eq.~\eqref{eq:latetime_expminus4rk} and obtain
\begin{align}
\begin{split}
\tanh \left( 2 r_k \right) & = 1 - 2 e^{-4 r_k} + \mathcal{O} \left( e^{-8 r_k} \right) \, , \\
& = 1 - 2 \frac{p_k^{-1}}{\gamma_{22}^2} + \mathcal{O} \left( x^4 \right) \, .
\end{split}
\end{align}
Since asymptotically $\cos \left( 2 \varphi_k \right) > 0$, we can remove the absolute value and use Eq.~\eqref{eq:latetime_sin_cos_2phik}. We then obtain
\begin{align}
\begin{split}
\label{eq:latetime_tanhcos}
\lvert \cos \left( 2 \varphi_k \right) \rvert \tanh \left( 2 r_k \right) & = 1 - 2 \frac{\gamma_{11}}{\gamma_{22}} + \mathcal{O} \left( x^4 \right) \, ,
\end{split}
\end{align}
where we used the expression of $p_k^{-1}$ in terms of the covariance-matrix elements in Eq.~\eqref{eq:purity_determinant}. Expanding the arcsine close to $1$, we obtain
\begin{equation}
\frac{4}{\pi^2} \arcsin^2 y = 1 - \frac{4 \sqrt{2}}{\pi} \sqrt{y - 1} + \mathcal{O} \left( y - 1 \right) \, .
\end{equation}
Combining this with the expansion~\eqref{eq:latetime_tanhcos}, we obtain
\begin{equation}
\frac{4}{\pi^2} \arcsin^2 \left[ \lvert \cos \left( 2 \varphi_k \right) \rvert \tanh \left( 2 r_k \right) \right] = 1 - \frac{8}{\pi} \sqrt{ \frac{\gamma_{11}}{\gamma_{22}} } + \mathcal{O} \left( x^3 \right) \, ,
\end{equation}
which, when inserted into the expression~\eqref{eq:Bell_ev} for the expectation value, gives Eq.~\eqref{eq:latetime_Bell}. Further expanding the purity finally yields Eq.~\eqref{eq:cases_Bell}.

\clearpage 
\bibliographystyle{JHEP}
\bibliography{refs}

\end{document}